# Less is more: error–distance scaling relation for data-efficient kilometer-scale downscaling of extreme heat

Ahmed Marey[1,2], Henry Lu[2], Abhishek Gaur[2], Sherif Goubran[3], Malek Aloui[1], Theodore Potsis[1], David Rolnick[4,5], Alex Hernandez-Garcia[4,6], Liangzhu Leon Wang[1,*],

[1] Centre for Zero Energy Building Studies, Department of Building, Civil and Environmental Engineering, Concordia University, Montreal, H3G 1M8 Canada

[2] Building and Climate Interface, Construction Research Centre, National Research Council Canada, Ottawa, ON, K1A 0R6, Canada

[3] Department of Architecture, School of Sciences and Engineering, The American University in Cairo, New Cairo 11835, Egypt

[4] Mila – Quebec Artificial Intelligence Institute, Montreal, QC, H2S 3H1, Canada

[5] School of Computer Science, McGill University, Montreal, QC, H3A 0G4, Canada

[6] Department of Computer Science and Operations Research, Université de Montréal, Montreal, QC, H3C 3J7, Canada

[*] Author to whom correspondence should be addressed: leon.wang@concordia.ca

## Abstract

Extreme heat is where urban adaptation needs kilometer-scale data the most, but the simulations training a downscaler can cost more than they save, and how much is needed has not been identified. We measured it with CASPER, a U-Net with a structure-preserving loss downscaling 32 km reanalysis to 1 km temperature, humidity and wind, across 24 configurations of one to eight months. Held-out error grows linearly with climatological distance to the training data, RMSE = 0.83 + 2.95 d, explaining 90% of its variance against 7% for volume and predicting unseen months in advance. On held-out extreme summer weeks CASPER preserves the fine-scale structure and cross-variable physics that matched-budget baselines degrade, and matches station observations during documented heat waves to within 1.8 K. Transfer to a new region degrades geographically; 11 days of local simulation cuts Vancouver's held-out error from 3.8 to 1.3 K. Training periods should span the target climate: the same accuracy for four times less simulation, putting kilometer-scale downscaling of extreme heat within reach of groups without large computing facilities.

## Introduction

Kilometer-scale atmospheric data is indispensable for urban climate adaptation [1], hydrological forecasting [2] and predicting extreme weather events [3]. The demand is highest for extreme heat, whose impacts fall on particular hours and locations within a city and are invisible in coarse-resolution fields; heat waves are the case this work is built around, with typical summer conditions kept as the control. Climate-resilient infrastructure must also be evaluated against detailed long-term projections spanning divergent warming pathways [4]. Dynamical downscaling with regional climate models such as the Weather Research and Forecasting (WRF) model resolves the interactions among topography, the land surface and atmospheric dynamics by solving the governing equations of atmospheric motion at high resolution [5], but the computational overhead is prohibitive [6]: producing a single month of training data for this study's domain required 5 days of computation on a supercomputer with 480 CPU cores, which puts large-scale scenario analysis and ensemble projections for uncertainty quantification out of reach [7].

Statistical downscaling infers empirical relationships between coarse-scale atmospheric patterns and local weather conditions [8]. Quantile mapping [9] and empirical regression [10] cannot capture the nonlinear dynamics and complex spatial dependencies of high-resolution atmospheric phenomena [11], and fail during extreme events, when linear assumptions break down and spatial structure dominates local impacts [12]. Machine learning has narrowed that gap [13], [14]: random forests [15], convolutional networks [16] and, in particular, U-Net architectures learn nonlinear mappings with a much better accuracy [17], while generative adversarial networks [18], [19] and diffusion models [20], [21] produce sharp, realistic fine-scale variability. Each family carries a cost. Adversarial training is prone to mode collapse, which often limits accurate representation of the distributional tails that define extreme events [22], [23].

Diffusion models train more stably but consume more data, and they treat bias correction as a separate stage: CorrDiff [24] adds a diffusion head to a deterministic regression backbone, and TAUDiff [25] applies quantile mapping as input preprocessing, rather than enforcing distributional agreement during training. Both generative families need large, diverse training sets – a conditional Wasserstein GAN for operational wind downscaling over Canada required a full year of paired forecasts [26]. The binding constraint is therefore not architecture but the generation of high-resolution training data, and it is amplified at kilometer scale. Published models are trained on continuous records of three to fifty years [6], [24], [27], and no public pre-computed paired dataset exists at 1 km, unlike coarser resolutions (10–25 km) at which reanalysis or global model output can serve as both input and target. Every group must generate its own simulations from scratch – a substantial barrier for meteorological services, especially in developing nations [28], for municipal governments and for groups without high-performance computing. Table 1 surveys what is reported: records from one year to nearly three decades, and no study reporting how skill would change if the record were shortened, which periods carry the information, or how much simulation a new deployment would need.

**Table 1: Training-data records reported in deep-learning downscaling.** High-resolution training record used by representative deep-learning downscaling studies, ordered by target resolution; the ratio is the linear resolution enhancement from the coarse input to the target grid. The final column records whether the study reports a data-volume sensitivity analysis – an explicit measurement of how skill depends on the amount, or the choice, of training data. None of the surveyed studies reports one: the length of the training record is stated but not justified.

| Study | Resolution (ratio) | Domain | High-resolution training record | Model family | Data requirement measured |
|---|---|---|---|---|---|
| Mardani et al. [24] | 25 → 2 km (12×) | Taiwan | 3 years hourly (24,154 fields, 2018-2020) | U-Net + corrective diffusion | No |
| Tomasi et al. [42] | 16 → 2.2 km (8×) | Italy | 15 years hourly (of a 2000-2020 archive) | Latent diffusion | No |
| Guevara et al. [26] | 20 → 2.5 km (8×) | Canada | 12 months of paired forecasts (7,150 forecast hours) | Conditional WGAN-GP | No |
| Pérez et al. [52] | 25 → 5.5 km (4.5×) | Europe | 29 years 3-hourly (1985-2013) | Swin transformer | No |
| Jha et al. [27] | 250 → 25 km (10×) | South Asia | 31 years | Residual CNN | No |
| Singh et al. [30] | 10 km → 300 m (33×) | Austin, USA | 9 years daily (2001-2009) | Iterative SRCNN | No |
| Chajaei and Bagheri [29] | 100 → 5 m (20×) | Amsterdam | 1 year (2017) | Gradient boosting | No |
| **This work (CASPER)** | **32 → 1 km (32×)** | **Montreal-Ottawa, +3 transfer domains** | **1-8 months; 1 month + 11 days for a deployed regional model** | **U-Net + structure-preserving loss** | **Yes** |

Four questions follow, and they are the questions any group planning a kilometer-scale simulation campaign for extreme heat has to answer before it starts. How much simulation is enough? How should the simulation periods be selected? How well the resulting model will generalize to conditions and places it has not seen, and can that be known before the computing time is spent? And how much additional simulation does it cost to move the model to a new region? Answering them requires treating the training set as a design variable rather than an inheritance, and it requires a downscaler that trains stably on months rather than years, so that the training set can actually be varied.

We therefore build on the U-Net, the backbone of deep-learning downscaling, in a deterministic configuration [29], and address its main weakness – spatially smoothed output – with a structure-preserving loss. We call the resulting framework CASPER (Context-Aware Structural Prior Enhanced Resolution). It takes high-resolution static geographic features – terrain elevation and land use – as primary input channels: the surface boundary conditions that modulate local atmospheric response through orographic forcing, differential heating and turbulence induced by surface roughness [30], [31], [32]. Combined with the multi-component loss, this design lets the model learn location-specific responses, capturing terrain-forced circulations [33] and urban heat island effects [34] that the 32 km NARR forcing leaves unresolved (Figure 1), without hard physical constraints [35].

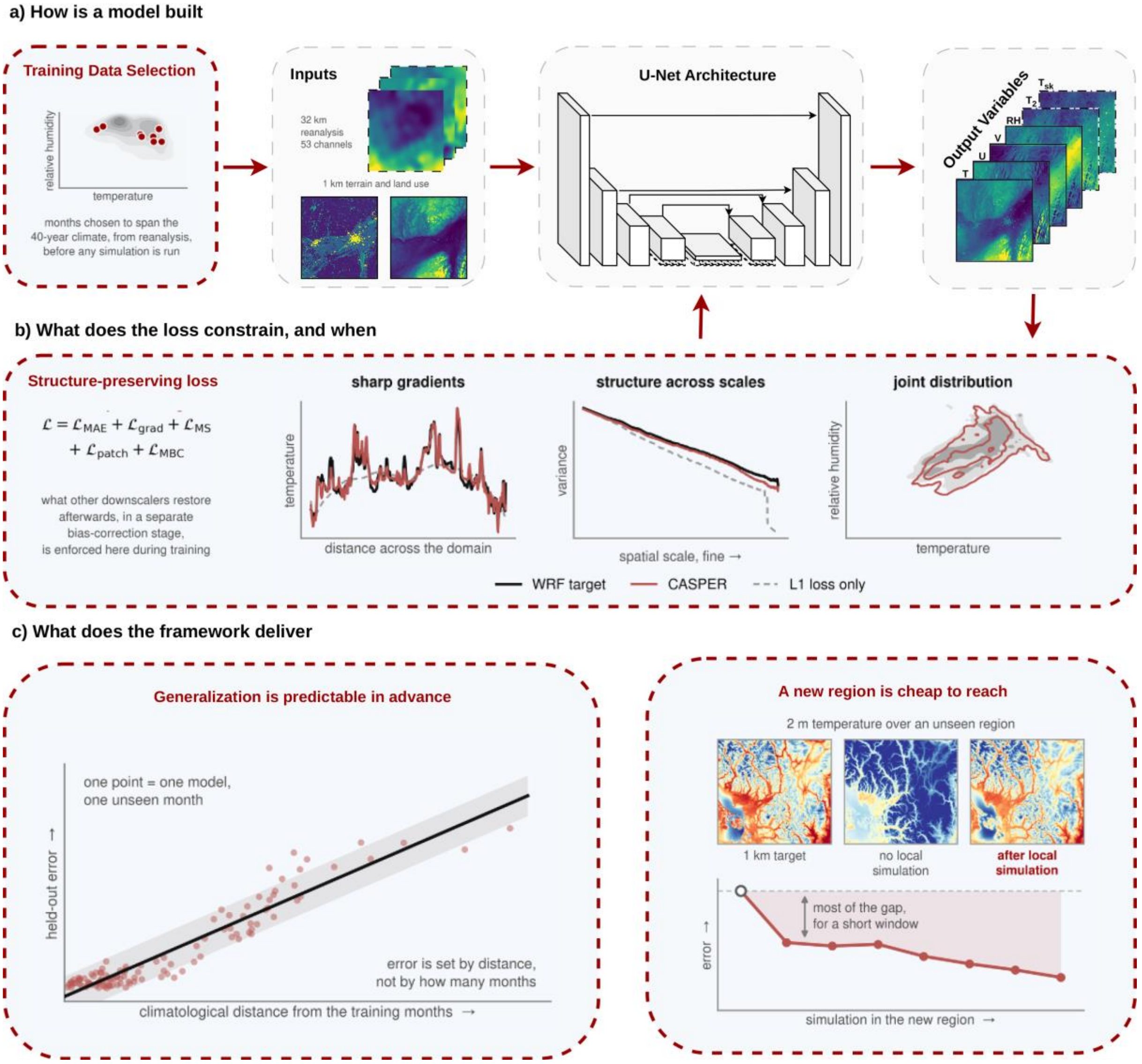

**Figure 1: The CASPER framework.** Panels outlined in maroon are this work's contribution; grey panels are standard components. (a) Candidate simulation periods are placed in climate space and the training months are chosen to span it, before any simulation is run. The network takes the 32 km reanalysis state as 53 coarse channels together with 1 km terrain elevation and land-use class, and returns the 1 km column: temperature, wind and humidity on 10 pressure levels plus 2 m and skin temperature. (b) The backbone is a conventional U-Net; the structure-preserving loss is what is added to it, acting during training rather than in a correction stage afterwards. Its terms hold the prediction to the target in three respects, shown for 2 m temperature: sharp gradients along a transect, variance across spatial scales, and the joint distribution with a second variable. A U-Net trained with an L1 loss alone, shown for contrast, smooths the gradients and loses fine-scale variance. (c) Left, held-out error rises with the climatological distance between an evaluation month and the training distribution, so error is governed by which months were simulated rather than by how many. Right, transfer to an unseen region – the 1 km target, the model applied with no local simulation, and the model after a short local window – with the error against the length of that window directly below them; the open marker is the no-simulation level.

Two design choices carry the framework, and Figure 1 marks them. First, where probabilistic approaches preserve spatial gradients and statistical properties by sampling [36], a structure-preserving loss reaches the same objectives deterministically: alongside the topographic forcing of the static input channels [37], CASPER is trained with a composite of gradient penalties [38], multiscale structural similarity [39], patch coherence [40] and multivariate bias correction [41], so sharp features and the joint distribution are enforced during training rather than restored by a separate post-processing stage, recovering at small budgets the distributional fidelity usually sought through sampling [42]. Second, most downscalers are univariate, which neglects the physical relationships between variables [43], [44] – thermal wind balance [45], Clausius-Clapeyron moisture-temperature coupling [46] – and risks meteorologically inconsistent fields. CASPER downscales the atmospheric column jointly: temperature, both wind components and relative humidity across 10 pressure levels (970-105 hPa), with 2 m and skin temperature, shared representations preserving consistency across variables [47]. Together these let a deterministic model trained on a few months retain the structure and cross-variable physics that make a measurement of the data requirement meaningful.

With the framework fixed, we use CASPER as an instrument rather than a product: because the architecture, loss and normalization never change across configurations, varying the training months alone isolates what each additional month buys. We train on budgets of one to eight months drawn from a 40-year climatology to span the joint distribution of near-surface temperature and relative humidity, and evaluate each configuration on months it never saw; the deployed model is then tested on held-out summer weeks that include the hottest of the 40-year record, and against station observations during documented heat waves. Where previous studies train on multi-year continuous records [24], [48], we ask which months carry the information and how far a model can be pushed before it fails – geographically as well as temporally, applying the Montreal-Ottawa model to Vancouver, Calgary and Toronto.

We answer all four. The first is the central contribution; the others follow from it.

- How much data is enough – a distance, not a number of months. Held-out error is set by the climatological distance between an evaluation target and the training distribution rather than by the volume of training data, and adding months that do not move the training distribution toward the conditions of interest leaves the error where it was. At a matched budget CASPER retains the fine-scale structure and cross-variable physics that interpolation, random-forest, U-Net and adversarial baselines degrade – the property that makes the measurement meaningful.
- How data should be selected – by statistical coverage, not by convention. Candidate simulation periods can be placed in a climate space computed from coarse reanalysis alone, ranked, and chosen to span the conditions of interest before any high-resolution simulation is committed. Temperature errors follow a thermodynamic distance and the winds a nearly orthogonal circulation distance, which turns the choice of months into a measurable design decision rather than a convention.
- How generalization can be measured and predicted. The relation between distance and error is tight enough, and stable enough on months the relation never saw, to predict the error of an unseen climate state within a stated uncertainty. The expected accuracy of a planned simulation campaign can therefore be quoted, with an interval, before the first hour of computing time is spent.
- How regional adaptation can be achieved efficiently. Zero-shot transfer to an unseen region degrades along a geographic axis that the climatological distance does not measure, and a short window of local simulation recovers most of that loss. The total simulation behind a deployed model can therefore be several times smaller than current practice assumes.

These results should be read in two parts. The data-selection framework is architecture-independent: placing candidate simulation periods in a climate space computed from coarse reanalysis, choosing them to span the target conditions, measuring held-out error against climatological distance, and re-measuring the coefficients for a new domain require only coarse input fields and a held-out score. Nothing in the procedure depends on the downscaler being CASPER, and it applies equally to a generative model or to a different pair of resolutions. What is specific to CASPER is the deterministic design and the structure-preserving loss: they set how low the baseline requirement can go, because a deterministic backbone trains stably at budgets where adversarial and score-based objectives do not and enforcing distributional agreement inside the loss removes the post-hoc correction stage generative pipelines need, but they do not change how the requirement is measured. The coefficients are specific too: measured for one domain, one reanalysis and one configuration, so what transfers is the procedure, not the constants.

## Results

### Error scaling with climatological distance

We selected candidate training periods from 40 years (1980–2020) of the North American Regional Reanalysis (NARR) [49], computing the domain-mean 2 m temperature and relative humidity of every three-hourly field over the exact model domain. These two variables define the climate space in which we place every training set and every evaluation target. From that space we drew eight candidate months, two per season, spanning the joint distribution over the whole year (Supplementary Figure S1). Holding architecture, loss function and normalization fixed across every configuration ensures that differences in skill reflect the training data alone. We built the model set on four principles. A budget ladder from one to eight months varies training volume. Stratified compositions at a fixed budget vary which months enter the training set while holding volume constant. Randomly drawn subsets, fixed before we examined any result, guard against post-hoc selection. A targeted ladder withholds both January months while the budget grows from four to six, populating the high-budget, high-distance regime that separates the effect of volume from the effect of climatological distance. In total the set comprises 24 configurations, of which the 23 that each withhold at least one month provide the held-out (model, month) pairs behind the scaling relation; Supplementary Tables S1 and S2 list every configuration with its design arm, month composition and evaluation periods.
Across the model set, held-out error depends far more on which months a model saw than on how many it saw (Figure 2). The clearest case needs no statistics: adding a second July to a July-trained model doubles its training data and leaves its error on January unchanged, at 26.7 and 26.6 K, because the second July widens the training distribution without moving it toward the target. Adding two May months to the two Julys instead reaches 14.3 K on the same target – a 12 K improvement for the same doubling of data.

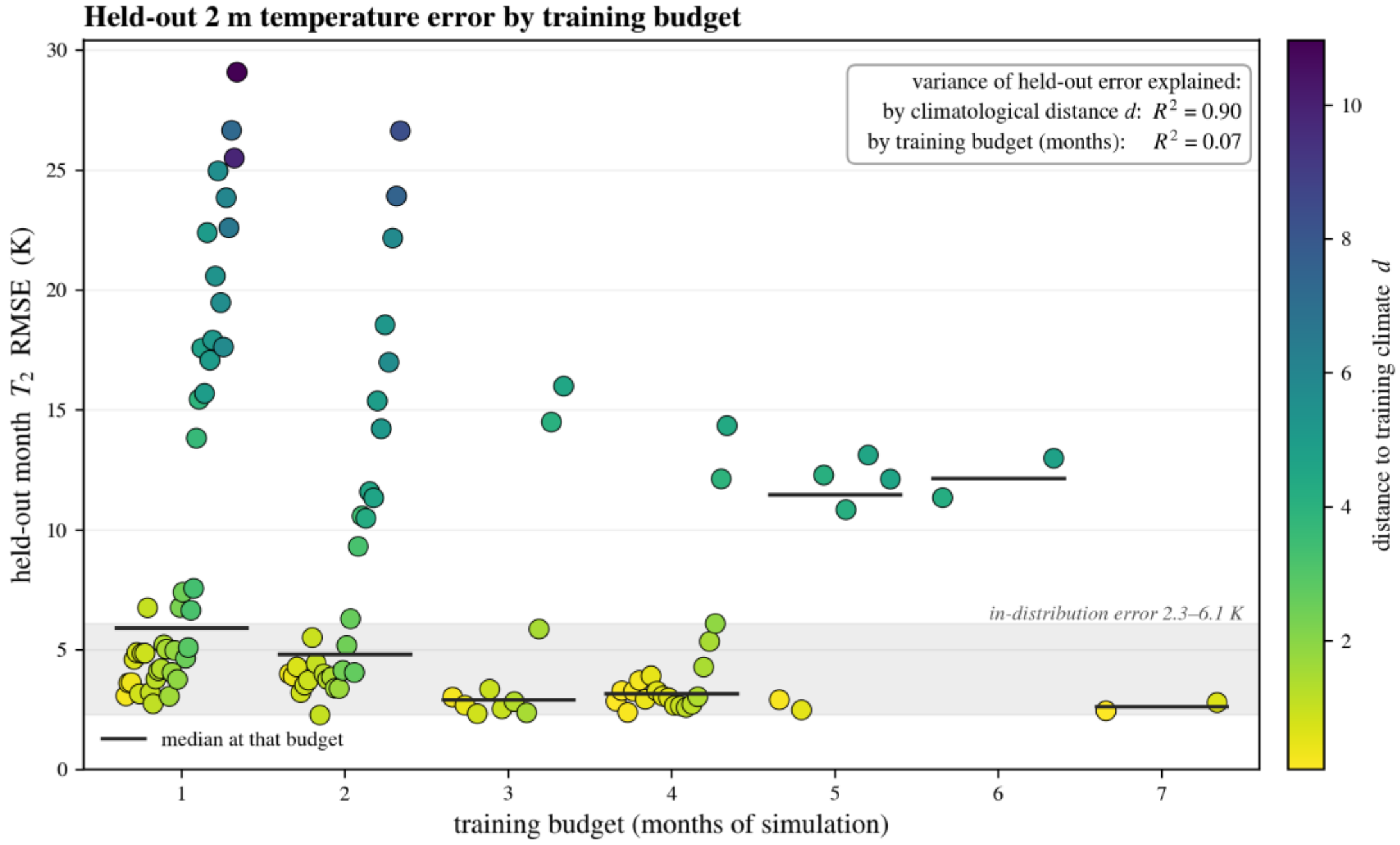

**Figure 2: Held-out error is set by climatological distance, not by training budget.** Each point is one (model, held-out month) pair (n = 112), placed at the model's training budget in months and coloured by the climatological distance between that month and the model's training distribution. Within each budget the points are sorted by distance and spread evenly across the column, so horizontal position encodes distance ordering only and carries no units; the dark tick marks the median at that budget. The same low-to-high distance spread recurs at every budget, so distance, not volume, organizes the error: distance explains 90% of the variance in held-out error against 7% for training budget. Medians at the sparsest budgets (6, 2 and 2 pairs at five, six and seven months) reflect which months were withheld rather than the budget itself. The full variance-explained comparison is given in Supplementary Fig. S2.

We designed the targeted ladder to test this directly. Holding out both January months while the budget grows from four to six months keeps the target far from the training distribution as volume increases. Held-out January error stays near 12 to 13 K across the ladder, 13.2 K at four months, 11.5 to 12.7 K at five and 12.2 K at six, while the distance to the training climate stays near or above four throughout. Adding months that do not move the training distribution toward the target does not reduce the error.
Held-out 2 m temperature RMSE grows linearly with the distance d between the target month's climate and the centre of the model's training distribution (Figure 3a), following RMSE = 0.83 + 2.95 d across 112 model–month pairs from the 23 configurations that each withheld at least one month, with a coefficient of determination of 0.90; a bootstrap over whole evaluation months gives a 90% confidence interval of 2.48 to 3.09 K per unit distance for the slope. Here d is measured in the same two-variable climate space used to select the months: it is the separation between the centre of the evaluation month's temperature–humidity cloud and the centre of the training cloud, with each axis scaled by the standard deviation of the training cloud, so a distance of one is one training-distribution standard deviation. Three cases fix the scale. A four-month model evaluated on the withheld May 1984 sits at d = 0.11 and scores 2.8 K; a six-month model evaluated on a January withheld from it sits at d = 4.7 and scores 13.0 K; a model trained on July 2008 alone and evaluated on that same January sits at d = 11.0 and scores 29.1 K. The larger budget is the worse model whenever the target lies further away. Bilinear interpolation of the coarse field degrades far more gently with distance (median 3.1

K, 5.2 K for the farthest January), so an out-of-coverage neural downscaler is worse than interpolation beyond a distance of about 0.8 and a coverage-selected model is better below it (Supplementary Figure S3): the large errors along the relation are extrapolation collapse, not the intrinsic difficulty of cold months.

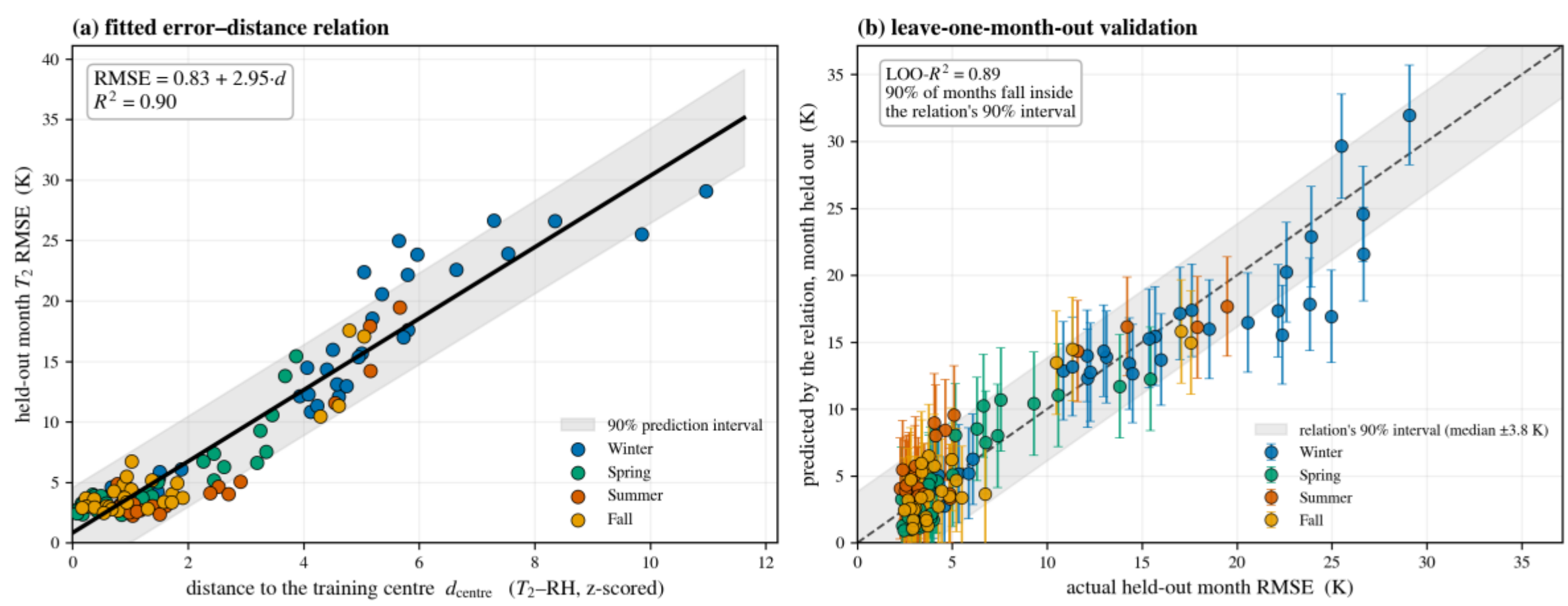

**Figure 3: The error–distance scaling.** a) Held-out month error against distance to the training climate, with the fitted relation and its 90% prediction interval; points are coloured by season. b) Leave-one-month-out validation: the relation is refitted with an entire month withheld and used to predict it, with each point carrying the prediction interval the relation itself quotes.

We then tested whether the relationship predicts as well as it fits. Because the 112 pairs come from only eight distinct climate states – an effective sample size of eight, with about half the held-out-error variance lying between months – we refit the relationship with an entire month withheld and use it to predict that month, repeating for all eight (Figure 3b). It retains a leave-one-month-out coefficient of determination of 0.89 on months it never saw, its 90% prediction interval attains 90% coverage overall, and the slope holds within every season. Its one weakness is the tail: the interval under-covers the January cold-extrapolation regime (78% versus 95% elsewhere, worst held-out error 8 K), so the relation is reliable for interpolation and shoulder-season extrapolation and should be read with an explicit cold-extrapolation caveat. It therefore predicts the error of an unseen climate state and reports how uncertain that prediction is – the property that makes it usable for planning a campaign rather than describing one.

Fitting the same relationship to every output variable separates them into two groups (Supplementary Fig. S4). Temperature errors track the thermodynamic distance closely: 2 m temperature gains 2.95 K per unit distance (coefficient of determination 0.90), skin temperature 2.83 K (0.89) and three-dimensional temperature 1.79 K (0.88). Humidity and wind errors are almost independent of it, with coefficients of 0.07, 0.38 and 0.22. Wind error is not unpredictable, however, but organized by a second axis nearly orthogonal to the first (r = 0.18): adding a circulation distance in domain-mean 544 hPa wind raises the zonal-wind coefficient from 0.38 to 0.72 in sample and from 0.09 to 0.63 under leave-one-month-out cross-validation, with meridional wind improving similarly, while humidity is predicted by neither (Supplementary Figure S5). The coordinate is therefore two-axis – thermodynamic for temperature, circulation for the winds – and the paper's headline relationship is its thermodynamic component.

## Matched-budget validation of the framework

We evaluate CASPER against multiple baselines on the extreme test set – four held-out warm-season weeks: the hottest–driest and hottest–wettest of the 40-year record, together with the two coolest weeks of the warm season that bracket the summer range (Supplementary Table S2) – with typical summer weeks as the control. Importantly, all baseline models were trained on the identical eight-month dataset, isolating the effect of architecture and loss from that of data selection; it is also where the case for a deterministic backbone is tested, since generative models are expected to struggle at such a small budget. Table 2 reports quadratic interpolation, random forest, a U-Net with L1 loss only, a conditional GAN and CASPER across all atmospheric variables on both test sets.

**Table 2: Matched-budget performance comparison.** All methods were trained on the identical eight-month dataset and scored on the same test splits, so differences reflect method rather than training data. The U-Net L1 baseline is architecturally identical to CASPER but trained with an L1 loss only and without the static geographic features. Values are root-mean-square error in physical units on the extreme and typical test sets, for 2 m temperature (T2, K), skin temperature (TSK, K), three-dimensional atmospheric temperature (T, K), relative humidity (RH, %), and the zonal and meridional wind components (U, V, m/s). Lowest error in each column is best; the lowest value per variable and test split is set in bold.

| Model | Test | $T_2(K)$ | $T_{sk}(K)$ | $T\ (K)$ | $RH\ (\%)$ | $U(m/s)$ | $V(m/s)$ |
|---|---|---|---|---|---|---|---|
| Quadratic Interpolation | Extreme | 2.58 | 4.70 | 2.66 | 8.15 | 4.02 | 4.75 |
| | Typical | 2.25 | 4.00 | 2.20 | 7.52 | 3.42 | **3.45** |
| Random Forest | Extreme | 2.69 | 4.26 | 2.58 | 7.18 | 4.04 | 4.80 |
| | Typical | 2.05 | 3.85 | 2.14 | **6.63** | 3.55 | 3.60 |
| U-Net L1 | Extreme | 2.66 | 4.23 | 2.53 | **6.79** | **3.90** | 4.74 |
| | Typical | 2.05 | 3.75 | **2.03** | 6.91 | **3.32** | 3.52 |
| GAN | Extreme | 2.41 | 4.11 | 2.74 | 6.82 | 4.04 | 4.74 |
| | Typical | 1.92 | 3.63 | 2.41 | 7.02 | 3.54 | 3.60 |
| CASPER | Extreme | **2.30** | **3.96** | **2.47** | 6.91 | **3.90** | **4.71** |
| | Typical | **1.86** | **3.51** | 2.16 | 6.76 | 3.35 | 3.46 |

The published downscaling models of Table 1 are typically trained on continuous records of three to fifty years [21], [24], [48], [50], [51]; none of them was retrained on our domain and budget, so we make no parity claim against them. Every baseline in Table 2, by contrast, we trained ourselves on the same eight months as CASPER: quadratic interpolation, random forest, an L1-only U-Net and a conditional GAN; a conditional diffusion baseline trained on those same months did not converge to physically valid fields and is discussed below rather than tabulated. The GAN converged but carries the signature of adversarial training at a small budget – low mean errors alongside spurious high-frequency spectral energy – consistent with the large-data appetite of generative downscalers. Since U-Net L1 optimizes mean errors alone, its RMSE is expected to match or better CASPER's, so spatial accuracy is assessed instead from the temporal mean across all six output variables over the 363×390 km domain (Figure 4 for the extreme test set; the typical test set is in Supplementary Fig. S6 and a mid-tropospheric level of the extreme set in Supplementary Fig. S7). Our results, along with Supplementary Note 3, cover both extreme and typical conditions, and our conclusions are drawn from both.

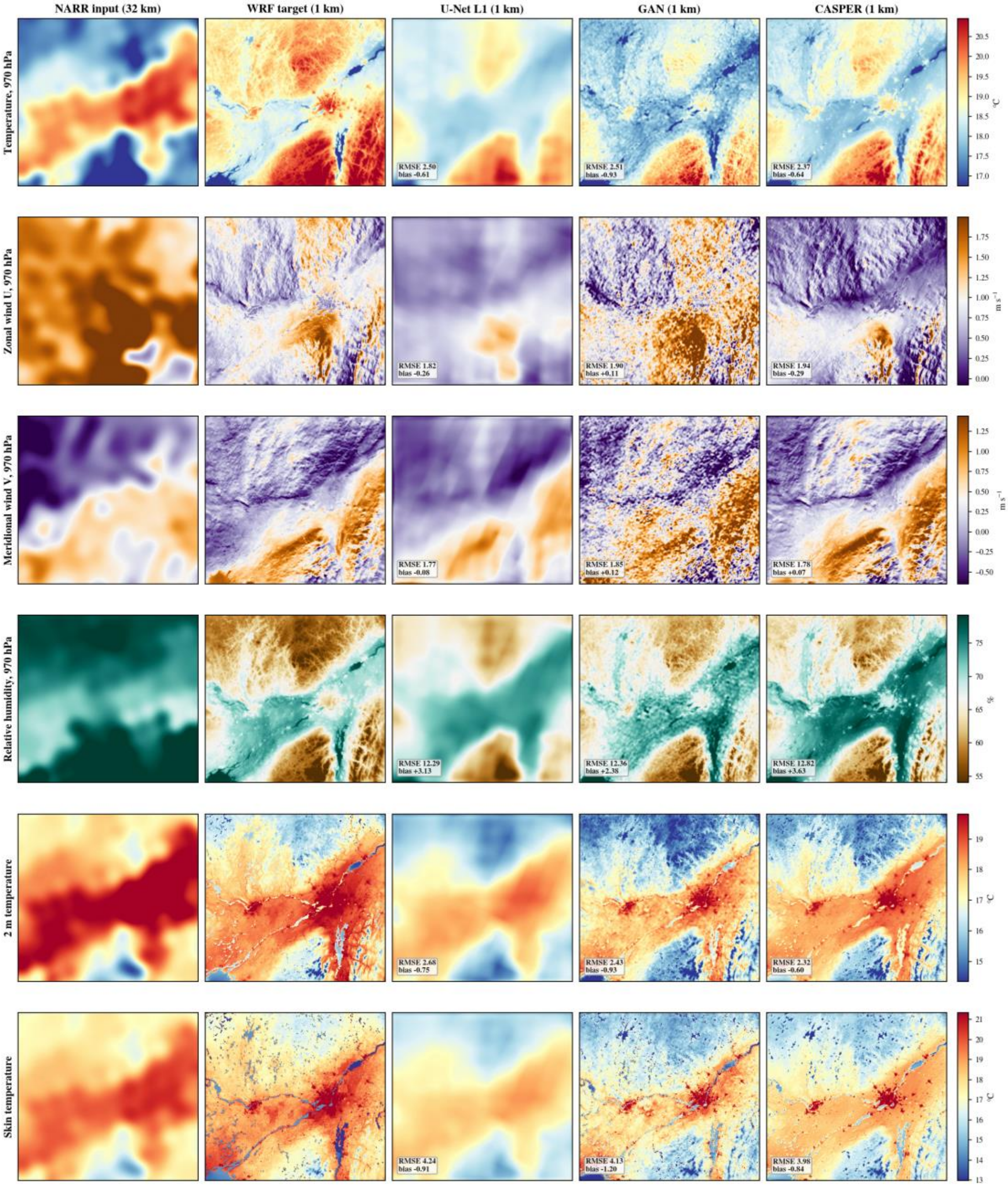

**Figure 4: Spatial prediction maps for the extreme test set (temporal mean over its 670 warm-season hours).** Columns correspond to NARR input, WRF training target, U-Net with L1 loss only, the conditional GAN, and CASPER predictions (left to right). Rows show the six predicted variables – temperature, zonal and meridional wind, and relative humidity at 970 hPa, then 2 m temperature and skin temperature (top to bottom) – with the root-mean-square error and bias of each model against WRF annotated. The 363x390 km domain covers the Montreal-Ottawa region at 1 km resolution. The same comparison for the typical test set is Supplementary Figure S6, and a mid-tropospheric level of the extreme test set is Supplementary Figure S7.

CASPER reproduces fine-scale features including the urban heat islands over Montreal and Ottawa that the L1 U-Net smooths away (a 1-2 K enhancement on the typical test set, Supplementary Figure S6), sharp temperature gradients along the St. Lawrence River Valley, and terrain-driven wind channeling. The L1 baseline in Table 2 omits both the multi-component loss and the static geographic inputs, so we separated the two: the surface-temperature gain is carried mostly by the static features (0.17 of the 0.19 K reduction on the typical set), whereas the fine-scale structural advantage is the loss – an L1 U-Net with identical static inputs still has a mean spectral error of 0.45 against CASPER's 0.27 (Supplementary Note 5.8). Spatial error patterns show a slight negative bias in extreme conditions and a slight positive bias in typical ones, both addressable at post-processing.

The eight-month budget does not cost the model fine-scale structure. On the extreme test set, power spectral analysis shows CASPER recovering variance across the resolved scales with a mean spectral error of 0.32, against 1.05 for an L1-trained U-Net, 0.44 for random forest and 1.07 for quadratic interpolation (Figure 5; the typical test set is shown in Supplementary Figure S8, where the corresponding values are 0.27, 0.98, 0.44 and 1.00). The conditional GAN reaches a lower mean, 0.29 on both test sets, and stays closer to the target than CASPER at the finest scales of 2 m and skin temperature, where CASPER loses variance; its spectra carry instead the spurious high-frequency energy that adversarial training introduces, visible as the upturn at the highest wavenumbers in every panel. The decisive range lies between 1 and 100 km, where the model must generate variability absent from the 32 km input; there CASPER reproduces the atmospheric energy cascade with spectral slopes that match WRF. Distributional agreement follows the same pattern, holding into the tails that represent heat waves and the coolest summer weeks (Supplementary Figures S9 and S10).

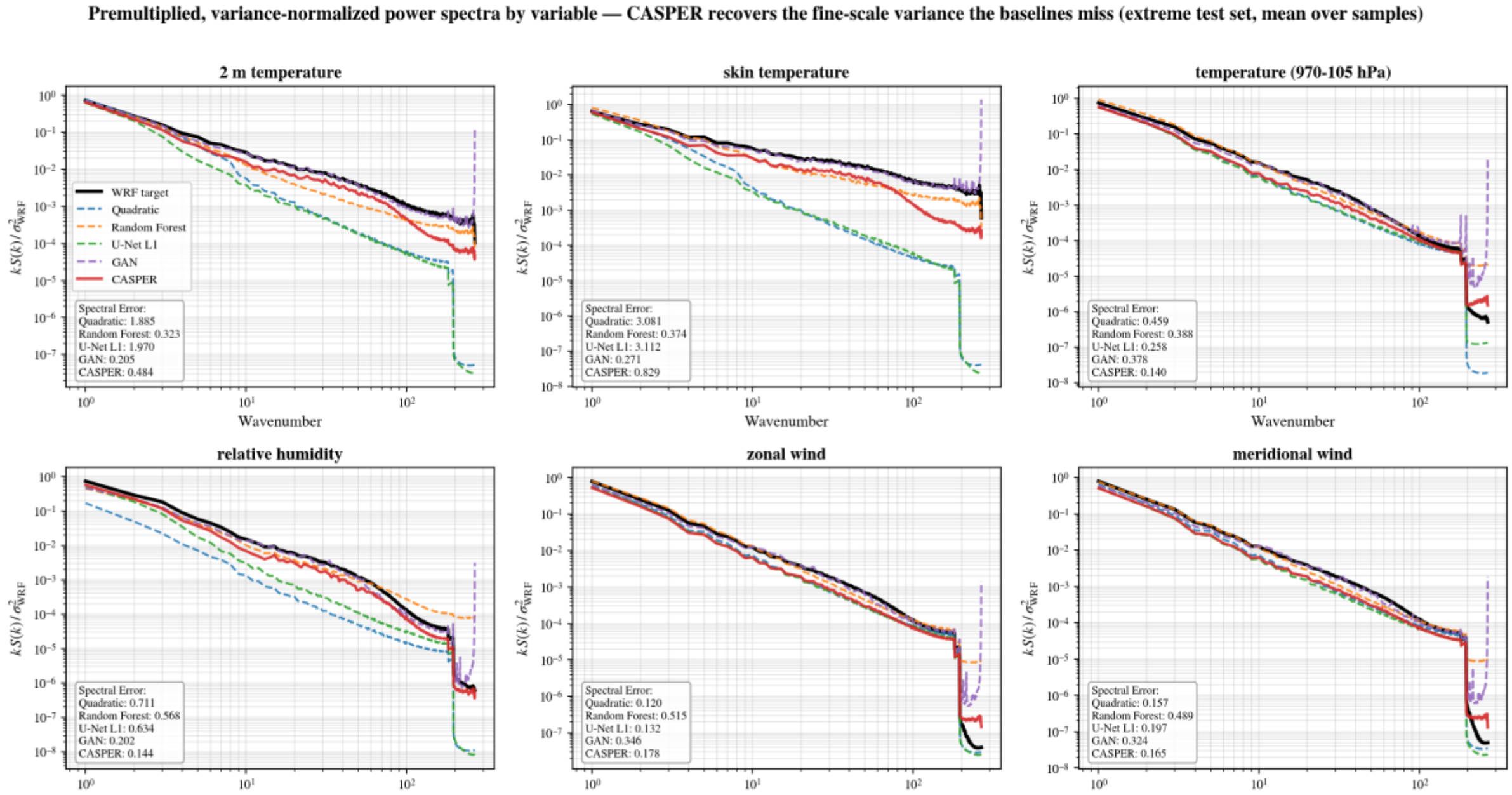

**Figure 5: Spatial structure validation via power spectral analysis.** Two-dimensional power spectral density S(k) as a function of wavenumber k for all six predicted variables (temperature, U-wind, V-wind, relative humidity, 2 m temperature, and skin temperature), as the mean over the extreme test set; the typical test set is shown in Supplementary Figure S8. Spectra are plotted premultiplied and variance-normalized, $k\,S(k) / \sigma^2$, where $\sigma^2$ is the resolved variance of the WRF target for that variable; the same constant scales every curve in a panel, so the six panels share one dimensionless axis, the $k^{-5/3}$ slope is flattened enough for the curves to separate, and the annotated spectral errors, computed in log space, are unchanged by the normalization. Each panel compares CASPER (red) against the WRF target (black), quadratic interpolation (blue), Random Forest (orange), U-Net with L1-only loss (green) and the conditional GAN (purple). Lower normalized spectral error, annotated per panel, indicates

better preservation of the atmospheric energy cascade across the mesoscale (10–100 km) and fine (1–10 km) ranges that downscaling must reconstruct.

The predicted variables retain the physical relationships between them. The 2 m temperature–humidity distribution follows the training target rather than admitting spurious combinations such as cold, saturated air, and the wind components retain the near-circular joint distribution expected of isotropic flow statistics (Supplementary Figures S11 and S12). Through the atmospheric column the model reproduces boundary-layer structure, including the 850–700 hPa jet and the mid-tropospheric moisture minimum, with the smallest errors in the mid-troposphere where reanalysis constraints are strongest (Supplementary Figures S13 and S14). These relationships are not imposed as constraints; they follow from predicting the whole column in one network under a loss whose multivariate term acts on the joint distribution. We did not train a separate network per variable, so the single network and the multivariate loss term are not separated experimentally here; the loss term itself is varied from zero to dominant across the ablation configurations of Supplementary Table S6.

**Case studies and model application**

Spatial and temporal means can conceal errors that cancel over time, so we examined two events end to end, pairing a single hour with the mean over the whole event (Figure 6). In the climatologically median hour of the typical event the 32 km input carries a single smooth warm anomaly, while the model resolves the St Lawrence valley, the Laurentian uplands and the urban areas of Montreal and Ottawa as distinct thermal structures, reproducing the WRF field to 1.52 K. During the hottest hour of the extreme test set the input reaches 33.5 °C as a near-uniform slab and the model recovers a field agreeing with WRF to 1.48 K. Averaged over the 119 hours of the typical event and the 168 of the extreme, the errors fall to 0.84 and 1.15 K, so hour-to-hour errors partly cancel rather than accumulate. The residuals concentrate along shorelines and steep terrain, where a 1 km field changes fastest.
Two heat events from the extreme test set are examined this way, each by the technique it suits: 12–18 August 2002 spatially, field by field against the 1 km WRF target, and 4–10 August 2003 temporally, hour by hour against station observations independent of the WRF fields the model was trained on. Both belong to the pre-specified test set – the hottest–driest and hottest–wettest weeks of the 40-year record. At the four in-domain stations it tracks most closely the model follows the observed diurnal cycle to 1.47–1.65 K with biases within ±0.6 K, and the median across all eleven stations in that window is 1.73 K. Agreement with WRF shows the model reproduces the simulation it was trained to emulate; agreement with instruments tests model and simulation together against the atmosphere itself.

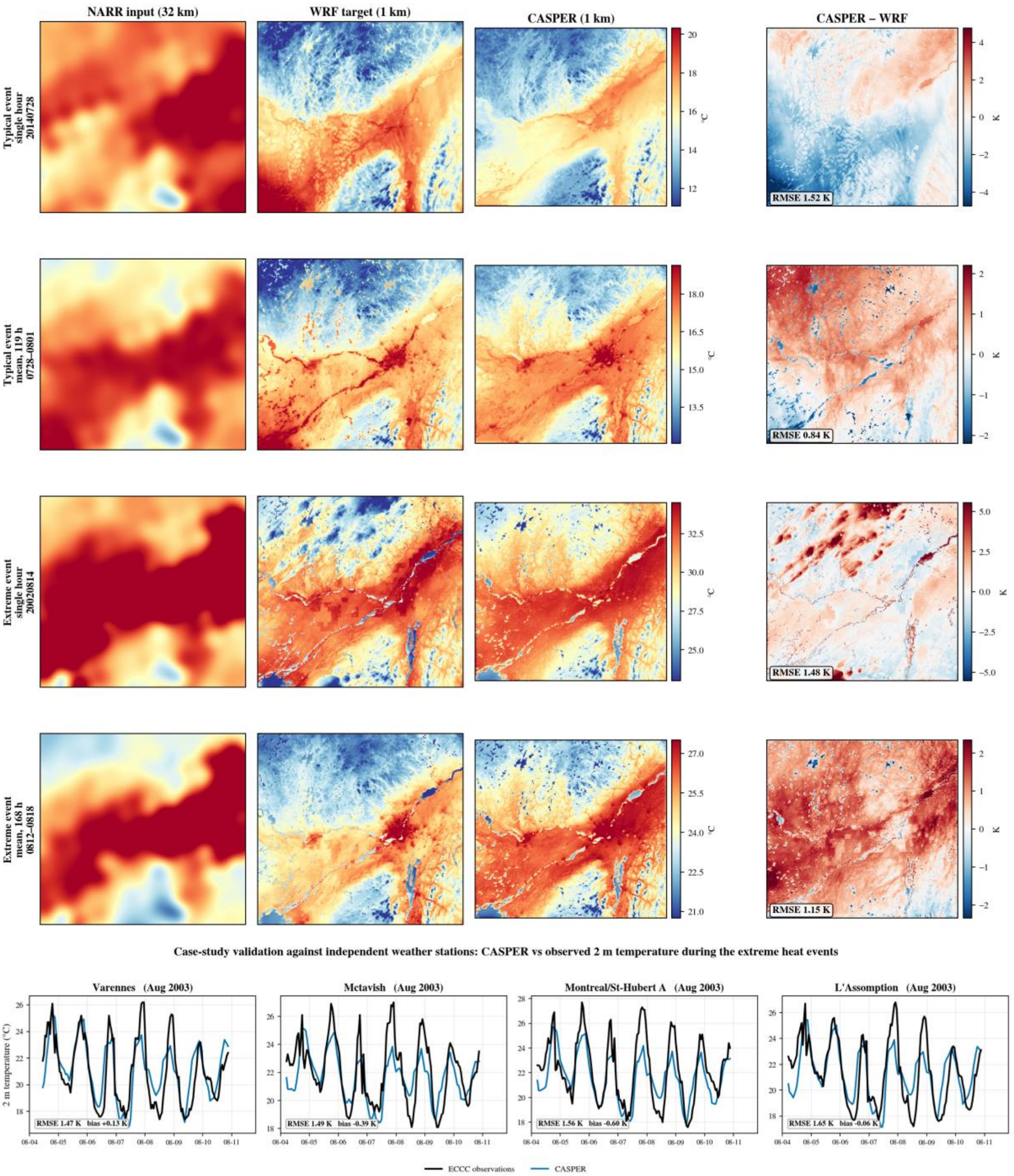

**Figure 6: Case studies - spatial fields and validation against independent weather stations.** Top: 2 m temperature for a typical event (first two rows, centred on 28 July 2014) and an extreme event (last two rows, centred on 14 August 2002); for each event the upper row shows a single hour and the lower row the mean over the whole event. Columns show the 32 km NARR input, the 1 km WRF target, the CASPER prediction and their difference, with the domain root-mean-square error annotated. Bottom: hourly 2 m temperature from CASPER (blue) against Environment and Climate Change Canada station observations (black) at four in-domain stations during the August 2003 heat event, an independent window outside the training months; the per-station root-mean-square error and bias are annotated and CASPER reproduces the observed diurnal cycle.

The scaling relation is built on held-out months within one domain, so transferring the model to a new region tests it against a second axis of dissimilarity: geography. We applied the eight-month model without retraining to Vancouver, Calgary and Toronto, three Canadian cities spanning Pacific coastal mountains, a prairie–mountain transition and a Great Lakes continental climate (Figure 7). Because these domains differ from the training region in terrain and land cover as well as climate, they probe the limit of what a climatological distance can anticipate.

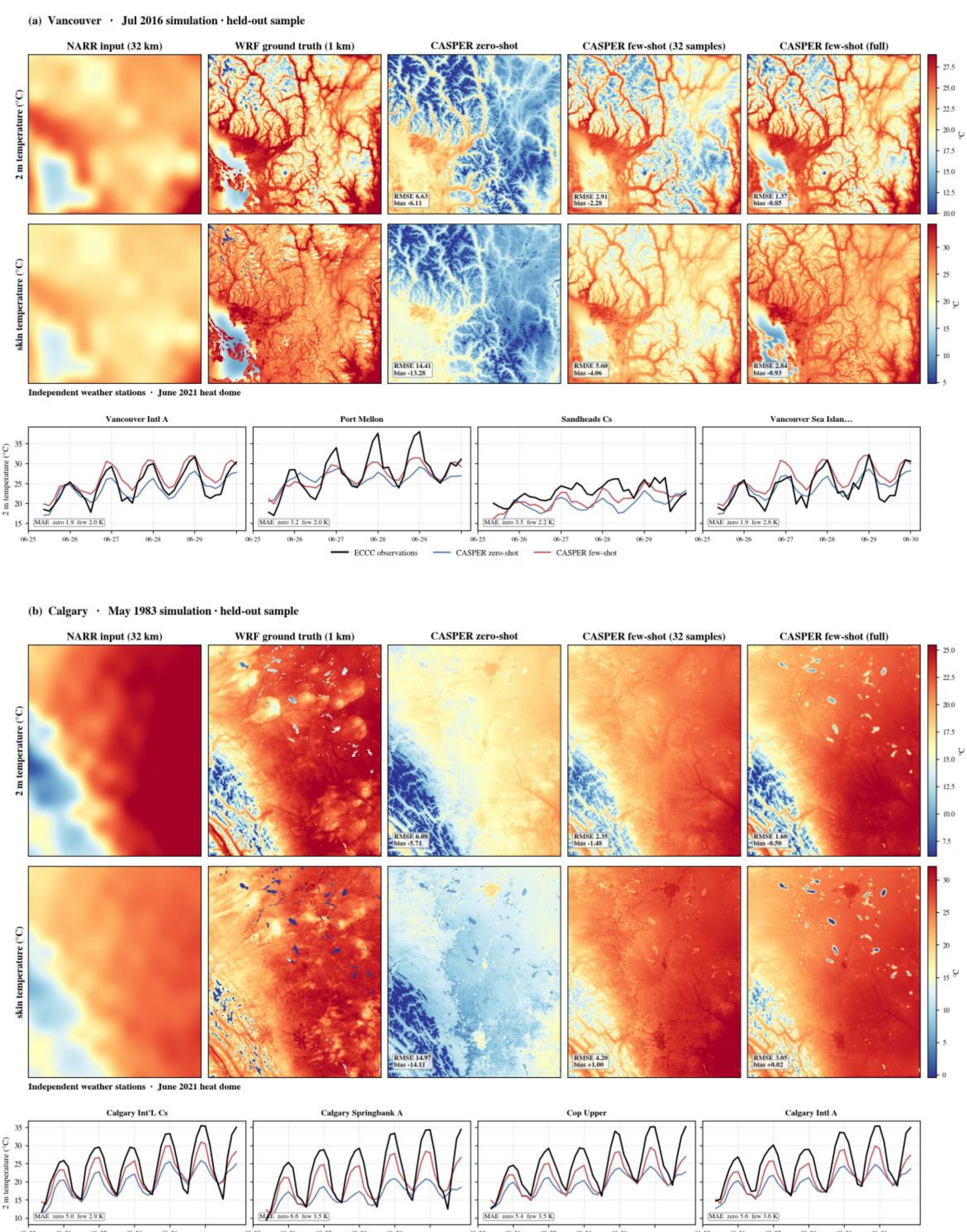

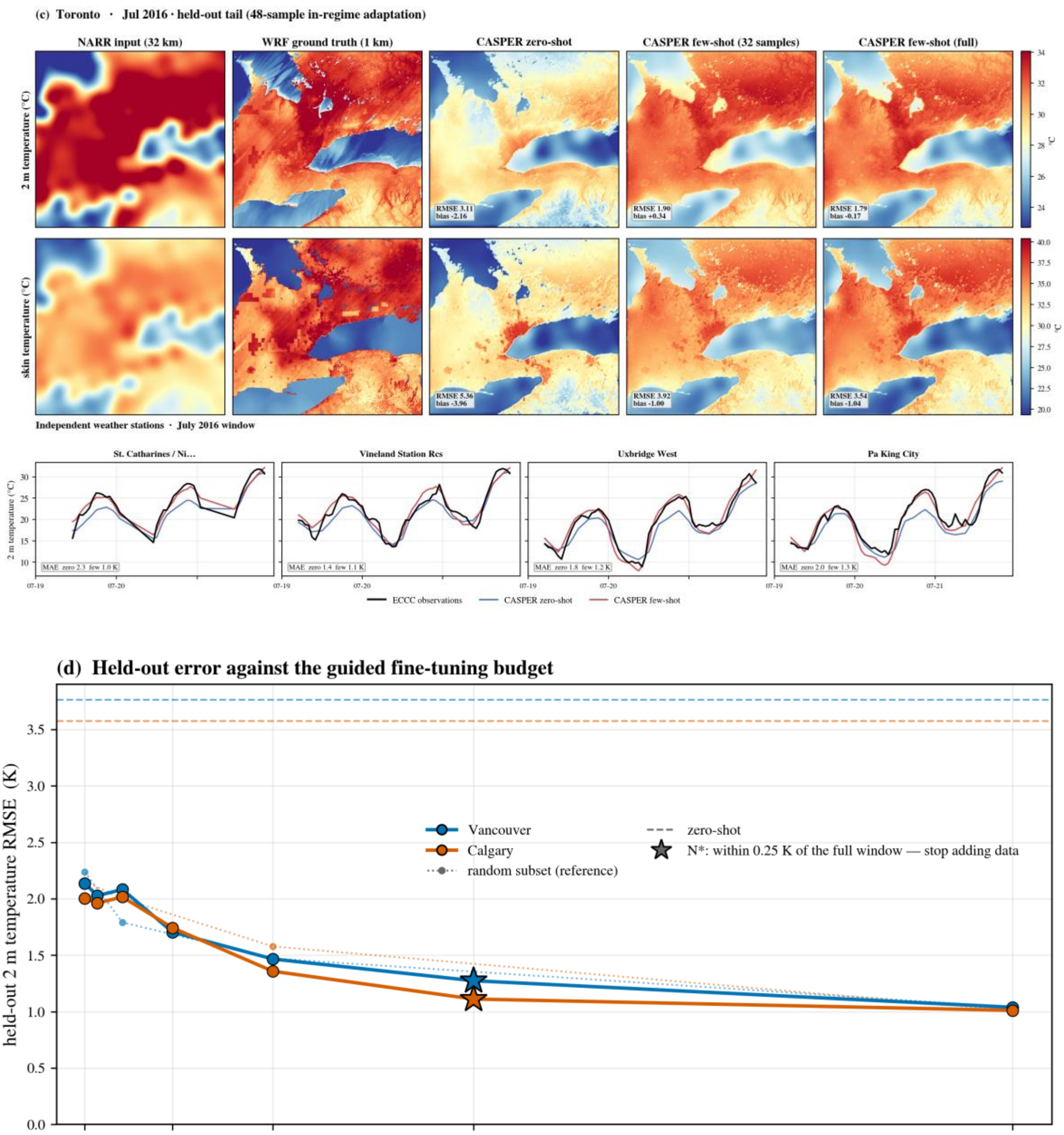

**Figure 7: Transfer by decoder adaptation to three out-of-domain cities.** Each panel is a held-out sample of that city's own 1 km simulation: (a) Vancouver (July 2016), (b) Calgary (May 1983) and (c) Toronto (July 2016). Within each panel the rows are 2 m temperature and skin temperature and the columns are the 32 km NARR input, the WRF 1 km ground truth, the eight-month model applied zero-shot, and the same model after decoder-only fine-tuning on 32 samples and on the full window (one month for Vancouver and Calgary; few-shot), with per-panel root-mean-square error and bias in degrees Celsius. Zero-shot prediction is strongly cold-biased over unseen terrain; local adaptation – one month for Vancouver and Calgary, the 48-sample window for Toronto – recovers the fine-scale structure, reducing 2 m temperature error from 6.6 to 1.4 K in Vancouver, 6.1 to 1.6 K in Calgary and 3.1 to 1.8 K in Toronto. Under each city's maps, hourly 2 m temperature from the zero-shot and few-shot models is compared against independent Environment and Climate Change Canada station observations at the stations each adapted model tracks best – during the June 2021 western heat dome for Vancouver and Calgary, and over the July 2016 simulation window for Toronto; adaptation

removes most of the zero-shot cold bias and follows the observed diurnal heat peaks. Panel (d) shows, for Vancouver and Calgary, the held-out 2 m temperature error, averaged over all held-out samples, against the fine-tuning budget when samples are chosen by guided coverage selection (farthest-point sampling over the target pool's climate features); the random-prefix ladder is drawn as a light reference, zero-shot as dashed lines, and stars mark N*, the smallest budget within 0.25 K of the full-window result, beyond which additional samples buy little; Toronto's 48-sample window is too short to resolve a plateau. The remaining 970 hPa fields and the June 2021 heat-dome maps are given in the Supplementary Information (Supplementary Figs. S15 and S16).

The three domains differ in how far they sit from the training region on the geographic axis, and that ordering decides how much adaptation buys. Measured on the static fields the model ingests, Toronto is closest – 0.72 times the training elevation range, two unseen land-use classes – then Calgary and Vancouver at 2.68 and 3.78 times the range with five and seven unseen classes. One month of local adaptation accordingly helps least where the domain is closest: Toronto's 2 m temperature error falls only from 3.1 to 1.8 K, against 6.1 to 1.6 K for Calgary and 6.6 to 1.4 K for Vancouver. The ordering is geographic rather than climatological – Vancouver is the closest of the three in climate space yet transfers worst – the same dissociation the station comparison shows, and part of the residual traces to unseen land-use classes such as wooded tundra. That argues for selecting training domains to span land cover and topography as we select months to span climate; removing terrain and land use degrades fine-scale reconstruction sharply, so explicit topographic encoding is what lets the model adapt at all. Degradation of this kind, recovered by local adaptation, is consistent with reports for neural downscalers evaluated outside their training domains [52].
We tested the model against independent station observations rather than against WRF, using two documented heat events across three city domains: the July 2010 Quebec heat wave, which killed 280 people, and the June 2021 western heat dome, which killed 619 in British Columbia and 66 in Alberta [53]. The 2021 event postdates every training month; the 2010 event lies outside the training months but within their span. Driving the model with public reanalysis alone, we compared 2 m temperature against Environment and Climate Change Canada records. Within the training domain no adaptation is needed: Montreal station temperatures are reproduced to 1.77 K mean absolute error across 11 stations. Applied unchanged to Vancouver and Calgary it degrades to 5.30 and 5.35 K (Supplementary Figure S16). That degradation is not ordered by climatological distance – the model lies at $d = 0.33$ from Vancouver and $d = 1.75$ from Calgary, yet the two errors are indistinguishable – so the transfer gap lies on a geographic axis the climate distance does not measure. Toronto, which the heat dome did not reach, was compared over its July 2016 window, where the adapted model tracks the four nearest stations to 1.0–1.3 K (Figure 7c).
A model deployed in a geographically distant region should therefore be adapted, and adaptation is cheap: 11 days of target-region simulation is enough. Its station mean absolute error falls to 3.23 K in Calgary, a 40 % reduction on the un-adapted model, and to 4.40 K in Vancouver, a 17 % reduction (Supplementary Figure S16). The gain concentrates where the zero-shot cold bias is largest: at the four Calgary stations of Figure 7b the error falls from 5.0–6.6 K to 2.9–3.6 K, whereas the Vancouver stations were already among the best tracked without adaptation and change little. Across four one-month fine-tuning seeds the recovery is stable, $0.98 \pm 0.04$ K held-out for both cities against 3.8 and 3.6 K zero-shot, so it is not a favourable-seed artifact. The budget behind the 11-day figure can also be planned: selecting samples by coverage of the target pool's climate features, mean held-out error falls from 2.1 K at 8 samples to 1.3 K at 256 in Vancouver and from 2.0 to 1.1 K in Calgary – within a quarter kelvin of the one-month result, beyond which additional samples buy little (Figure 7d). Those 256 samples are 11 days' worth of hourly fields, so converting the budget into simulation time assumes they are run as short contiguous blocks. Toronto, already the closest domain, sits at that point from the start: no budget below its full 48-sample window improves materially on zero-shot (2.5 to 2.2 K). What adaptation cannot supply is the extreme itself: the residual error concentrates in the hottest hours, which a fine-tuning month of ordinary conditions does not contain.

The distance that governs training does not govern fine-tuning. When samples are added to an already-trained model, selecting them to minimise distance to the target – the choice the relation appears to recommend – produced the worst overall error of any strategy we tested, in every configuration and across three random seeds. The reason is geometric. A training distribution that already holds hundreds of samples cannot be recentred by a few dozen more unless those few are extreme, so minimising the distance pushes the selection past the target and trains the model on its tail. Selecting samples that span the target distribution instead gave the lowest overall error in every configuration; at the smallest budgets the distance-minimizing selection was better on the extremes while remaining worse overall. The practical rule is therefore split in two: use distance to decide which periods to simulate before training, and coverage of the target distribution to decide which samples to add afterwards. The relation predicts what a trained model will do; it does not prescribe what to add to one.

Because a model can be adapted after the fact, the quantity a group pays is not the size of the training set but the total simulation behind the deployed model. Measured that way, a small model with targeted adaptation beats a larger one without it. Adapting the one-month model to the held-out January with 32 selected samples reaches 3.99 ± 0.40 K, against 14.55 K for the four-month model applied directly – a lower error from 1.04 months of simulation than from four. The same holds spatially: one month plus 32 selected Vancouver samples reaches 2.37 K where the eight-month model without adaptation gives 4.00 K, at roughly an eighth of the simulation.

This is possible because the samples are chosen from reanalysis alone: placing candidate hours in the climate space needs only the coarse input fields, so the selection is made before any high-resolution simulation is committed. That accounting assumes those hours are produced as short contiguous blocks, since high-resolution fields come from continuous integrations with spin-up rather than isolated time steps. The advantage is over an un-adapted model, not over adaptation in general: given the same 32 samples the larger base remains ahead, at 3.16 ± 0.08 K in-domain and 1.87 K at Vancouver. A wider training distribution still sets a lower floor, and where that floor matters the simulation has to be spent up front; where it does not, it can be spent far more sparingly than current practice assumes.

## Discussion

The bottleneck in deploying machine-learning downscaling is the cost of generating high-resolution training data, and the field has addressed it by assuming that more simulation is better. We find instead that error is set by how far the target climate lies from the training distribution. This formalizes, for kilometre-scale downscaling, the domain-shift principle that a model's error grows with distributional distance from its training data; what is new is that the distance is computable from reanalysis before any simulation and is calibrated to predict an unseen month's error with a stated interval. Extremes are valuable because they extend that distribution, not because hard examples teach more [54].

The relation is applied as follows. Compute the joint temperature–humidity climatology of the target domain and place the candidate simulation periods and the conditions of interest in that space. Ranking candidate periods by distance costs nothing and needs no simulation because the coordinates come from reanalysis alone; converting a distance into an expected error in kelvin is a further step that requires the local slope and intercept. Where that error is unacceptable, the remedy is to simulate months that span the target, not to simulate more of them. The assessment therefore costs nothing to run before committing supercomputer time, which is exactly when the decision must be made.
The multivariate loss term is what makes this regime workable: where conventional bias correction calibrates region-specific transfer functions after the fact [8], it enforces distributional agreement during training, so a group can train on a short, targeted simulation campaign without building a separate post-processing workflow. Static geographic features are what make transfer possible at all: removing them degrades performance sharply, and the model adapts to a new region by passing that region's terrain and land cover through the same learned response functions. This is why transfer quality tracks geographic resemblance to the training domain, and why selecting training domains for geographic and climatological diversity matters more than extending them in time. Predicting the full column jointly, rather than one surface variable at a time, preserves the relationships between variables: moisture and temperature follow the Clausius–Clapeyron relation [46], and the vertical structure retains realistic lapse rates that single-level methods cannot guarantee.
Spectral energy tracks WRF through the mesoscale, temperature fronts co-locate with moisture discontinuities, and wind shear preserves rotational structure, consistent with quasi-geostrophic theory [55] and frontogenesis [56] – all of it obtained from the training objective alone, without hard physical constraints. Convolutional U-Nets remain well suited to this problem. Transformers [50] and state-space architectures [51] model long-range dependencies more flexibly, but they scale quadratically with sequence length [57] and require more data – a poor trade in a regime where the binding constraint is simulation cost rather than model capacity.
Generative models, including GANs [26] and diffusion models [21], [24], remain the natural route to ensembles and calibrated uncertainty, but they are poorly matched to the small-data regime this paper targets. We tested this directly at matched budget: a conditional diffusion downscaler trained on the identical eight-month dataset failed to converge to a usable model. Its per-step denoising loss decreased, but the reverse-diffusion sampler diverged to temperatures far outside any physical range, so it could not be scored against WRF and is excluded from the matched-budget comparison – the behaviour expected of score-based training, which requires large and diverse datasets for stable sampling. We therefore chose a deterministic model because the mapping is largely deterministic once high-resolution boundary conditions are supplied, and structure-preserving losses recovered the distributional properties usually attributed to sampling, without the separate post-hoc correction stage that current diffusion pipelines require [24], [25].
Several limitations bound these results. The framework is deterministic and produces point predictions, which suits operational use [29] but leaves uncertainty unquantified for individual fields; ensemble extensions would address this but might sacrifice data efficiency [7]. The relation itself rests on eight climate states in a single domain, driven by one reanalysis and one model configuration, so its coefficients should be re-measured rather than transported; what transfers is the procedure, not the constants. Transfer also degrades where climate or terrain fall outside the training range, which meta-learning [58] and few-shot adaptation [59] could mitigate.

This changes what a downscaling project must commit to in advance. Rather than assembling a multi-year archive before training begins, a group can simulate a handful of months chosen to span the conditions it cares about, train in at most a few days on a single consumer GPU, and know beforehand where the resulting model will be reliable – lowering the barrier for services without large computing facilities [28]. Measuring the relationship between training data and error, rather than assuming it, tells a group planning a simulation campaign how much to simulate, which periods to choose, and where the resulting model will fail, before the first hour of computing time is spent.

## Methods

### Problem Formulation and Study Domain

We formulate statistical downscaling as a supervised learning problem that maps coarse-resolution atmospheric fields from the North American Regional Reanalysis (NARR) at 32 km horizontal resolution to high-resolution Weather Research and Forecasting (WRF) model simulations at 1 km. The primary study domain encompasses the Montreal and Ottawa regions of Eastern Canada, characterized by complex topography, including the St. Lawrence River Valley, extensive urban development, and the Laurentian Mountains (Supplementary Figure S17a). The downscaled output comprises a 363×390 pixel grid at 1 km resolution. Our model predicts three-dimensional multivariate atmospheric fields including temperature (T), horizontal wind components (U and V), and relative humidity (RH) across 10 vertical pressure levels spanning the troposphere from 970 hPa to 105 hPa, along with two critical surface variables: 2-meter temperature (T2) and skin temperature (TSK). This vertical discretization captures the essential atmospheric structure from the planetary boundary layer to the upper troposphere, with higher vertical resolution near the surface, where gradients are most substantial and meteorological impacts are most significant.

To assess the spatial transferability of the learned downscaling relationships, we evaluated model performance on three additional Canadian regions exhibiting distinct climatological regimes and topographic characteristics: Vancouver on the Pacific coast with complex mountainous terrain and maritime influence (Supplementary Figure S17b), Calgary at the prairie-mountain transition zone subject to dramatic chinook wind events (Supplementary Figure S17c), and Toronto influenced by Great Lakes dynamics and urban heat island effects (Supplementary Figure S17d). These test domains contributed no training data to the zero-shot evaluation, providing a rigorous assessment of zero-shot spatial generalization.

### Data Acquisition and Preprocessing

NARR provides three-hourly atmospheric state estimates on a 32 km Lambert conformal grid covering North America. From this dataset, we extract five three-dimensional variables: zonal wind $U$, meridional wind $V$, temperature $T$, relative humidity $RH$, and pressure $P$ at 10 selected pressure levels, plus surface pressure $P_{sfc}$, yielding 51 dynamic atmospheric channels. Vertical level selection prioritized meteorologically significant pressure surfaces while reducing computational cost, with higher sampling density near the surface (970, 953, 909, 826, 711 hPa) where boundary layer processes dominate, and coarser spacing in the free atmosphere (580, 435, 288, 175, 105 hPa) sufficient for capturing synoptic-scale flow patterns. Two static geographic features derived from WRF preprocessing system geogrid files (terrain elevation H and land use category index LU) are normalized and concatenated as additional input channels, bringing the total input dimensionality to 53. These static features encode crucial surface boundary conditions that modulate local weather through orographic forcing, surface roughness, and thermal properties.

WRF model version 4.3 simulations at 1 km resolution provide the high-resolution training target for training and evaluation (see Supplementary Note 5.2 and Supplementary Table S3 for the complete WRF configuration). From these simulations, we extract four three-dimensional variables at the same 10 pressure levels used for input, plus surface variables, yielding 42 output channels. This configuration captures the complete three-dimensional atmospheric state while maintaining computational tractability for the deep learning framework.

The preprocessing pipeline applies bilinear spatial interpolation to NARR variables to upscale from 32 km to 1 km grid spacing using the WRF Preprocessing System (WPS), and nearest-neighbor interpolation to the categorical land-use field to preserve discrete classification boundaries. All continuous variables undergo standardization (Equation 1) using statistics computed exclusively from the training split:

$$\tilde{x} = (x - \mu_{train})/\sigma_{train} \quad (1)$$

**Model Architecture**

Our architecture employs a U-Net design with multi-head self-attention mechanisms, processing the full 363×390 spatial domain without patching to preserve long-range correlations essential for coherent atmospheric features (Supplementary Figure S18). The network comprises a contracting encoder path that progressively downsamples spatial resolution while increasing feature dimensionality, a bottleneck that incorporates global context via attention, and an expanding decoder path that reconstructs high-resolution predictions via learned upsampling and skip connections from corresponding encoder levels (see Supplementary Note 5.5 and Supplementary Table S4 for complete architectural specifications).

The encoder consists of six downsampling stages. Each encoder stage contains a ResidualBlock that implements two sequential 3×3 convolutions with GroupNorm, SiLU activations, and dropout (rate 0.1), followed by 3×3 strided convolution for spatial downsampling. Residual connections employ 1×1 convolutions when channel dimensions change, enabling gradient flow through the deep architecture. GroupNorm provides batch-size-independent normalization, critical for stable training with our small batch size, with the group count fixed at 32, ensuring 8-32 channels per group across all feature dimensions.

Multi-head self-attention blocks are placed at the four deepest encoder levels and at the bottleneck, but each is gated to act only where the spatial resolution has fallen to 32 pixels or below, so for the 363×390 input attention is active at the two deepest encoder levels and the bottleneck, where quadratic attention is computationally tractable. At these coarse scales, attention mechanisms aggregate global context, which is crucial for representing synoptic-scale atmospheric patterns such as jet stream position, frontal system orientation, and large-scale pressure gradients. The attention computation follows a standard scaled dot-product formulation with query, key, and value projections split across multiple heads to capture diverse long-range dependencies.

The decoder mirrors the encoder structure, with six upsampling stages: each consists of 2× bilinear upsampling followed by a 3×3 convolution for feature refinement, concatenation of skip connections from the corresponding encoder level, and processing through ResidualBlocks. Skip connections transfer fine-scale spatial information lost during encoding directly to the decoder, enabling reconstruction of sharp gradients and detailed mesoscale features. A final 1×1 convolution projects the 256-channel decoder output to the required 42 output channels representing all predicted atmospheric variables and vertical levels.

To encode absolute spatial position, which is essential for learning location-dependent phenomena such as coastal effects, lake breezes, and orographic precipitation, we concatenate an 8-channel sinusoidal positional encoding to the input. This encoding employs four spatial frequencies $2^0$ through $2^3$, with paired sine and cosine components computed on normalized spatial coordinates $y, x \in [-1,1]$according to $sin(2^i \pi y)$and $cos(2^i \pi x)$for frequency index $i$. The multi-frequency representation captures spatial patterns across scales from domain-wide gradients to local variations.

All convolutional layers use Kaiming initialization which promotes stable gradient magnitudes during early training. The complete architecture contains about 625 million trainable parameters.

**Multi-Component Loss Function**

Standard pixel-wise regression losses such as L1 or L2 produce spatially blurred predictions with systematic misalignment of sharp features – a fundamental limitation for meteorological applications where steep gradients define frontal boundaries, convergence zones, and orographic wind acceleration. To address this, we developed a multi-component loss function (Equation 2) balancing point-wise accuracy with explicit preservation of spatial structure:

$$L_{total} = w_1 L_{MAE} + w_2 L_{grad} + w_3 L_{MS} + w_4 L_{patch} + w_5 L_{MBC} \tag{2}$$

The mean absolute error term provides baseline point-wise accuracy (Equation 3):

$$L_{MAE} = \frac{1}{N} \sum | \hat{y} - y | \tag{3}$$

where $\hat{y}$ and $y$ denote predicted and target values, respectively, $N$ is the total number of elements across all variables and atmospheric pressure levels. We employ L1 (absolute error) rather than L2 (squared error) to be more robust to outliers inherent in extreme weather events, as the absolute error metric penalizes large deviations linearly rather than quadratically, reducing sensitivity to rare but physically important extremes.

The gradient loss penalizes differences in spatial derivatives to preserve sharp features. We compute spatial gradients in x and y directions using Sobel operators, yielding gradient components. $\partial\hat{y}/\partial x$, $\partial\hat{y}/\partial y$ and gradient magnitude $| \nabla\hat{y} | = \sqrt{(\partial\hat{y}/\partial x)^2 + (\partial\hat{y}/\partial y)^2}$. The gradient loss employs smooth L1 loss (Huber loss) for robustness (Equation 4):

$$L_{grad} = \frac{1}{3}\left( L_{Huber}\left(\frac{\partial\hat{y}}{\partial x}, \frac{\partial y}{\partial x}\right) + L_{Huber}\left(\frac{\partial\hat{y}}{\partial y}, \frac{\partial y}{\partial y}\right) + L_{Huber}(| \nabla\hat{y} |, | \nabla y |) \right) \tag{4}$$

This formulation preserves both directional gradient information and edge strength, which are critical for weather downscaling, where atmospheric phenomena manifest as spatial gradients: temperature fronts, convergence-divergence patterns, and topographically induced wind acceleration.

The multi-scale loss ensures pattern consistency across spatial scales by computing structural similarity at four resolutions: the native 1 km grid and the half-, quarter- and eighth-resolution fields obtained by 2×, 4× and 8× average pooling. The four scales carry weights $w_s$ of 0.4, 0.3, 0.2 and 0.1, so the native resolution carries the largest weight. At each scale, we compute the structural similarity index (Equation 5):

$$SSIM_s(\hat{y}, y) = \frac{(2\mu_{\hat{y}}\mu_y + C_1)(2\sigma_{\hat{y}y} + C_2)}{(\mu_{\hat{y}}^2 + \mu_y^2 + C_1)(\sigma_{\hat{y}}^2 + \sigma_y^2 + C_2)} \tag{5}$$

where $\mu$, $\sigma$, and $\sigma_{\hat{y}y}$ denote local means, standard deviations, and covariance computed over 11×11 Gaussian windows, with stability constants $C_1 = 0.01^2$ and $C_2 = 0.03^2$. The multi-scale loss aggregates across scales (Equation 6):

$$L_{MS} = 1 - \sum_{s=1}^{4} w_s SSIM_s \tag{6}$$

This multi-resolution approach ensures both large-scale synoptic patterns (100+ km) and fine-scale mesoscale features (1-10 km) are accurately reproduced.
The patch loss addresses global spatial misalignment by enforcing local coherence. We randomly sample 32 patches of 16×16 pixels per training sample and compute normalized correlation within each patch, averaged across all patches. This local coherence constraint prevents the model from producing globally shifted predictions while allowing regional flexibility, a common failure mode in dense spatial regression tasks.
$K = 1200$ $N = 42$ $Y_{pred}, Y_{true} \in R^{K \times N}$The multivariate bias correction term preserves the physical relationships among variables by combining energy distance and quantile mapping. We randomly sample 1,200 spatial locations uniformly across the domain and extract, separately for each of the ten pressure levels and for the two surface variables, the corresponding channels at those points, forming sample matrices. The energy distance component measures multivariate distributional similarity through pairwise sample distances (Equation 7):

$$D_E^2 = \frac{2}{K^2} \sum_{i,j} \left\| y_{pred,i} - y_{true,j} \right\| - \frac{1}{K^2} \sum_{i,j} \left\| y_{pred,i} - y_{pred,j} \right\| - \frac{1}{K^2} \sum_{i,j} \left\| y_{true,i} - y_{true,j} \right\| \quad (7)$$

where $\|\cdot\|$ denotes Euclidean distance in the $N$-dimensional variable space. This metric captures multivariate correlation structure while being computationally tractable.
The quantile mapping component ensures distributional matching for each variable independently. For each output channel, we compute the prediction and target quantiles at 99 evenly spaced levels from 0.01 to 0.99, then minimize the L1 distance between them, averaged across channels and quantile levels. The complete MBC loss combines these components (Equation 8):

$$L_{MBC} = w_{energy} D_E^2 + w_{qmap} L_{qmap} \quad (8)$$

An ablation study across 15 loss configurations identified optimal weights (see Supplementary Note 5.7 and Supplementary Table S5 for the loss-component weights and complementary implementations, and Supplementary Note 5.8, Supplementary Table S6 and Supplementary Figures S19–S22 for the detailed ablation results). The ablation was carried out with the earlier four-month warm-season training configuration; the selected weights were retained unchanged for all models trained in this work. MAE and multi-scale terms receive moderate weights to balance spatial quality with point-wise accuracy, while patch loss and MBC use smaller weights to avoid over-constraining local predictions and unstable training. The gradient loss was the highest to preserve sharp gradients in the atmospheric fields. These weights remain fixed throughout training rather than employing progressive schedules, simplifying the training procedure.

### Training Procedure and Evaluation

We trained the model using the AdamW optimizer with an initial learning rate of $10^{-4}$, weight decay of $10^{-4}$, and a cosine annealing schedule with a minimum learning rate of $10^{-6}$ (see Supplementary Note 5.9 and Supplementary Table S7 for all hyperparameters). Limited by GPU memory constraints, we used a batch size of 2 with gradient accumulation over 2 steps to achieve an effective batch size of 4. Gradient clipping with a maximum norm of 1.0 prevented training instabilities caused by extreme weather outliers in the data.

$10^{-4}$The model was trained for 100 epochs, requiring approximately 72 hours on a single NVIDIA RTX 3090 GPU with 24 GB of memory. Early stopping monitored validation loss with a patience of 15 epochs and a minimum improvement threshold of $10^{-4}$. Training and validation loss remain close at every training budget (Supplementary Figure S23). We selected the best model based on the minimum validation loss, with checkpoints saved every 5 epochs to protect against hardware failures. We did not use mixed-precision training due to numerical stability concerns with the gradient-based loss components at our small batch size.
We evaluated model quality using multiple complementary metrics targeting different aspects of downscaling performance. We quantified point-wise accuracy through mean absolute error and root mean square error per variable (Equations 9 and 10):

$$MAE = \frac{1}{N}\sum \mid \hat{y} - y \mid, RMSE = \sqrt{\frac{1}{N}\sum(\hat{y} - y)^2} \quad (9, 10)$$

$y(x, y)$Spatial structure was assessed through a two-dimensional power spectral density computed via Fast Fourier Transform. For a 2D field y(x, y), the power spectrum quantifies energy distribution across spatial wavenumbers (Equation 11):

$$P(k) = \mid F\{y\}(k) \mid^2 \quad (11)$$

where $F$ denotes the 2D Fourier transform, and $k$ is the wavenumber magnitude. We quantified spectral fidelity between predicted and target spectra through normalized root-mean-square deviation in logarithmic space. This metric captures pattern similarity at different spatial scales corresponding to synoptic versus mesoscale features, with lower values indicating better preservation of multi-scale atmospheric structure. Physical consistency was verified through temperature-humidity and U-V winds joint distributions and gradients, confirming thermodynamic relationships and validating the dynamical balance, along with vertical wind shear profiles assessing a realistic boundary-layer structure.
We compared CASPER against four baseline methods: quadratic interpolation of 32 km NARR fields to 1 km resolution, Random Forest regression with 100 trees trained independently per output channel and per pixel, a standard U-Net trained with L1 loss only, and a conditional GAN of comparable capacity. All baselines used identical training data and evaluation protocols, ensuring fair comparison. We additionally attempted a conditional diffusion baseline, a one-stage denoising-diffusion model with a denoiser of comparable capacity, trained on the identical eight-month dataset; its reverse-diffusion sampling did not converge to physically valid fields at this data budget, so it is documented in Supplementary Note 5.10 rather than included in Table 1. To assess spatial generalization, we applied the Montreal-Ottawa-trained model without modification to the Vancouver, Calgary, and Toronto domains – a stringent zero-shot test across diverse geographic regions and climatic zones.

## Training data selection and the climate feature space

We characterized the domain climate from 40 years of NARR (1980–2020) [60]. For every three-hourly field we computed the domain-mean 2 m temperature and relative humidity over the exact 1 km model grid, taking the mean across all grid points of the model domain rather than over a bounding box, so that the statistic matches the quantity the network actually receives. These two variables define the feature space used throughout: every training set is a cloud of points in it, and every evaluation target is another cloud. We chose temperature and humidity because they are available in any reanalysis product, which keeps the procedure reproducible for any region.

The eight months selected on this criterion are January 1981, May 1984, January 1999, September 1999, July 2002, May 2005, July 2008 and September 2016. After quality control these yielded 5,902 hourly training samples (Supplementary Table S2). During training we monitored a validation set formed by a random 10% split with a fixed seed, used only for early stopping. Because temporally adjacent hours are correlated, this split is not an independent test; the independent tests are the held-out months, which no configuration saw during training, and the transfer cities. Overfitting could not produce the held-out results: a memorized training set cannot yield an error that falls on a leave-one-month-out-validated scaling relation across 23 independently trained models, nor generalize to a region absent from training. Regularization during training comprised dropout of 0.1, weight decay of $10^{-4}$, and early stopping.
Distance in this space is the standardized-Euclidean distance between the centre of the evaluation target's cloud and the centre of the training cloud, with each axis scaled by the standard deviation of the training cloud. Scaling by the training cloud rather than by the climatology is deliberate: it expresses distance in units of what the model has actually seen, so a wide training distribution is correctly treated as closer to any given target than a narrow one. We compared this metric by cross-validation against alternatives, including Mahalanobis distance, nearest-neighbour distances and distances measured beyond the edge of the training support under several definitions of that support (Supplementary Note 5.3).

**Design of the model set**

The model set was built to separate the effect of training volume from the effect of climatological distance, which are otherwise confounded because larger training sets are usually also closer to any given target. Four principles govern it. A budget ladder of nested subsets, in which each model's months are a superset of the previous model's, varies volume from one to eight months. Stratified compositions at fixed budget vary which months enter while holding volume constant. A group of randomly drawn subsets, generated from a fixed seed before any results were examined, guards against post-hoc selection of favourable compositions. Finally, a targeted ladder withholds both January months while the budget grows from four to six, populating the high-budget, high-distance region of the design that is otherwise empty and that alone distinguishes a volume effect from a distance effect. Every configuration shares the same architecture, loss function, optimizer, schedule and normalization statistics, so differences in skill are attributable to the training data alone.

**Fitting and validating the error–distance scaling**

For each trained model we evaluated every month absent from its training set, giving one (model, month) pair per evaluation, and regressed the resulting root-mean-square error on distance by ordinary least squares. Because the pairs are drawn from only eight distinct climate states, they are not independent: several models share an evaluation month, so a conventional coefficient of determination overstates how well the relationship predicts a new climate state. We therefore validate by blocking on month, refitting the relation with all pairs from one month withheld and using it to predict them, repeating for each of the eight months in turn. We report the resulting out-of-sample coefficient of determination alongside the in-sample value. We also report calibration: for each withheld month we compute the 90% prediction interval implied by the month-held-out fit and record the fraction of withheld months whose observed error falls inside it. A relationship that predicts accurately but understates its own uncertainty would be unsafe to use for planning, so both quantities are necessary.

**Observational validation**

We validated the model against station observations rather than against WRF for two documented heat events spanning three city domains: 4–10 July 2010 in Quebec, and 25–30

June 2021 in British Columbia and Alberta. The 2021 event postdates every training month; the 2010 event falls outside the training months but within their span. For each event we downloaded public NARR fields, generated boundary conditions with the WRF Preprocessing System, and ran the model in inference mode without any WRF target. Hourly 2 m temperature records came from the Environment and Climate Change Canada historical climate archive. We retained only stations lying within the model domain and within 3 km of a grid cell centre, deduplicated records across overlapping city queries, aligned observation and model times in UTC, and sampled the predicted field at each station's coordinates. This left 11 stations in Montreal, 17 in Vancouver and 20 in Calgary. We report mean absolute error, bias and Pearson correlation per station and average across stations. Station measurements represent a point while a model value represents a 1 km cell mean, so a residual representativeness error is unavoidable and is not removed; and because the model emulates WRF rather than the atmosphere, the station comparison additionally reflects WRF's own near-surface bias, making agreement with observations a stricter test than agreement with WRF. We also quantified how far each target region lies from the training domain on the geographic axis, using only the static fields the network ingests. For each domain we compared the spread of normalized terrain elevation with that of the training domain and counted the land-use classes present in the target but absent from training. This measure is independent of the climatological distance used for the error–distance scaling and of the station records used to evaluate skill, so the three lines of evidence are not circular.

### Few-shot adaptation and selection experiments

To test how cheaply a trained model can be adapted to a new target, we fine-tuned it on a small number of samples drawn from the target region or month. For the transfer figure we additionally fine-tuned the eight-month model along two budget ladders: random prefixes of the seed-0 permutation (8-128 samples), and a guided ladder whose samples are chosen by farthest-point (coverage) sampling over the pool's domain-mean temperature and humidity, at budgets of 8-256 samples. Every ladder draws from the same pool as the one-month fine-tune, so all budgets remain disjoint from the held-out evaluation samples; the plateau budget N* is the smallest whose held-out error lies within 0.25 K of the full-window result. The encoder path was frozen and only the decoder path was updated, which limits the number of free parameters and suits the small sample counts involved; optimization used Adam at a learning rate of 1e-4 for at most 30 epochs, with early stopping on a validation split disjoint from both the fine-tuning samples and the test split.
To ask which samples should be selected, we compared four strategies for choosing N fine-tuning samples from a candidate pool: random selection; coverage, which spans the pool by farthest-point sampling; nearest, which takes the samples closest to the target distribution centre; and a distance-minimizing strategy that greedily selects whichever sample most reduces the climatological distance between the augmented training distribution and the target. Each configuration was repeated for three random seeds, and every configuration shared one fixed test split, evaluated once at the end. We report error over the whole test split and over its most distant fifth, the part of the target the model is extrapolating into. Full results are given in Supplementary Note 5.4 and Supplementary Table S8.

### Data Availability

NARR reanalysis data are publicly available from NOAA/NCEP through the NCAR Research Data Archive at https://rda.ucar.edu/datasets/ds608.0/. WRF simulation data and preprocessed training/test datasets will be made available upon request. The eight training months and the 40-year climatological analysis used to select them are documented in the paper. Station observations are from the Environment and Climate Change Canada historical climate archive.

### Code Availability

Source code for model architecture, training procedures, multi-component loss implementation, and evaluation scripts is available at https://github.com/UMBE-LAB/CASPER-Context-Aware-Structural-Prior-Enhanced-Resolution. Trained model weights and inference code will be released upon publication. The documentation includes instructions for reproducing training across the eight selected months and for applying the model to new geographic regions.

### Competing interests

The authors declare no competing interests.

### Acknowledgements

This research was supported by the Natural Sciences and Engineering Research Council of Canada (NSERC) Discovery Grants Program [RGPIN-2024-06297], the Fonds de recherche du Québec – Nature et technologies (FRQNT) Doctoral Research Scholarships Program, the Canada First Research Excellence Fund (CFREF) IMPACT Project on "Transforming Built and Urban microclimates: Advancing Resilience Science for Vulnerable Populations in a Decarbonized and Electrified Canada," and the Canada First Research Excellence Fund (Volt-Age) SEED project on "Creating Electrified and Decarbonized Healthy Urban Microclimate around Building Clusters through Climate-Resilient Solutions." We thank the Digital Research Alliance of Canada for providing the computational resources to run the Weather Research and Forecasting (WRF) simulations.

### Authors Contributions

All authors contributed to the writing of the manuscript. A.M. proposed the idea with L.W., led the project, managed experiments and prepared the codebase. H.L. and A.G. generated the training data, prepared the preprocessed data as well as provided weather and climate domain expertise. L.W. secured funding, facilitated connections with partners in addition to providing guidance and supervision on the scientific implementation. S.G. led the editing and revisions and ensured scientific validity. M.A. and T.P analyzed the results, led the literature review process and assisted on revisions. A.HG. and D.R. assisted in the experiments, ablation studies and scaling as well as supervised the models training and evaluation, providing machine learning and artificial intelligence domain expertise.

# Supplementary Information for: Less is more: error–distance scaling relation for data-efficient kilometer-scale downscaling of extreme heat

Ahmed Marey[1,2], Henry Lu[2], Abhishek Gaur[2], Sherif Goubran[3], Malek Aloui[1], Theodore Potsis[1], David Rolnick[4,5], Alex Hernandez-Garcia[4,6], Liangzhu Leon Wang[1,*],

[1] Centre for Zero Energy Building Studies, Department of Building, Civil and Environmental Engineering, Concordia University, Montreal, H3G 1M8 Canada

[2] Building and Climate Interface, Construction Research Centre, National Research Council Canada, Ottawa, ON, K1A 0R6, Canada

[3] Department of Architecture, School of Sciences and Engineering, The American University in Cairo, New Cairo 11835, Egypt

[4] Mila – Quebec Artificial Intelligence Institute, Montreal, QC, H2S 3H1, Canada

[5] School of Computer Science, McGill University, Montreal, QC, H3A 0G4, Canada

[6] Department of Computer Science and Operations Research, Université de Montréal, Montreal, QC, H3C 3J7, Canada

[*] Author to whom correspondence should be addressed: leon.wang@concordia.ca

**Supplementary Note 1: Related Work and Context**

Statistical downscaling has emerged as a cost-effective alternative to dynamical downscaling for generating high-resolution atmospheric predictions from coarse global model outputs [1]. Early efforts focused on relatively simple architectures such as convolutional neural networks for precipitation downscaling [2] and random forests for temperature interpolation [3]. More recent work has explored deeper architectures, including U-Nets [4], which have proven particularly effective for image-to-image translation tasks in computer vision [5].

Recent studies demonstrate the extent of data requirements for achieving strong performance at large resolution differences. Jha et al. [6] downscale ERA5 2-m temperature using 31 years of training data, with deeper residual models achieving SSIM of 0.96 and PSNR of 34 dB. Rampal et al. [7] downscale daily precipitation over New Zealand from ERA5 to VCSN using 33 years of training, improving explained variance from 0.35 to 0.52. The challenge of data efficiency becomes particularly acute for very high resolution ratios. DeepSD demonstrated 8× spatial-resolution enhancement in precipitation using convolutional networks trained on 25 years of data [8]. Baño-Medina et al. [9] showed similar results for temperature and relative humidity downscaling using 30 years of reanalysis data. Recent work on super-resolution climate downscaling [10] achieved 12× resolution enhancement but required training on continuous 22-year climate reanalysis data. These substantial data requirements present practical limitations for operational weather services and research groups with limited access to high-resolution numerical weather prediction outputs. Similar data efficiency has been demonstrated in urban wind and temperature prediction, where localized training strategies with geometric features enabled accurate 3D predictions [11], [12].

Generative models have recently attracted attention for atmospheric downscaling due to their ability to capture uncertainty and generate realistic fine-scale variability [13]. Generative adversarial networks (GANs) have been applied to precipitation downscaling [14], [15], [16], demonstrating improved spatial structure compared to deterministic models. Diffusion models [17] offer improved training stability and have shown promise for probabilistic weather downscaling [18], [19], [20]. However, both types of models have their limitations related to training instability and/or practical limitations for operational purposes.

Several architectural innovations have emerged to improve the preservation of spatial structure in downscaling applications. Attention mechanisms [21] enable models to capture long-range spatial dependencies relevant for synoptic-scale weather patterns. Stochastic weight averaging [22] and ensemble methods [29] improve the quantification of prediction uncertainty. Multi-scale loss functions [23] encourage models to preserve atmospheric variability across the hierarchy of spatial scales from synoptic systems to mesoscale features. However, these techniques have primarily been evaluated on problems with moderate resolution ratios and extensive training datasets.

The present work addresses several gaps in existing literature. First, we measure how a downscaler's held-out error depends on its training data, and show that error follows a predictable error–distance scaling in a two-variable climate space, so which months are simulated governs skill more than how many. This data efficiency stems from recognizing that atmospheric downscaling relationships are largely deterministic, with most variance in high-resolution fields explainable by coarse-scale inputs and static terrain features [24], [25], [26]. Second, we introduce a composite loss function that combines spatial gradient preservation, multi-scale structural consistency, local patch coherence, and multivariate distribution, specifically designed for atmospheric applications where sharp meteorological features coexist with smooth synoptic-scale patterns. Third, we validate model performance on a challenging 32:1 resolution ratio (32 km → 1 km) across multiple atmospheric variables and vertical levels, demonstrating robust generalization to both typical conditions and extreme weather events. Finally, we provide a comprehensive physical consistency analysis examining multivariate relationships, vertical structure, and spatial-scale-dependent behavior to ensure meteorologically realistic predictions beyond simple point-wise accuracy metrics.

## Supplementary Note 2: Training-Data Design and the Error–Distance Scaling

### 2.1 Training Data Selection Strategy

Supplementary Figure S1 places the hourly training states within the 40-year climatology of the model domain; Supplementary Table S1 lists the 24 training configurations and their exact month compositions, and Supplementary Table S2 the training and held-out evaluation periods.

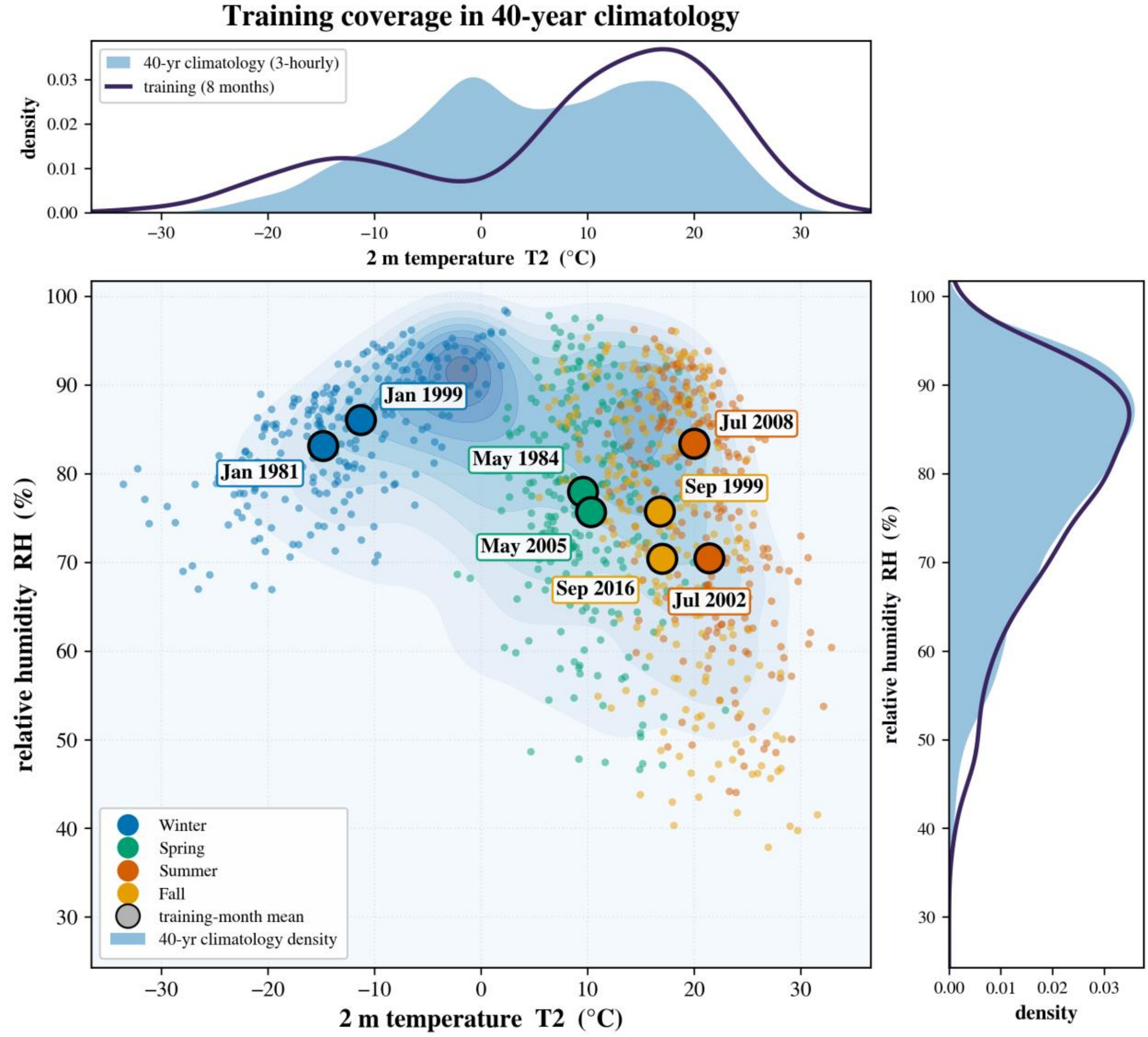

**Supplementary Figure S1: Training coverage within the 40-year climatology.** Hourly training states, coloured by season, and the eight training-month means are shown against the 1980–2020 three-hourly climatology of the model domain (blue density), in the climate space of domain-mean 2 m temperature and relative humidity. Marginal distributions compare the training states with the full climatology on each axis. The eight months span essentially the full 40-year temperature range, while the climatology extends to lower humidity than the training data.

**Supplementary Table S1: The 24 training configurations and their exact month compositions.** Each row is one independently trained model. The eight candidate months (two per season, spanning the 40-year joint temperature-humidity distribution) are January 1981,

May 1984, January 1999, September 1999, July 2002, May 2005, July 2008 and September 2016.

| Configuration | Budget (months) | Training months |
|---|---|---|
| b1_f | 1 | Sep 1999 |
| b1_sp | 1 | May 1984 |
| b1_su | 1 | Jul 2002 |
| b1_w | 1 | Jan 1981 |
| sw_k1_a | 1 | Jul 2008 |
| sw_k1_b | 1 | Sep 2016 |
| sw_k2_a | 2 | Sep 1999, Sep 2016 |
| sw_k2_b | 2 | May 1984, May 2005 |
| b2_cold | 2 | Jan 1981, Jan 1999 |
| b2_mixed | 2 | Jan 1981, Jul 2002 |
| b2_warm | 2 | Jul 2002, Jul 2008 |
| sw_k3_a | 3 | May 1984, Jul 2002, Sep 2016 |
| sw_k3_b | 3 | Jan 1999, May 2005, Jul 2008 |
| sw_k4_a | 4 | Jan 1981, May 2005, Jul 2008, Sep 2016 |
| sw_k4_b | 4 | Jan 1999, Jul 2002, Jul 2008, Sep 2016 |
| b4_balanced | 4 | Jan 1981, May 1984, Sep 1999, Jul 2002 |
| b4_cold | 4 | Jan 1981, Jan 1999, Sep 1999, Sep 2016 |
| b4_warm | 4 | May 1984, Jul 2002, May 2005, Jul 2008 |
| tc_k5_a | 5 | May 1984, Jul 2002, May 2005, Jul 2008, Sep 2016 |
| tc_k5_b | 5 | May 1984, Sep 1999, Jul 2002, May 2005, Jul 2008 |
| tc_k6 | 6 | May 1984, Sep 1999, Jul 2002, May 2005, Jul 2008, Sep 2016 |
| sw_k7_a | 7 | Jan 1981, Jan 1999, Sep 1999, Jul 2002, May 2005, Jul 2008, Sep 2016 |
| sw_k7_b | 7 | Jan 1981, May 1984, Jan 1999, Jul 2002, May 2005, Jul 2008, Sep 2016 |
| b8_all | 8 | Jan 1981, May 1984, Jan 1999, Sep 1999, Jul 2002, May 2005, Jul 2008, Sep 2016 |

**Supplementary Table S2: Training and testing sets from 40-year climatology.**

| Split | Category | Start Date | End Date | Type |
|---|---|---|---|---|
| Training & Data Budgeting | 8 months, two per season | 1/1/1981 | 1/31/1981 | Winter |
| | | 5/1/1984 | 5/31/1984 | Spring |
| | | 1/1/1999 | 1/31/1999 | Winter |
| | | 9/1/1999 | 9/30/1999 | Fall |
| | | 7/1/2002 | 7/31/2002 | Summer |
| | | 5/1/2005 | 5/31/2005 | Spring |
| | | 7/1/2008 | 7/31/2008 | Summer |
| | | 9/1/2016 | 9/30/2016 | Fall |
| Testing | Extreme | 8/12/2002 | 8/18/2002 | Hottest Driest |
| | | 8/4/2003 | 8/10/2003 | Hottest Wettest |
| | | 6/6/1988 | 6/12/1988 | Coldest Driest |
| | | 5/26/2003 | 6/1/2003 | Coldest Wettest |

| | | | |
|---|---|---|---|
| Typical | 7/28/2014 | 8/3/2014 | Typical Summer |
| | 6/26/2000 | 7/2/2000 | Typical Summer |
| | 8/18/2008 | 8/24/2008 | Typical Summer |
| | 8/20/2001 | 8/26/2001 | Typical Summer |

## 2.2 Predictors of Held-Out Error and the Interpolation Baseline

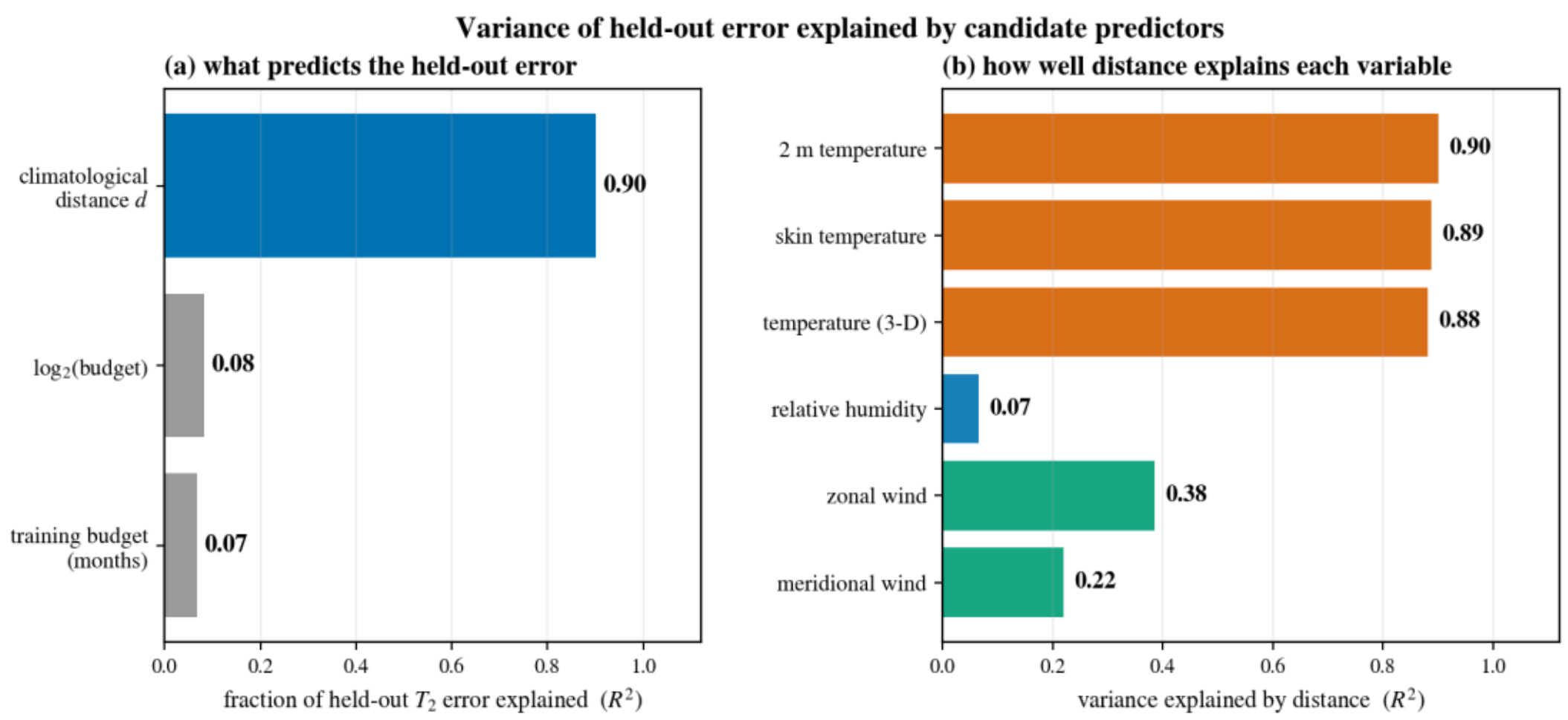

**Supplementary Figure S2: Variance of held-out error explained by candidate predictors.** (a) For the held-out 2 m temperature error, the fraction of variance explained by training budget, log-budget, and climatological distance. (b) For each of the six output variables, the fraction of error variance explained by T2-RH climatological distance. Distance dominates for temperature and is weak for humidity and wind.

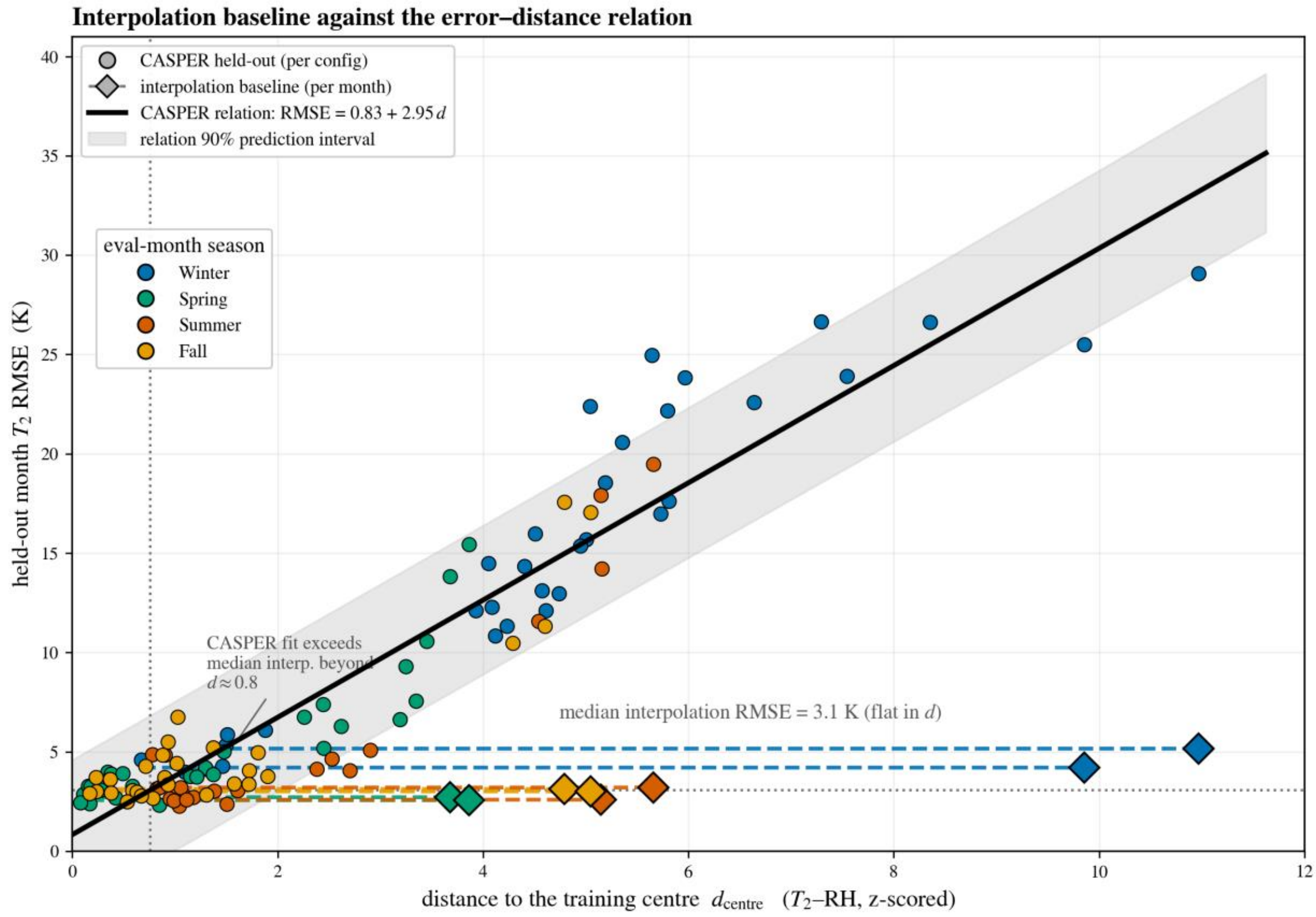

**Supplementary Figure S3: The neural downscaler's large held-out errors are extrapolation collapse, not intrinsic difficulty.** CASPER held-out 2 m temperature RMSE per configuration and held-out month (circles) against climatological distance, with the fitted relation and its 90% prediction interval, overlaid with a trivial bilinear interpolation of each target month's 32 km field to 1 km (diamonds, one per month). Interpolation degrades only gently with distance (median 3.1 K, rising to 5.2 K for the farthest January), so an out-of-coverage neural downscaler exceeds interpolation beyond a distance of about 0.8, whereas a coverage-selected model is more accurate below it. The order-of-magnitude error growth along the relation therefore reflects neural extrapolation out of the training regime rather than the intrinsic difficulty of cold months, and motivates selecting training months to span the target.

### 2.3 Variable-Resolved Error–Distance Scaling

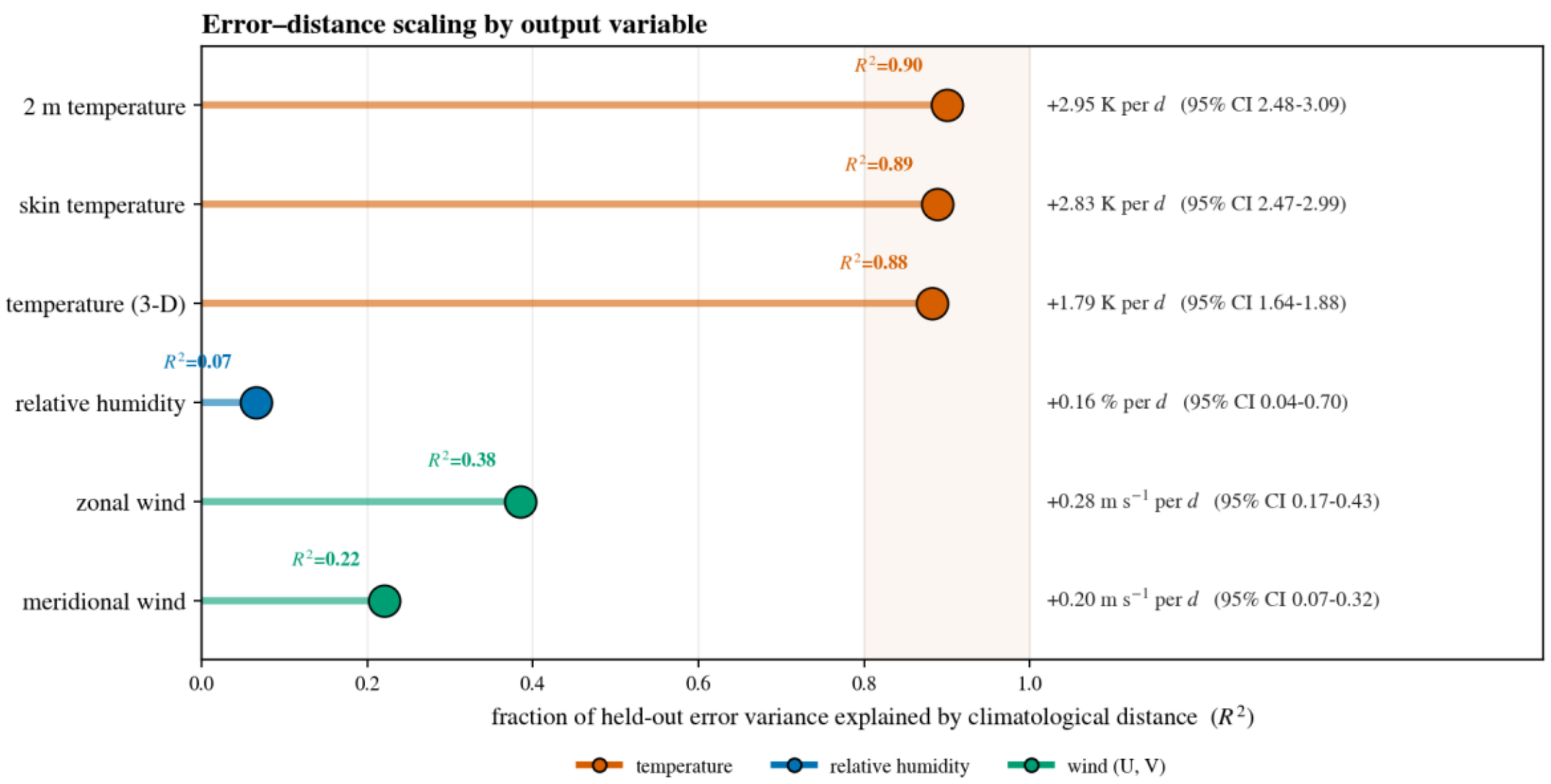

**Supplementary Figure S4: The error–distance scaling resolved by output variable.** Error added per unit of climatological distance for each of the six predicted variables, in physical units and grouped by unit, with 95% confidence intervals from bootstrap resampling blocked by evaluation month; the coefficient of determination (variance explained by distance) is annotated on each bar. All three temperature variables scale strongly with distance (1.8-3.0 K per unit distance, $R^2$ = 0.88-0.90), while relative humidity and the wind components are nearly flat ($R^2 \leq 0.38$), so the scaling is fundamentally a temperature relation.

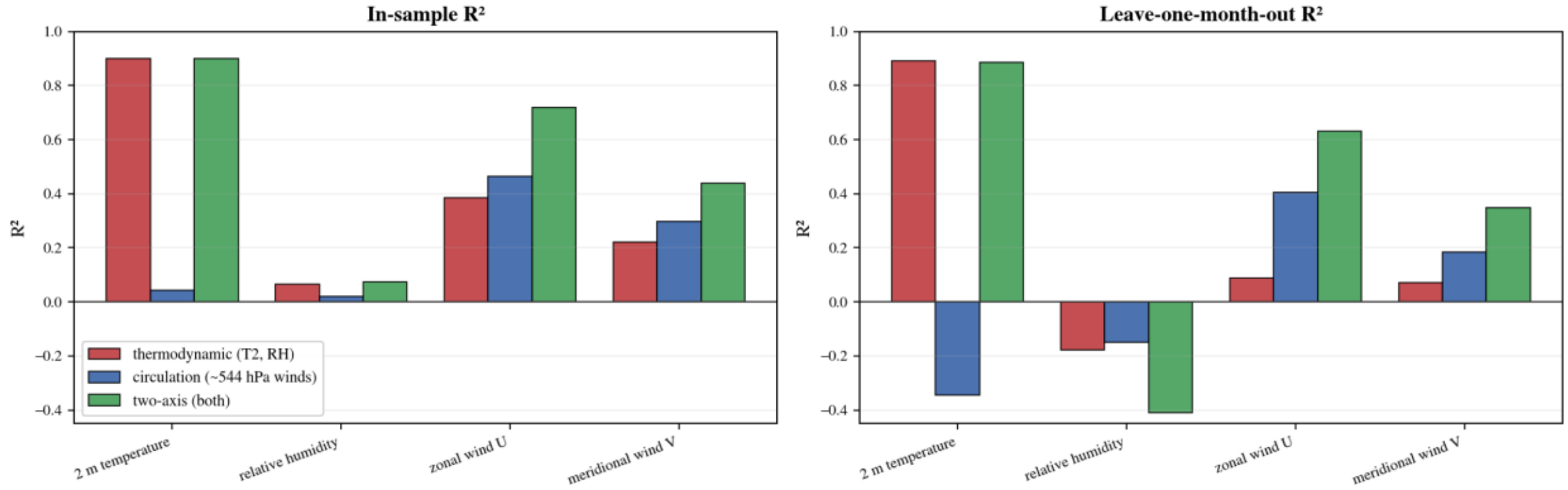

**Supplementary Figure S5: A two-axis distance coordinate.** Coefficient of determination for held-out error regressed on a thermodynamic distance (standardized-Euclidean separation in domain-mean 2 m temperature and relative humidity), a circulation distance (the same reduction in domain-mean mid-tropospheric ~544 hPa zonal and meridional wind, near-orthogonal to the thermodynamic distance, r = 0.18), and both combined, for the four continuous predicted variables, in-sample (left) and under leave-one-month-out cross-validation (right). Surface-temperature error follows the thermodynamic axis, wind error the circulation axis (zonal-wind cross-validated $R^2$ rises from 0.09 to 0.63 when the circulation axis is added), and humidity error is predicted by neither.

## Supplementary Note 3: Matched-Budget Evaluation

### 3.1 Spatial Performance on Typical and Extreme Conditions

The comparisons in this note cover the deterministic family - quadratic interpolation, random forest, an L1-only U-Net and CASPER - because their purpose is to isolate what the structure-preserving loss and the static geographic inputs add to a U-Net backbone. The conditional GAN is evaluated in the main text (Table 2 and Figures 4 and 5) on the extreme test set, where the relevant question is not architectural enhancement but whether adversarial training is stable at this data budget; Supplementary Figures S6 and S8 repeat those two main-text comparisons, GAN included, for the typical test set. Supplementary Figure S6 presents the mean spatial fields of all six output variables over the typical test set at 970 hPa and the surface, comparing CASPER and the matched-budget baselines against the WRF target.

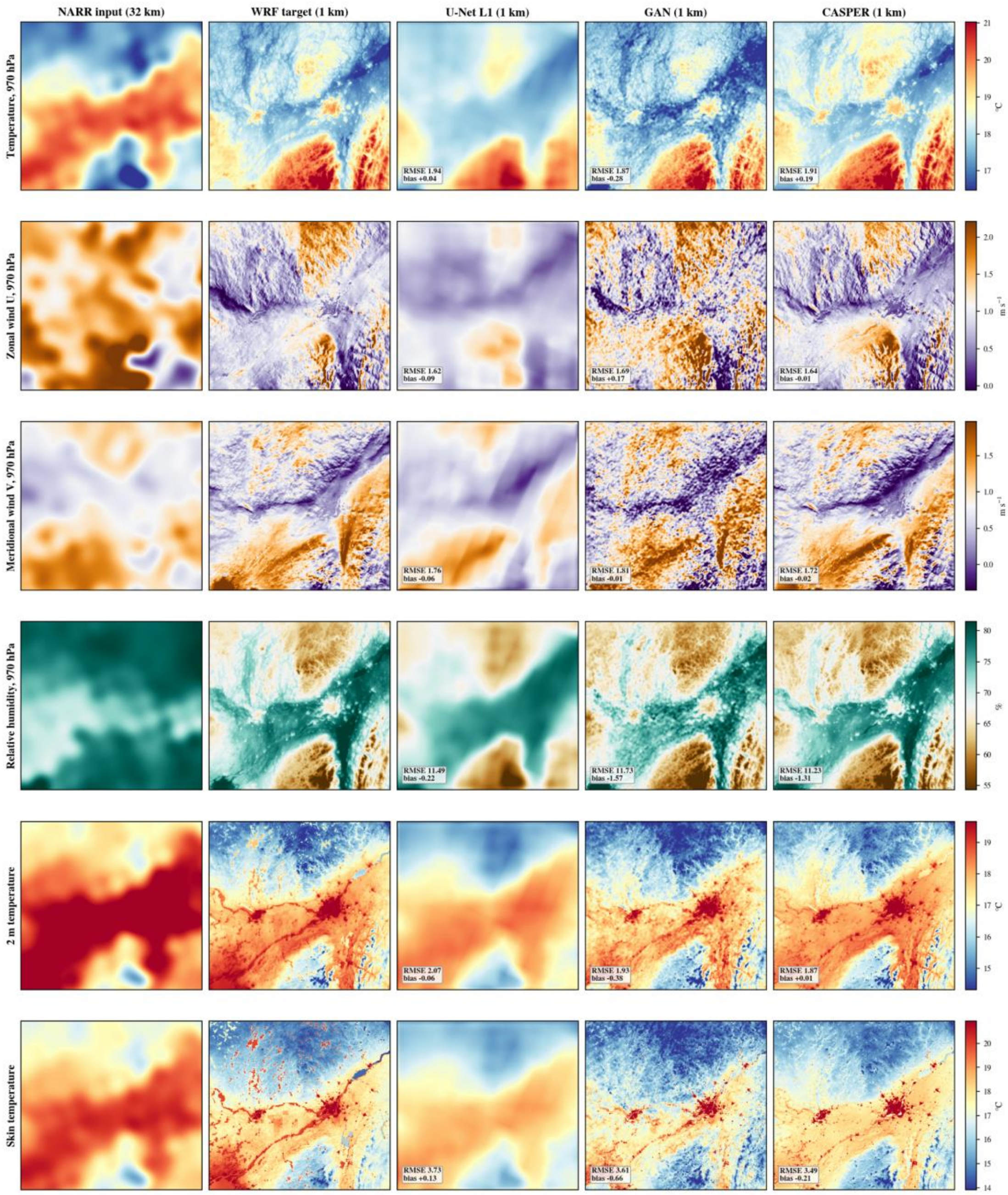

**Supplementary Figure S6: Spatial comparison over the typical test set (temporal mean).** Columns show the 32 km NARR input, the 1 km WRF target, the U-Net with L1 loss only, the conditional GAN and CASPER; rows show temperature, zonal and meridional wind and relative humidity at 970 hPa, then 2 m and skin temperature, with the root-mean-square error and bias of each model against WRF annotated. The layout is that of main-text Figure 4, which shows the extreme test set.

Model robustness under extreme atmospheric conditions is assessed using the extreme dataset containing samples representing climatological extremes in temperature and relative humidity. These cases challenge the model's ability to simulate rare conditions that other models usually struggle to reproduce, highlighting the advantage of the curated training selection (Supplementary Figure S7).

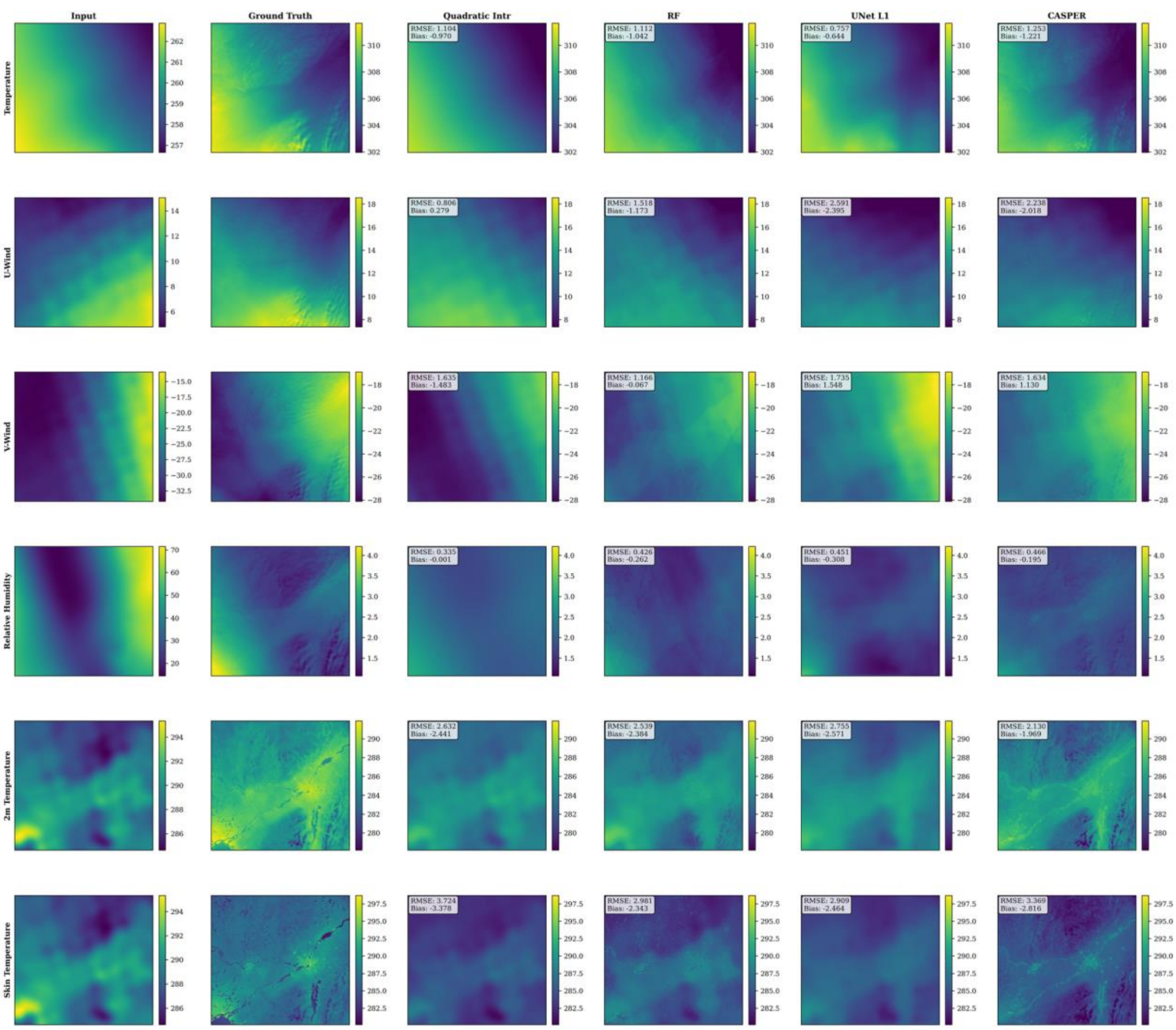

**Supplementary Figure S7: Spatial comparison over the extreme test set at a mid-tropospheric level.** Mean fields at output level 6 of 10 (580 hPa) for the three-dimensional variables, together with 2 m and skin temperature. Columns show the 32 km NARR input, the 1 km WRF ground truth, and four of the matched-budget methods (quadratic interpolation, random forest, U-Net L1, CASPER), with RMSE and bias against WRF annotated on each method panel; the three-dimensional temperature is WRF potential temperature.

### 3.2 Power Spectra and Spatial Structure Preservation for Typical Conditions

Preservation of atmospheric variability across spatial scales from synoptic systems (>1000 km) through mesoscale (10-100 km) to fine scales (<10 km) is essential for capturing realistic weather patterns. Supplementary Figure S8 presents power spectra for all six predicted variables over the typical test set, on the same axes as main-text Figure 5.

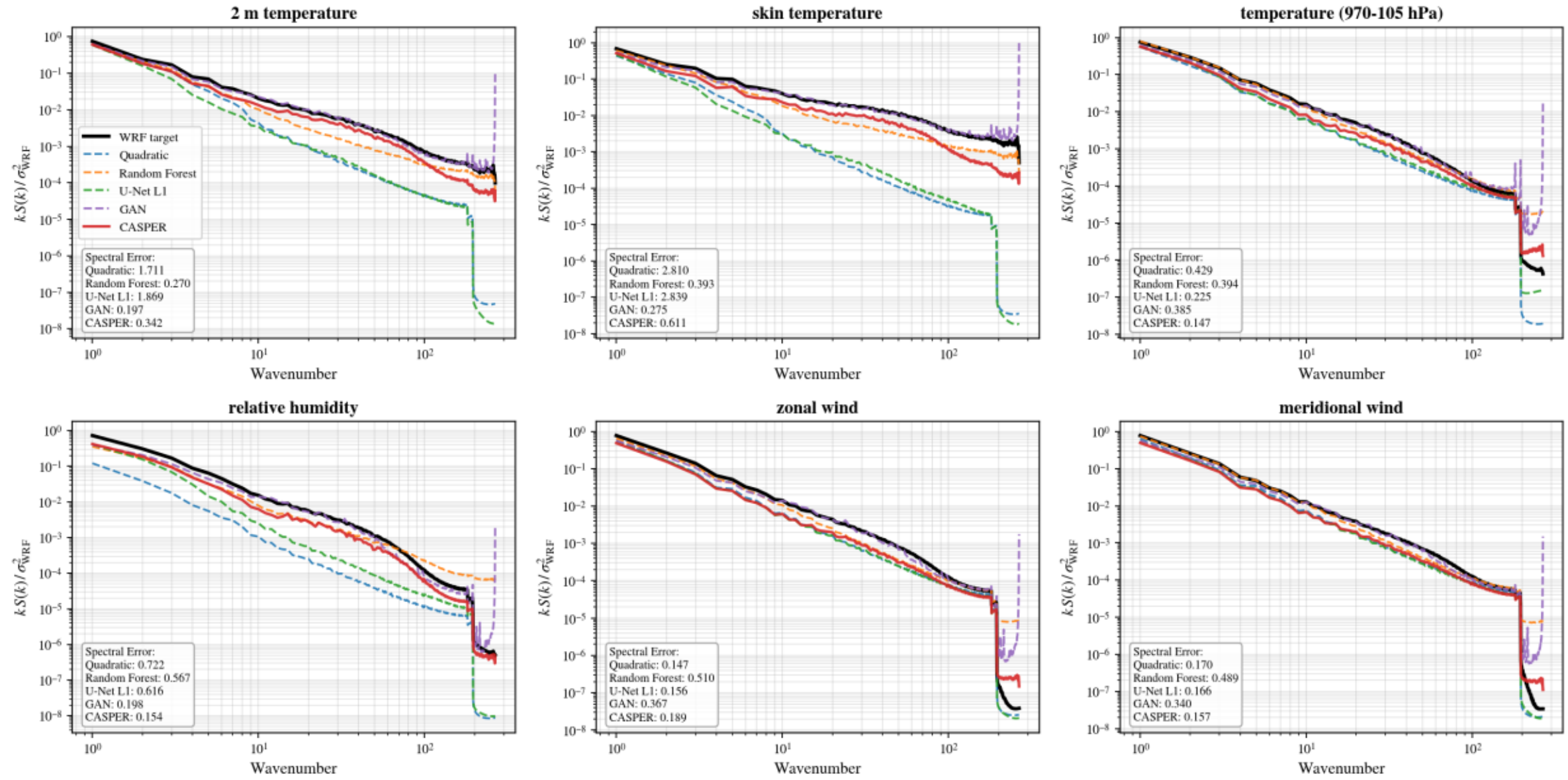

**Supplementary Figure S8: Power spectra for the typical test set.** As in main-text Figure 5, for the typical test set: two-dimensional power spectral density S(k) against wavenumber k for the six output variables, as the mean over the full typical test set, plotted premultiplied and variance-normalized (k S(k) / σ², with σ² the resolved variance of the WRF target). Each panel compares CASPER against the WRF target, quadratic interpolation, random forest, the U-Net with L1-only loss and the conditional GAN, with the normalized spectral error of each method annotated.

### 3.3 Statistical Distribution Fidelity

Supplementary Figures S9, S11 and S13 provide the distributional, multivariate and vertical validation for the typical test set; each is paired below with its extreme-test-set counterpart.

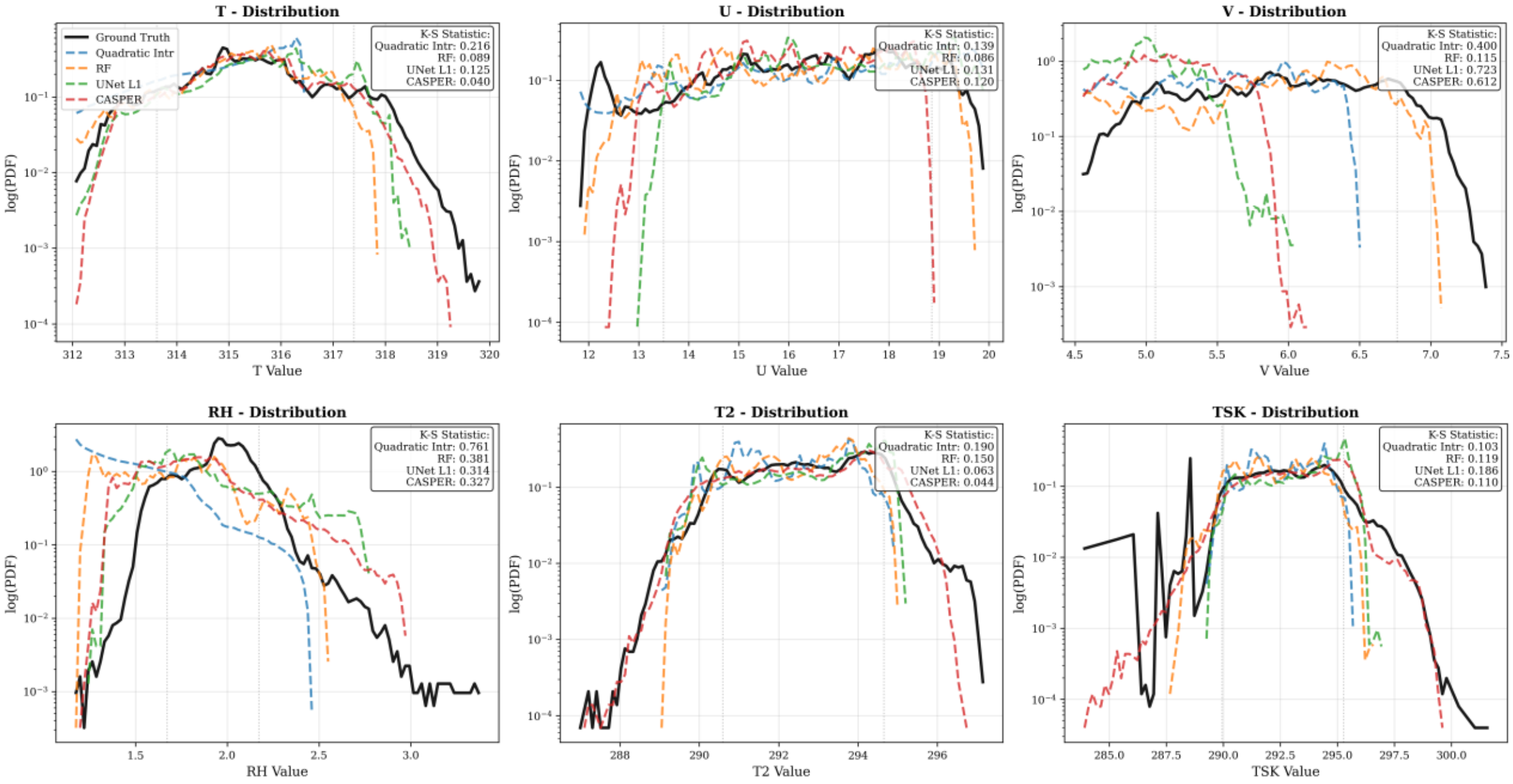

**Supplementary Figure S9: Statistical fidelity for the typical test set. Probability density functions (logarithmic scale) for the six output variables, comparing the WRF ground truth with quadratic interpolation, random forest, U-Net L1 and CASPER; the**

**Kolmogorov–Smirnov statistic of each method is annotated per panel. Distributions here are computed on a 50-sample subset at a single output level, so the annotated statistics are not directly comparable with the full-test-set means quoted in the main text.**

Accurate representation of probability distributions across the full range is essential for risk assessment and climate adaptation applications during extreme events. Supplementary Figure S10 examines distributional fidelity through probability density functions.

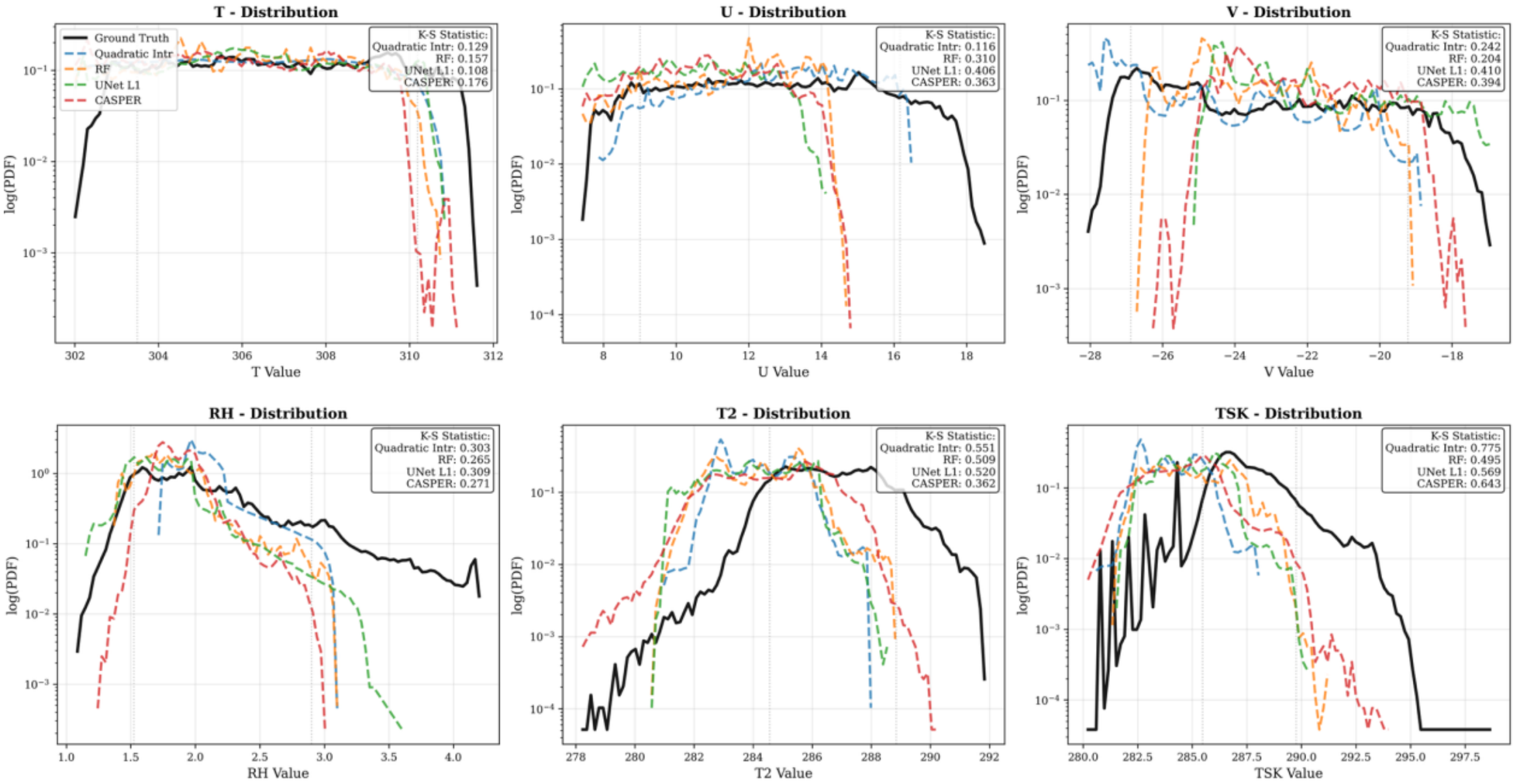

**Supplementary Figure S10: Statistical fidelity for the extreme test set.** As in Supplementary Figure S9, for the extreme test set.

### 3.4 Physical Consistency Across Variables and Vertical Levels

Atmospheric variables exhibit coupled relationships, constrained by thermodynamic conditions and dynamical balances that trained models must preserve. Supplementary Figure S11 examines multivariate relationships to validate that the model learns fundamental atmospheric coupling.

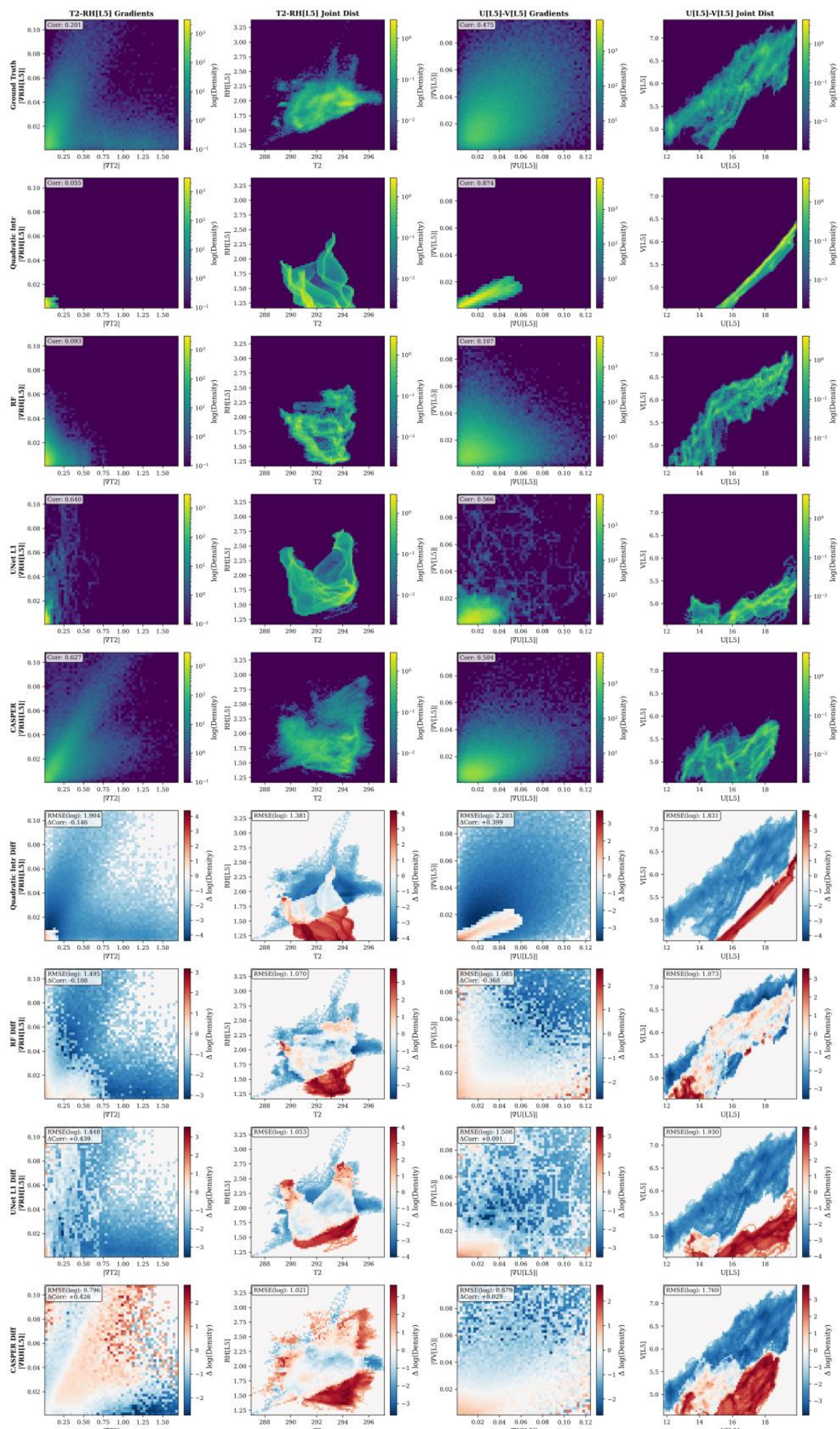

**Supplementary Figure S11: Multivariate physical consistency for the typical test set.** Cross-variable spatial gradients (columns 1 and 3) and joint distributions (columns 2 and 4) for the temperature–humidity and wind–wind pairs: the upper rows compare the WRF ground truth with each matched-budget method, and the lower rows show each method's difference from the WRF density, with the log-density RMSE annotated on every difference panel and the change in cross-variable correlation on the gradient panels. L5 in the panel titles denotes that same mid-tropospheric level.

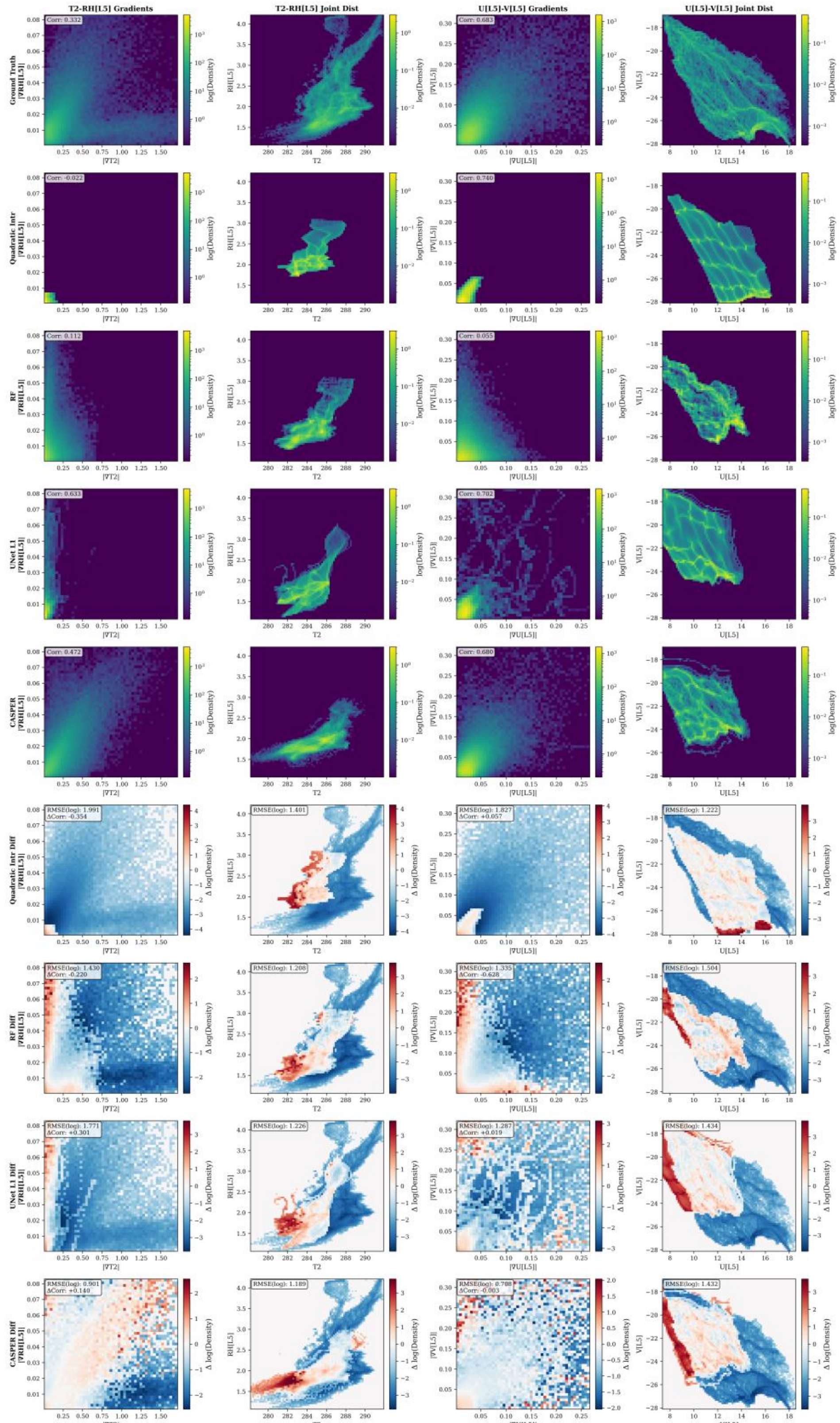

**Supplementary Figure S12: Multivariate physical consistency for the extreme test set.** As in Supplementary Figure S11, for the extreme test set.

Atmospheric predictions must also maintain realistic vertical structure across all pressure levels to ensure physical consistency for downstream applications such as radiation calculations and aviation forecasting, especially during extreme events. Supplementary Figure S13 validates predictions across the tropospheric column from surface to upper levels.

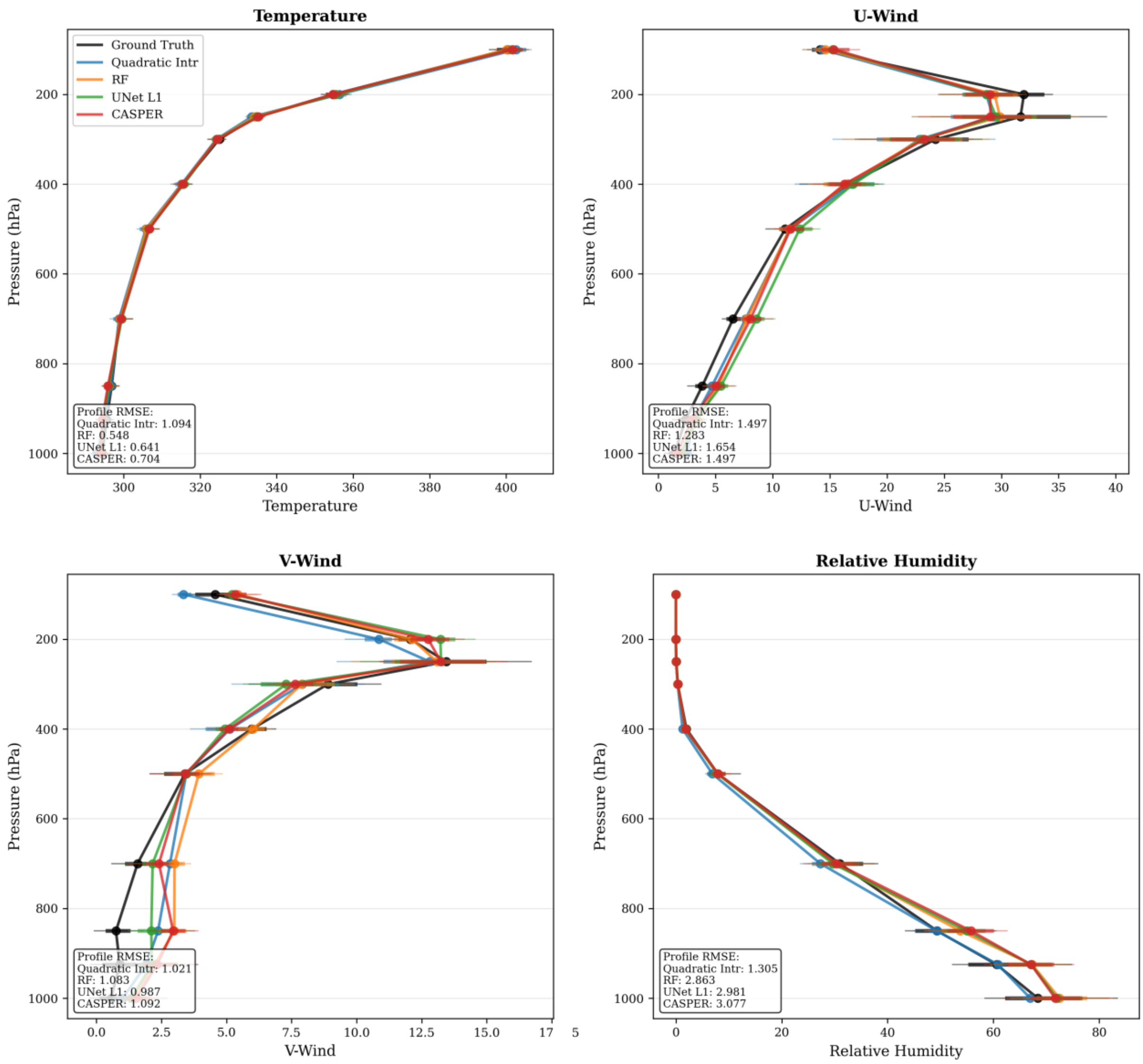

**Supplementary Figure S13: Vertical profile validation for the typical test set.** Domain-mean profiles of temperature (WRF potential temperature), zonal and meridional wind, and relative humidity across the output pressure levels, for the WRF ground truth and the four matched-budget methods; horizontal bars show the spread across samples and the profile RMSE of each method is annotated.

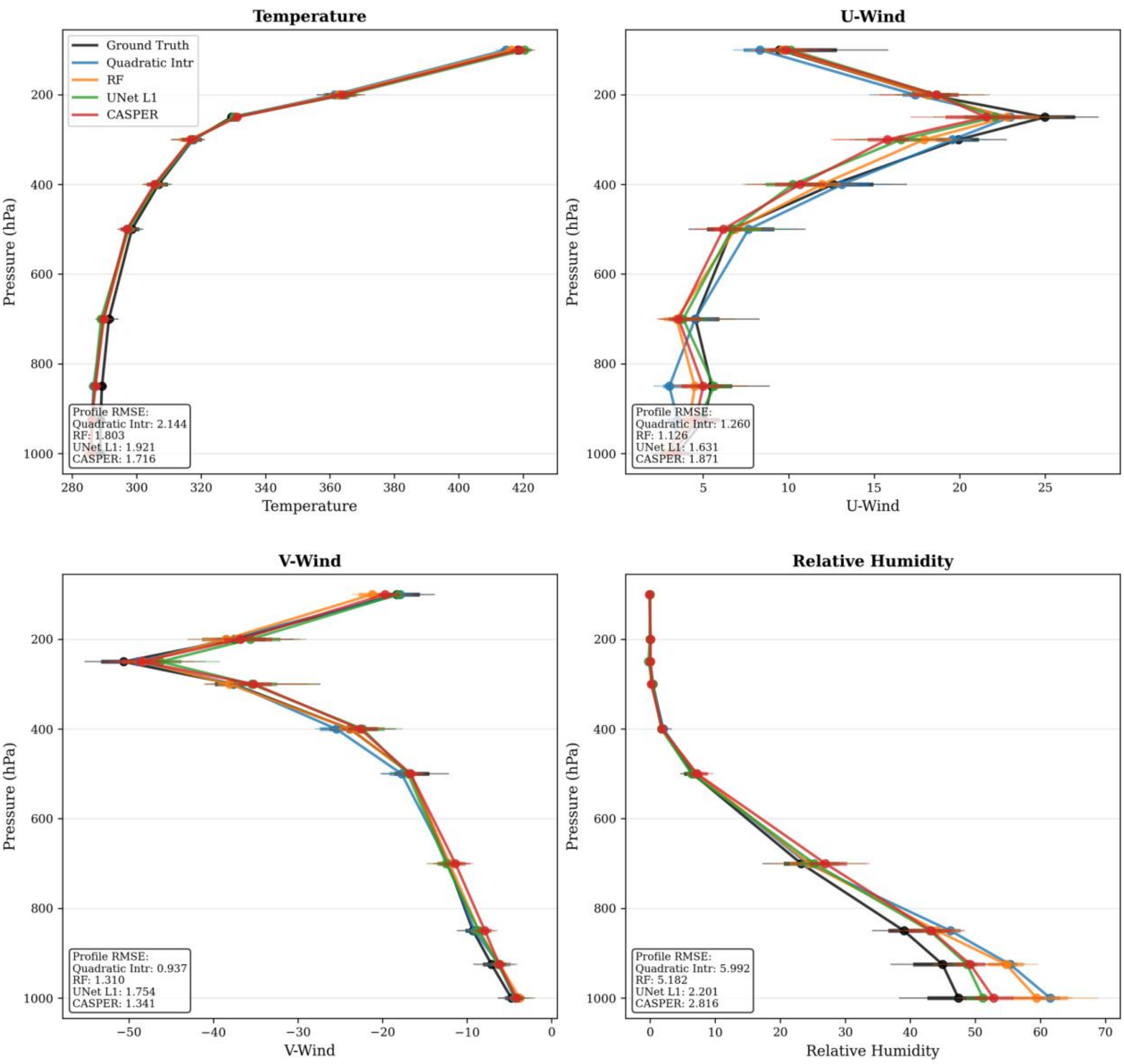

**Supplementary Figure S14: Vertical profile validation for the extreme test set.** As in Supplementary Figure S13, for the extreme test set.

## Supplementary Note 4: Case Studies and Transfer

Supplementary Figures S15 and S16 extend the main-text transfer analysis: Supplementary Figure S15 shows the remaining 970 hPa variables for the three transfer cities, and Supplementary Figure S16 the zero-shot versus few-shot comparison on the June 2021 western heat dome.

**(a) Vancouver · Jul 2016 simulation · held-out sample**

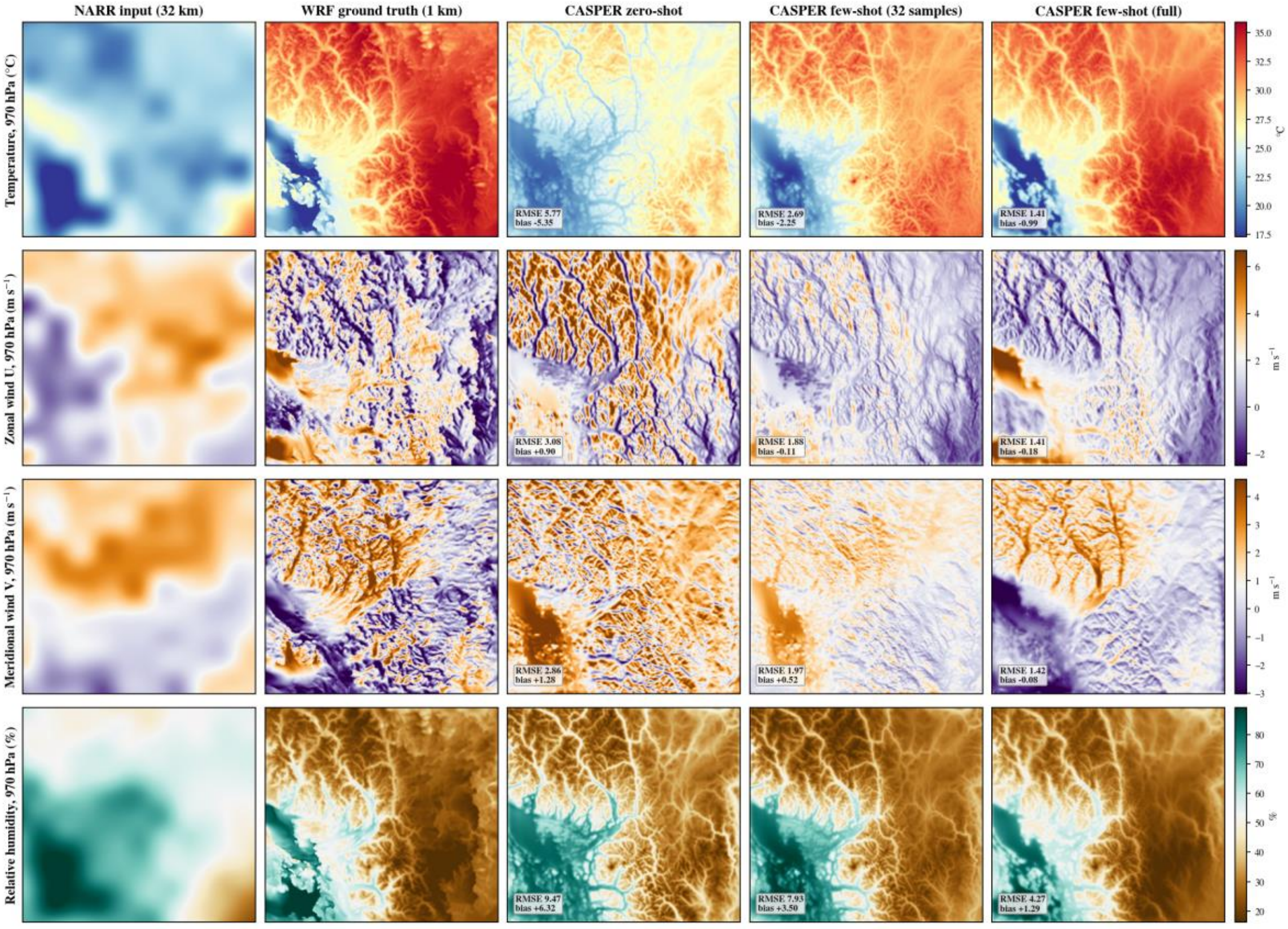

**(b) Calgary · May 1983 simulation · held-out sample**

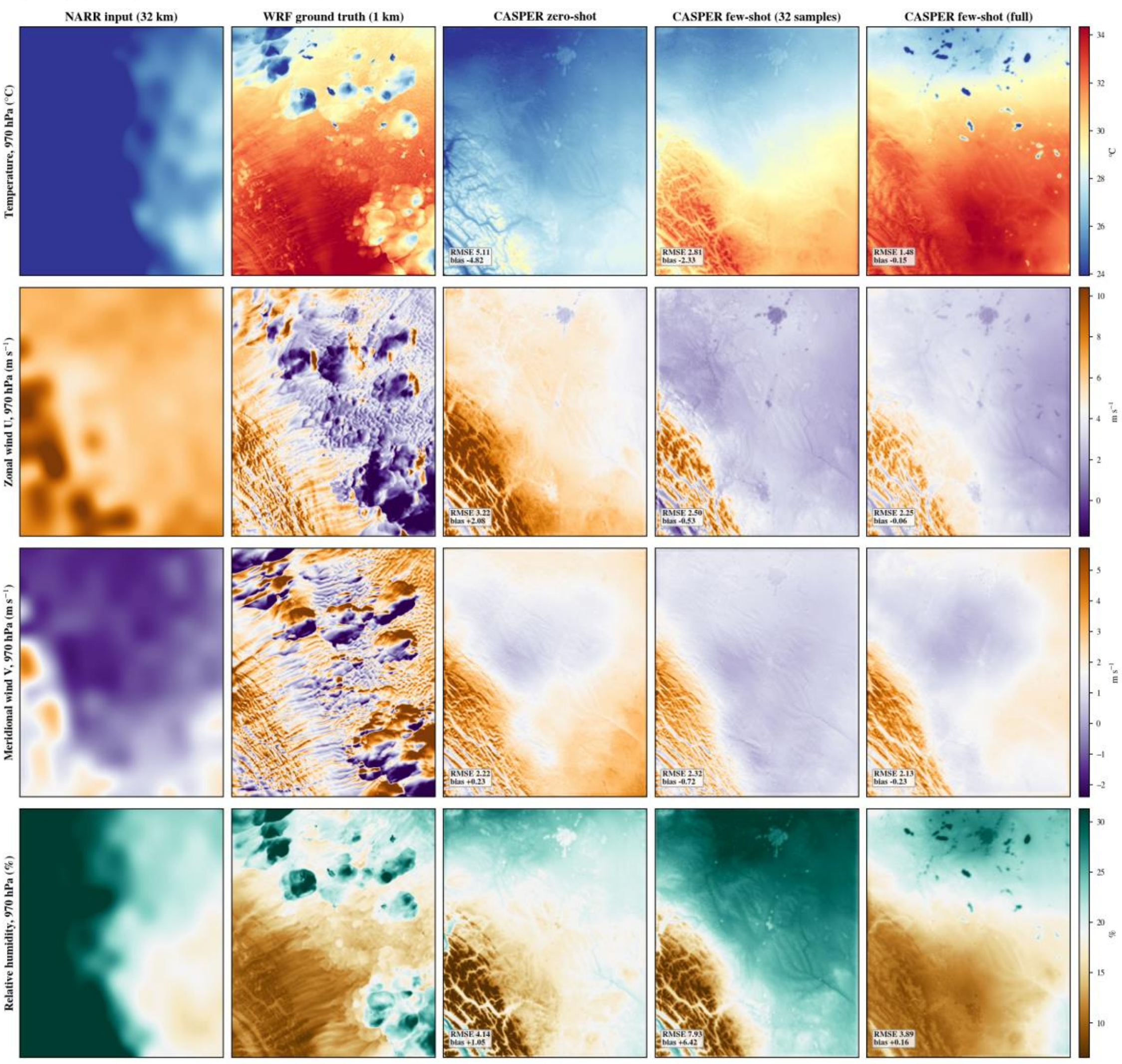

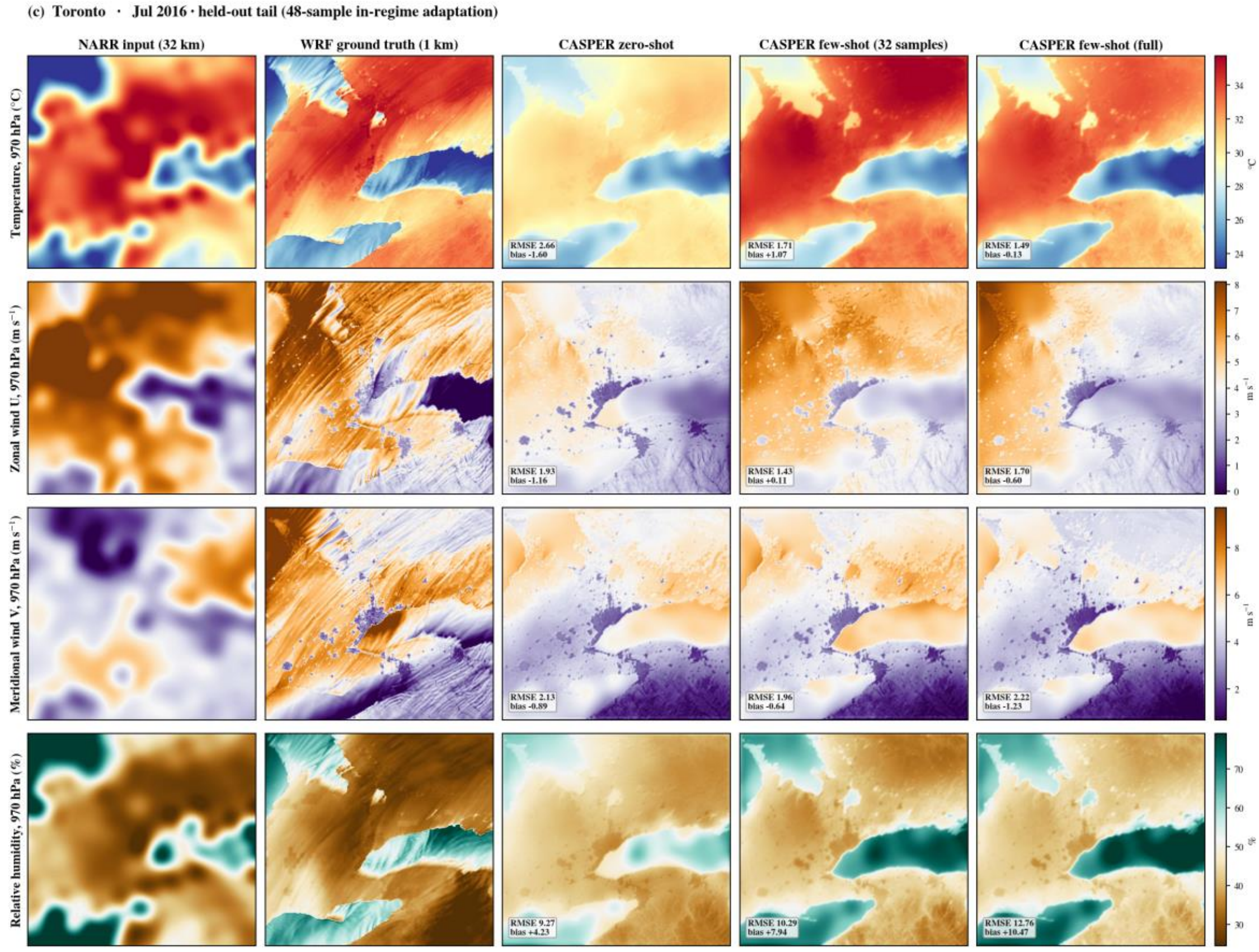

**Supplementary Figure S15: Transfer by decoder adaptation, remaining variables.** As in the main-text transfer figure but for the 970 hPa temperature, zonal and meridional wind, and relative humidity, for (a) Vancouver, (b) Calgary and (c) Toronto, each on a held-out sample of its own 1 km simulation. Columns are the 32 km NARR input, the WRF 1 km ground truth, CASPER zero-shot, and the same model after decoder-only fine-tuning on 32 samples and on the full adaptation window; per-panel root-mean-square error and bias are annotated.

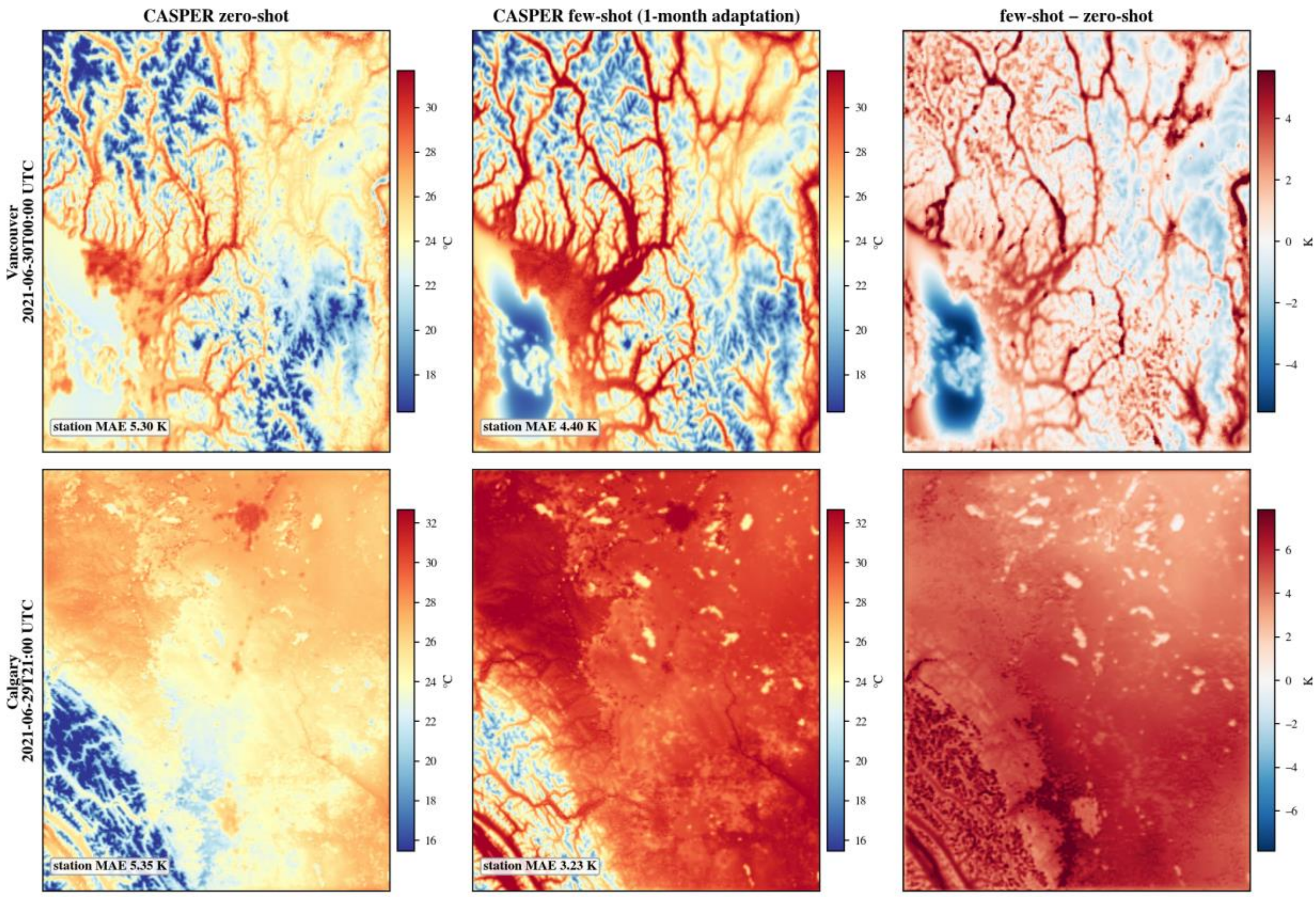

**Supplementary Figure S16: Extreme-event transfer - CASPER zero-shot versus few-shot on the June 2021 western heat dome, for Vancouver and Calgary.** Columns show the zero-shot 2 m temperature, the few-shot 2 m temperature, and their difference at the peak heat-dome hour; the mean absolute error against independent ECCC station observations is annotated. Adaptation warms the cold-biased zero-shot fields, reducing station error from 5.30 to 4.40 K in Vancouver and 5.35 to 3.23 K in Calgary.

## Supplementary Note 5: Methods

### 5.1 Study Domains

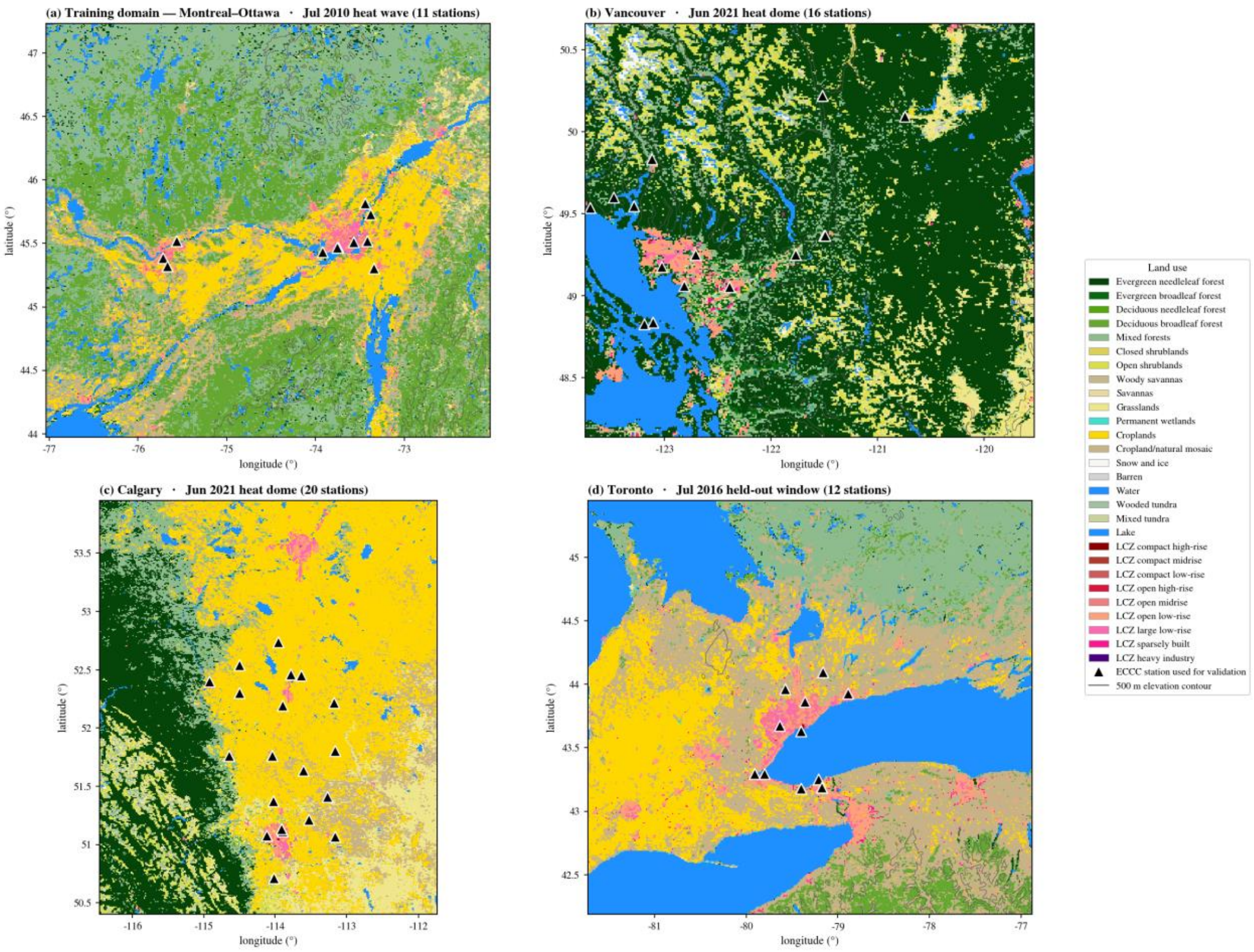

**Supplementary Figure S17: Training and testing domains with validation weather stations.** Land use of the four 1 km domains: (a) the Montreal–Ottawa training domain and the (b) Vancouver, (c) Calgary and (d) Toronto transfer domains, with the ECCC stations used for validation marked as triangles — the July 2010 heat wave for the training domain (11 stations), the June 2021 western heat dome for Vancouver and Calgary (stations with resolvable coordinates: 16 of 17 and all 20), and the July 2016 window for Toronto (the 12 stations inside the 3 km co-location gate). The grey line is the 500 m elevation contour; the Toronto grid is the shoreline-georeferenced reconstruction described in Methods.

### 5.2 WRF Model Configuration

High-resolution target fields for model training and evaluation were generated using the Weather Research and Forecasting (WRF) model version 4.3 [27]. WRF simulations were initialized and forced with North American Regional Reanalysis (NARR) data at 32 km horizontal resolution and 3-hourly temporal resolution [28]. The model domain covers the Montreal-Ottawa region with 363×390 grid points at 1 km horizontal spacing. The vertical coordinate system employs 40 pressure levels, which were then reduced to 10 for computational efficiency, spanning approximately 970 hPa near the surface to 105 hPa in the upper troposphere. The simulation follows validated setup in the same study domain [29]. Supplementary Table S3 summarizes the complete physics configuration, and Supplementary Figure S17 shows the training and testing domains of our study area.

**Supplementary Table S3: WRF model physics configuration.**

| Component | Configuration |
|---|---|
| Boundary conditions | NARR 32 km, 3-hourly |
| Microphysics | WRF Single – Moment 3 |
| Cumulus parameterization | Kain-Fritsch |
| Planetary boundary layer | BouLac |
| Surface layer | Eta similarity |
| Land surface model | Unified Noah |
| Longwave radiation | RRTM |
| Shortwave radiation | Dudhia |

### 5.3 Choice of Distance Metric

The error–distance scaling requires a distance between a model's training distribution and an evaluation target in the two-variable climate space. Several definitions are defensible, so we selected one by cross-validation rather than by assumption. Each candidate was used to refit the relation, which was then scored by leave-one-month-out cross-validation, blocking on evaluation month because the model–month pairs derive from only eight distinct climate states. This comparison was run on a pilot set of 45 model-month pairs from eight configurations; the 0.89 quoted in the main text is the same statistic over the full set of 112 pairs. Distance to the centre of the training cloud, standardized per axis by the training cloud's standard deviation, achieved an out-of-sample coefficient of determination of 0.902 on the pilot set. Distances measured beyond the edge of the training support scored higher (0.934 for the mean over target samples, 0.938 for the 90th percentile), and nearest-neighbour distances scored slightly lower (0.887 for the single nearest training sample, 0.892 for the mean of the two nearest). The fraction of the target lying outside the training support, which discards magnitude information, performed markedly worse at 0.657, and a Mahalanobis distance using the training covariance performed worst of all at 0.731, because the sign of the temperature–humidity correlation flips across the eight months, making the pooled covariance unstable to invert.

We retain the centre distance for the reported relation. Although edge-based distances fit marginally better, their fitted intercept implies an error floor at zero distance that is not physically credible, and the centre distance is the quantity a practitioner can compute directly from monthly climatological statistics. Adding a second predictor did not help: combining centre and edge distances improved the in-sample fit but reduced out-of-sample performance, so the single-term relation is retained.

The edge-based definitions also require a definition of the training support itself. We compared a per-axis bounding box, the convex hull, a union of balls whose radius was set from the training cloud's own nearest-neighbour distances, and the highest-density region of a kernel density estimate. The simple bounding box performed best. Tighter definitions classify almost no evaluation target as lying wholly inside the support, because a few of its samples always fall outside, so the distance never reaches zero and degenerates towards a nearest-neighbour distance.

### 5.4 Selection-Strategy Experiment

Held-out January 1981 was the target, with geography held fixed so the only gap between model and target is climatological. Two base models (one and four training months) were fine-tuned on N samples chosen by each strategy, for three random seeds (Supplementary Table S8). Values are mean ± s.d. over seeds of the 2 m temperature RMSE (K) on the fixed test split, and on its most distant fifth.

### 5.5 Network Architecture

The fundamental architecture follows the U-Net encoder-decoder paradigm [5], with several domain-specific modifications for atmospheric downscaling. The encoder comprises six hierarchical levels with progressively increasing feature dimensionality as detailed in Supplementary Table S4.

**Supplementary Table S4: Model architecture configuration parameters.** Feature dimensions represent channel counts at each of the six encoder-decoder levels.

| Component | Specification | Value |
|---|---|---|
| Encoder levels | Number of downsampling stages | 6 |
| Decoder levels | Number of upsampling stages | 6 |
| Base channels | Initial feature dimensionality | 256 |
| Channel multipliers | Per-level expansion factors | [1, 2, 3, 4, 5, 6] |
| Feature dimensions | Channel counts by level | [256, 512, 768, 1024, 1280, 1536] |
| Attention heads | Multi-head attention configuration | 8 |
| Attention levels | Encoder-decoder stages with attention | [4, 5, 6] |
| Attention threshold | Maximum spatial dimension for attention | 32×32 |
| GroupNorm groups | Normalization group count | 32 |
| Dropout rate | Spatial dropout probability | 0.1 |
| Positional encoding | Number of frequency octaves | 4 |
| Total parameters | Learnable weights (millions) | 625.7 |

Beginning from the 53 atmospheric input channels, the architecture incorporates eight additional channels of sinusoidal positional encoding, yielding an initial 61-channel tensor that enters the first encoder block operating at full 363×390 spatial resolution. The encoder expands features via convolutional operations, with channel counts following the sequence [256, 512, 768, 1024, 1280, 1536], as specified in Supplementary Table S4. Each downsampling operation, a 3×3 convolution with stride 2, halves the spatial resolution while the channel count increases with depth, enabling the network to capture increasingly coarse-scale atmospheric patterns across the spatial hierarchy.

Positional encoding provides explicit spatial information essential for learning position-dependent atmospheric processes such as terrain-forced flows and land-sea contrasts. The implementation employs sinusoidal functions at four frequency octaves. For each octave $i \in \{0,1,2,3\}$, the model computes frequency $f_i = 2^i$ and generates two encoding channels via $\sin(f_i \pi y_{\text{norm}})$ and $\cos(f_i \pi x_{\text{norm}})$, where $y_{\text{norm}}$ and $x_{\text{norm}}$ represent normalized pixel coordinates spanning the interval [-1, 1]. This multi-frequency representation enables learning both large-scale geographic patterns captured by low-frequency components and fine-scale position-dependent features represented by high-frequency components, which are crucial for modeling topographic influences and circulation patterns that vary systematically across the domain.

Each encoder and decoder level contains ResidualBlocks [30] that preserve gradient flow through the deep network architecture. A ResidualBlock implements the following computational sequence: (i) 3×3 convolution with padding to maintain spatial dimensions, configured without bias terms since subsequent normalization layers include affine transformations; (ii) GroupNorm with 32 groups providing batch-independent normalization suitable for small batch training [31]; (iii) SiLU activation function [32] defined as $f(x) = x \cdot \sigma(x)$ where $\sigma$ represents the sigmoid function; (iv) Dropout2d [33] with rate 0.1 providing spatial regularization by randomly zeroing entire feature maps during training; (v) second 3×3 convolution followed by GroupNorm; and (vi) residual addition with 1×1 projection convolution applied when input and output channel dimensions differ. This design enables training of very deep networks exceeding 100 layers while maintaining stable gradient magnitudes throughout the optimization process.

$\text{Attention}(Q,K,V) = \text{softmax}(QK^T/\sqrt{d_k})V\, d_k$Self-attention mechanisms operate at the four deepest encoder and decoder levels and at the bottleneck, where computational costs remain tractable due to reduced spatial dimensions. The attention module activates only when feature map spatial dimensions fall below 32×32 pixels; larger feature maps bypass attention operations to manage GPU memory requirements during training. With eight attention heads, the mechanism computes scaled dot-product attention following where represents the per-head dimension, calculated as the total number of channels divided by 8 [21]. This architectural choice allows the network to aggregate global atmospheric context at coarse spatial scales where long-range teleconnections and large-scale circulation patterns dominate, while relying on convolutional operations for efficient local feature extraction at finer spatial scales where computational costs would otherwise become prohibitive.

The decoder implements a symmetric structure mirroring the encoder through six upsampling stages. Each decoder level begins with 2× bilinear upsampling of features from the previous coarser level, followed by channel-wise concatenation with skip connections originating from the corresponding encoder level. These skip connections represent a fundamental component of U-Net performance, providing the decoder with high-resolution spatial information inevitably lost during the encoding bottleneck compression and enabling sharp reconstruction of fine-scale atmospheric features. The concatenated feature tensors pass through ResidualBlocks structurally identical to encoder blocks but parameterized with independently learned weights. The final decoder level produces 256-channel features at full spatial resolution (363×390 pixels), which a subsequent 1×1 convolution projects to the 42 output channels representing the predicted atmospheric state across all variables and vertical levels.

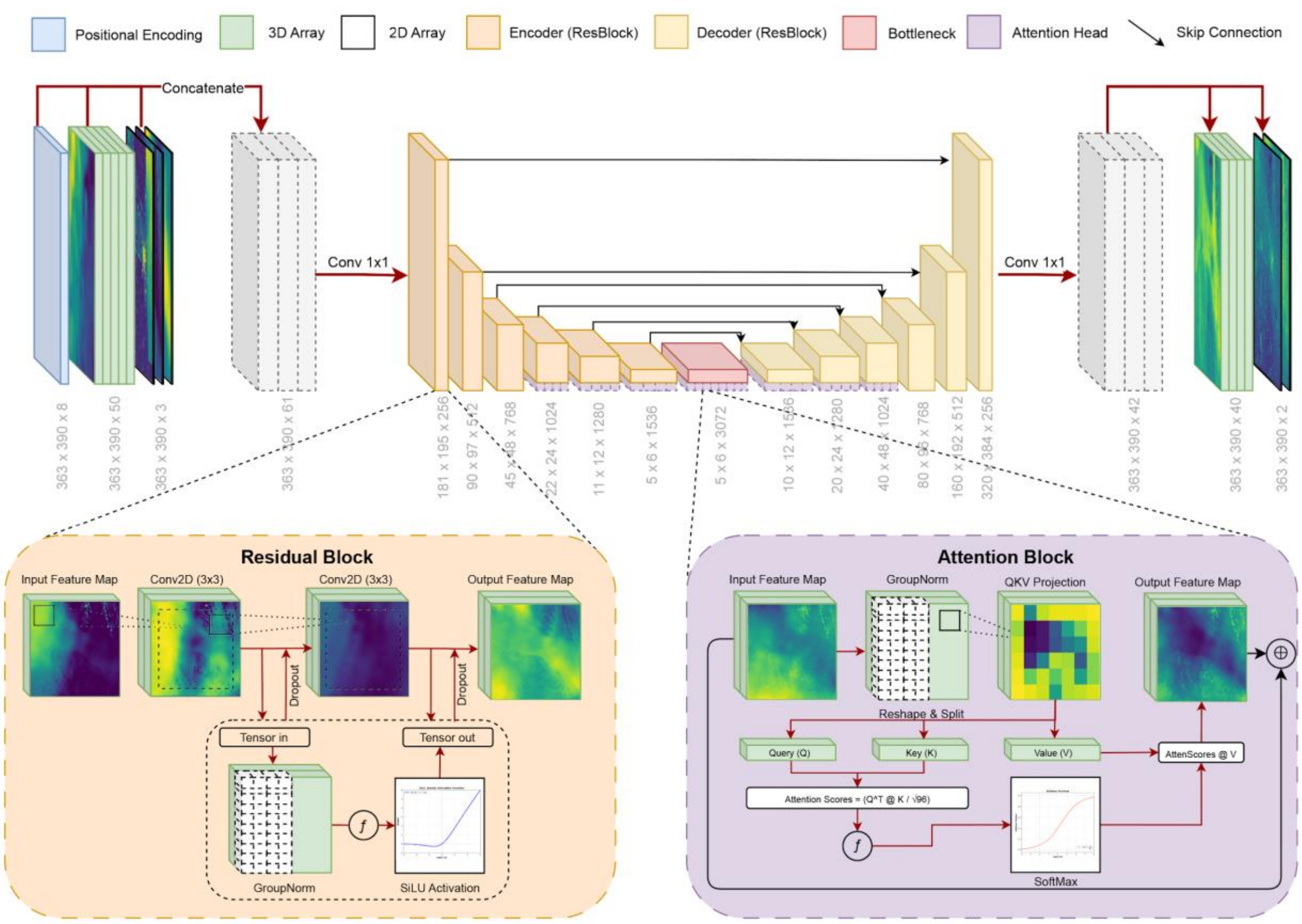

**Supplementary Figure S18: CASPER model architecture and training design.** The six-level U-Net encoder-decoder with skip connections (black arrows) maps a 53-channel input to 42 output channels. Orange blocks (ResidualBlocks, detailed at bottom-left) use two sequential 3×3 convolutions with GroupNorm and SiLU activation; purple blocks (multi-head self-

attention, detailed at bottom-right) operate at the three deepest encoder levels and the bottleneck, providing global context for synoptic-scale patterns.

### 5.6 Parameter Budget

The network contains 625.7 million trainable parameters, distributed as follows: decoder path 173.0 M (27.6%), bottleneck 169.9 M (27.2%), encoder path 100.2 M (16.0%), learned upsampling 83.8 M (13.4%), strided-convolution downsampling 53.7 M (8.6%), and encoder and decoder attention 22.6 M each (3.6% each). The bottleneck is a single block operating on 3,072 channels with 8 attention heads, which is why so large a share of the parameters sits at the coarsest resolution.

### 5.7 Loss Function Design

The training objective combines multiple loss components designed to preserve different aspects of atmospheric structure and spatial fidelity. Unlike conventional single-objective losses that optimize exclusively for point-wise accuracy, the composite loss function ensures the model simultaneously learns accurate field values and realistic spatial patterns characteristic of atmospheric flows. The complete loss function specification, including weights and mathematical formulations, appears in Supplementary Table S5.

**Supplementary Table S5: Composite loss function weights and component descriptions.**

| Loss Component | Weight | Purpose |
|---|---|---|
| L1 Reconstruction | 0.5 | Point-wise accuracy |
| Gradient Preservation | 2.0 | Sharp meteorological features |
| Multi-scale Structural | 0.5 | Hierarchical spatial consistency |
| Patch Coherence | 0.1 | Local smoothness |
| Multivariate Bias Correction (MBC) | 0.1 | Multivariate distributional fidelity |

The L1 formulation provides robustness to outliers while maintaining sensitivity to systematic biases, with a weighting of 0.5 in the composite objective function, as shown in Supplementary Table S5.

Sharp meteorological features such as temperature fronts, wind shear zones, and moisture gradients require explicit penalties on spatial derivatives. The gradient loss component computes spatial derivatives via normalized Sobel operators, implementing 3×3 convolutional kernels defined as:

$$S_x = \frac{1}{8}\begin{bmatrix} -1 & 0 & 1 \\ -2 & 0 & 2 \\ -1 & 0 & 1 \end{bmatrix}, \quad S_y = \frac{1}{8}\begin{bmatrix} -1 & -2 & -1 \\ 0 & 0 & 0 \\ 1 & 2 & 1 \end{bmatrix}$$

For each atmospheric field channel, the model computes x-direction and y-direction spatial derivatives.

To manage GPU memory constraints when processing 42 output channels across large spatial domains, the implementation processes atmospheric variables in chunks of 32 channels, computing gradients independently for each chunk before aggregating results. This chunked processing strategy enables training on consumer-grade GPUs with 24GB memory. The gradient loss receives a weight of 2.0 in the composite objective function.

Atmospheric processes inherently span multiple spatial scales, from synoptic weather systems extending thousands of kilometres to mesoscale features on the order of tens of kilometres. The multi-scale structural loss enforces consistency across this spatial hierarchy by computing structural similarity at four progressively coarser resolutions. Starting from native resolution (363x390 pixels), the model generates three additional scales via 2x2 average pooling: half resolution (181x195), quarter resolution (90x97), and eighth resolution (45x48). The multi-scale loss aggregates these scale-specific components with weights [0.4, 0.3, 0.2, 0.1] that progressively de-emphasize coarser scales. This weighting scheme ensures the model preserves both large-scale synoptic circulation patterns and fine-scale terrain-forced features that emerge at native resolution. The multi-scale loss receives a weight of 0.5 in the composite objective function.

Local spatial coherence prevents checkerboard artifacts and ensures smooth transitions between neighboring atmospheric features. The patch loss randomly extracts 32 patches of size 16×16 pixels from each prediction-target pair during training. For each patch, the loss computes normalized correlation between predicted and target fields. When the spatial dimensions fall below 16×16 pixels, the implementation computes global correlation rather than patch-based statistics. This patch loss component, weighted at 0.1 in the composite objective, provides regularization encouraging local spatial consistency without imposing the excessive smoothness characteristic of global correlation penalties.

The Multivariate Bias Correction (MBC) loss [34] enforces statistical fidelity across coupled atmospheric variables at each vertical level. Unlike point-wise losses that treat variables independently, MBC loss preserves the multivariate distributional structure, which is essential for maintaining physical relationships among temperature, wind, and humidity. The MBC loss operates independently at each of the ten pressure levels, computing two complementary metrics that, together, ensure distributional consistency.

The first component employs energy distance [35], a metric for comparing multivariate distributions. This metric quantifies the discrepancy between joint distributions of all atmospheric variables at a given level, capturing correlations and dependencies that univariate metrics cannot detect. To manage computational costs, energy-distance calculations subsample 1,200 spatial points per level.

The second component uses quantile-mapping distance, which measures how well the predicted quantiles align with the target quantiles across the full distribution for each variable. For 99 evenly spaced quantile levels from 0.01 to 0.99, the loss computes the mean absolute difference between the predicted and target quantile values. This ensures the model reproduces not just the mean and variance but the complete distributional shape, including tails representing extreme events, which are critical for weather applications where rare extremes often matter most.

The total training objective combines these five components as:

$$\mathcal{L}_{\text{total}} = w_1\mathcal{L}_{\text{L1}} + w_2\mathcal{L}_{\text{grad}} + w_3\mathcal{L}_{\text{multi-scale}} + w_4\mathcal{L}_{\text{patch}} + w_5\mathcal{L}_{\text{MBC}}$$

with weights [$w_1$, $w_2$, $w_3$, $w_4$, $w_5$] = [0.5, 2.0, 0.5, 0.1, 0.1] determined through systematic experimentation.

**5.8 Ablation Study**

To understand the contribution of each loss component and identify optimal weighting configurations, we conducted systematic ablation experiments that varied the weights of the L1 (w_L1), gradient (w_Grad), multi-scale structural (w_Struct), patch (w_Patch), and MBC (w_MBC) loss terms. Supplementary Table S6 summarizes 15 model configurations designed to test specific hypotheses about loss component interactions, internal MBC metric balancing (quantile mapping Q vs. energy distance E), and the necessity of static geographic features. Supplementary Figures S19–S22 present quantitative comparisons across spatial accuracy, spectral fidelity, and distributional preservation metrics on the typical test set. This ablation was carried out with the earlier four-month warm-season training configuration; the selected weights were retained unchanged for every model trained in this work.

**Supplementary Table S6: Ablation study configurations testing loss component contributions and optimal weighting.**

| Model ID | w_L1 | w_Grad | w_Struct | w_Patch | w_MBC | Internal MBC Split | Purpose |
|---|---|---|---|---|---|---|---|
| M0 (U-Net L1) | 1 | 0 | 0 | 0 | 0 | N/A | Baseline without static features: U-Net L1 model |
| M1 | 1 | 0 | 0 | 0 | 0 | N/A | Pure L1 reconstruction: establishes systematic over-smoothing baseline |
| M2 | 1 | 0.5 | 0.5 | 0.1 | 0 | N/A | Structure-preserving losses only: tests spatial sharpness without distributional correction |
| M3 | 1 | 0.5 | 0.5 | 0.1 | 0.1 | Q=0.8, E=0.2 | Minimal distributional correction: tests if light MBC improves extreme value tails |
| M4 | 1 | 0.5 | 0.5 | 0.1 | 0.5 | Q=0.8, E=0.2 | Moderate distributional correction: balances spatial structure and statistical fidelity |
| M5 | 0.5 | 0.5 | 0.5 | 0.1 | 1 | Q=0.8, E=0.2 | Maximum distributional fidelity: tests if extreme MBC weight over-constrains predictions |
| M6 | 1 | 0.5 | 0.5 | 0.1 | 0.2 | Q=0.2, E=0.8 | Energy distance emphasis: tests multivariate coupling over marginal distribution matching |
| M7 (CASPER) | 0.5 | 2 | 0.5 | 0.1 | 0.1 | Q=0.8, E=0.2 | Final CASPER configuration: emphasizes sharp meteorological features through enhanced gradient preservation |
| M8 | 0.1 | 1 | 1 | 0 | 1 | Q=0.8, E=0.2 | Physics-prioritized: tests if reducing point-wise accuracy improves physical consistency |
| M9 | 1 | 0.5 | 0.5 | 0.1 | 0.1 | Q=0.5, E=0.5 | Balanced MBC components with conservative weight: tests equal energy/quantile importance |
| M10 | 1 | 0.5 | 0.5 | 0.1 | 0.5 | Q=0.5, E=0.5 | Balanced MBC components with aggressive weight: strong distributional correction equally weighted |
| M11 | 1 | 1 | 0 | 0 | 0 | N/A | Sequential ablation step 1: isolates gradient loss contribution to boundary sharpness |
| M12 | 1 | 1 | 1 | 0 | 0 | N/A | Sequential ablation step 2: adds multi-scale structural consistency across atmospheric hierarchy |
| M13 | 1 | 1 | 1 | 1 | 0 | N/A | Sequential ablation step 3: adds patch coherence for local spatial smoothness |
| M14 | 1 | 1 | 1 | 1 | 1 | Q=0.5, E=0.5 | Sequential ablation step 4: complete loss with all components for comprehensive evaluation |

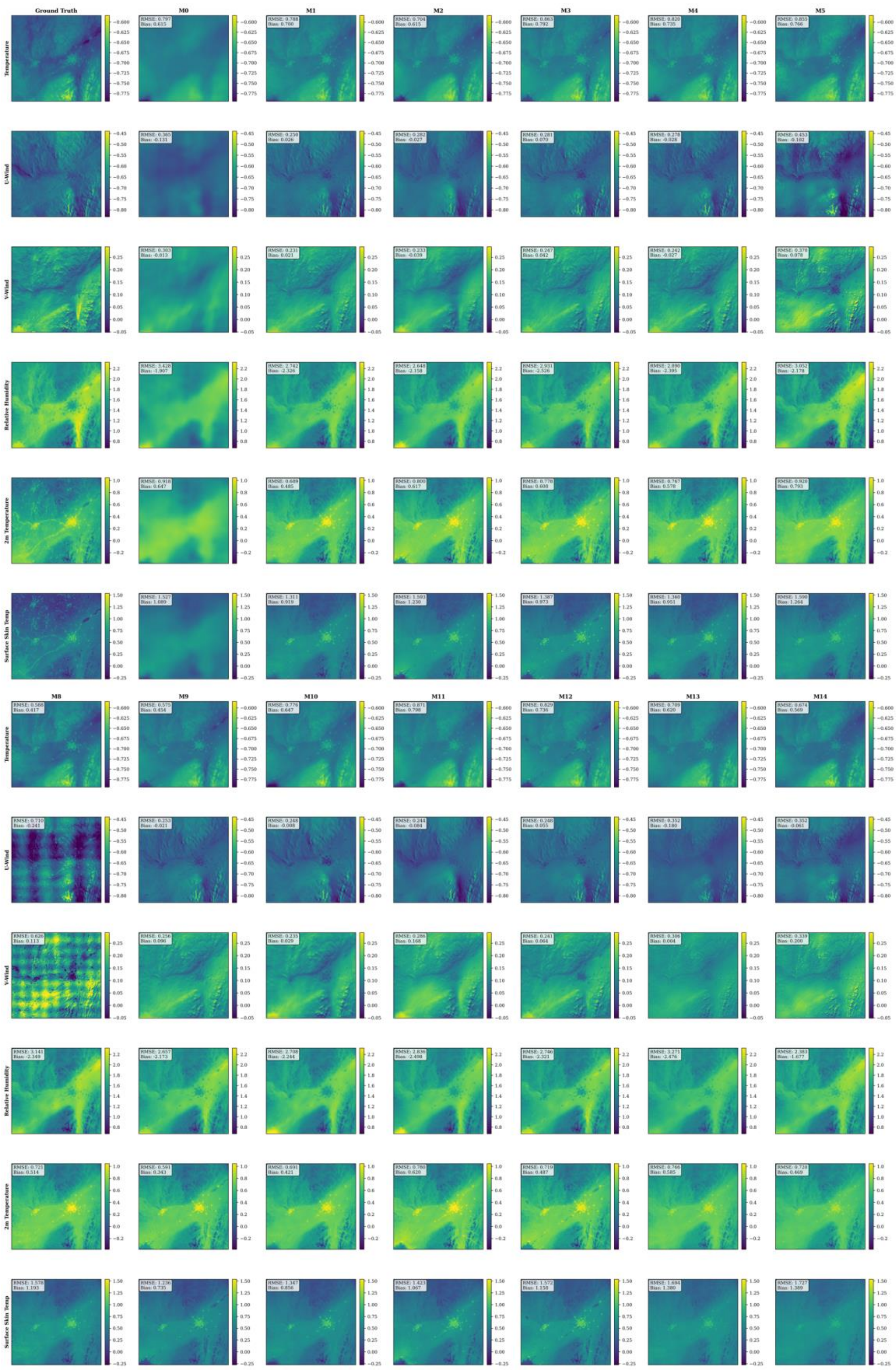

**Supplementary Figure S19: Spatial predictions from Level 970 hPa for 3D variables and surface variables.** Maps showing the predictions of 15 models with different weights for the loss function components compared to the WRF target for six atmospheric variables – temperature, U-wind, V-wind, relative humidity, 2 m temperature, and skin temperature –

(from top to bottom). Domain: 363×390 pixels at 1 km resolution covering the Montreal-Ottawa region.

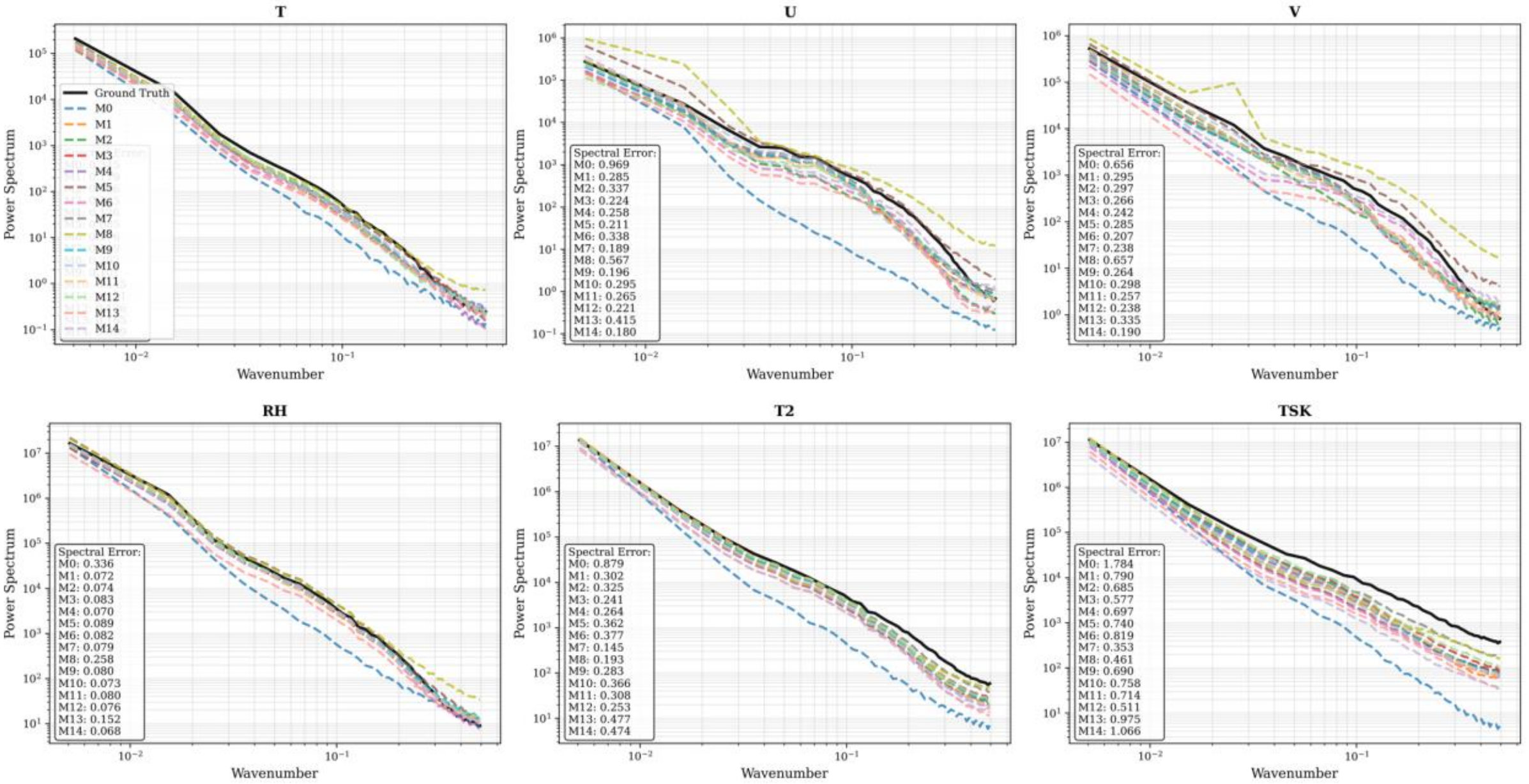

**Supplementary Figure S20: Spatial structure validation.** 2D power spectra (power spectral density vs. wavenumber) for temperature, U-wind, V-wind, relative humidity, 2-m temperature, and skin temperature. Comparison across 15 models with different weights for the loss function components with WRF target.

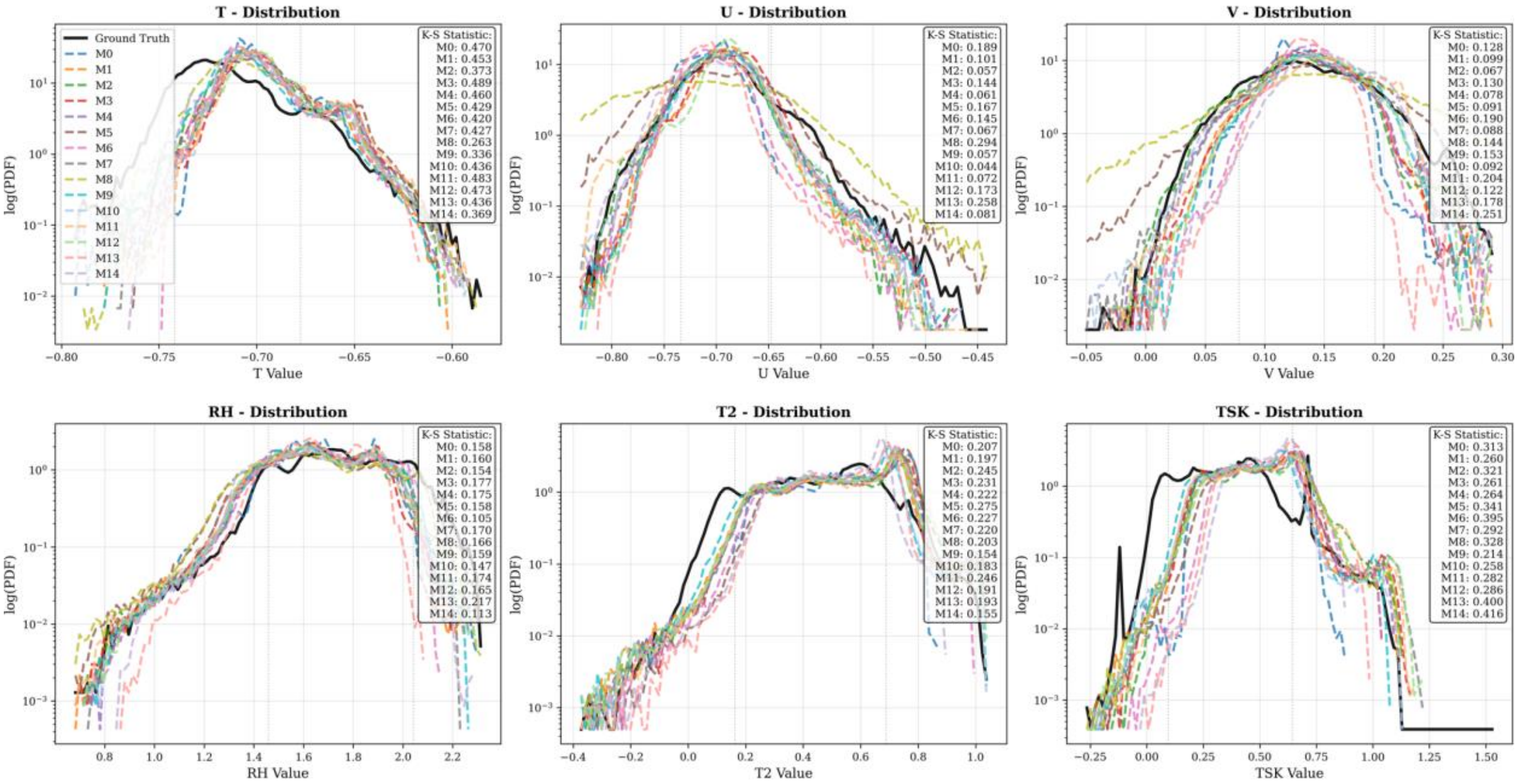

**Supplementary Figure S21: Statistical fidelity validation.** Probability density functions (log scale) for temperature, U-wind, V-wind, relative humidity, 2-m temperature, and skin temperature comparing 15 models with different weights for the loss function components with WRF target. The logarithmic y-axis emphasizes the tails of the distribution, which represent extreme events. Kolmogorov-Smirnov (KS) statistics quantify distributional agreement, with lower values indicating better fidelity.

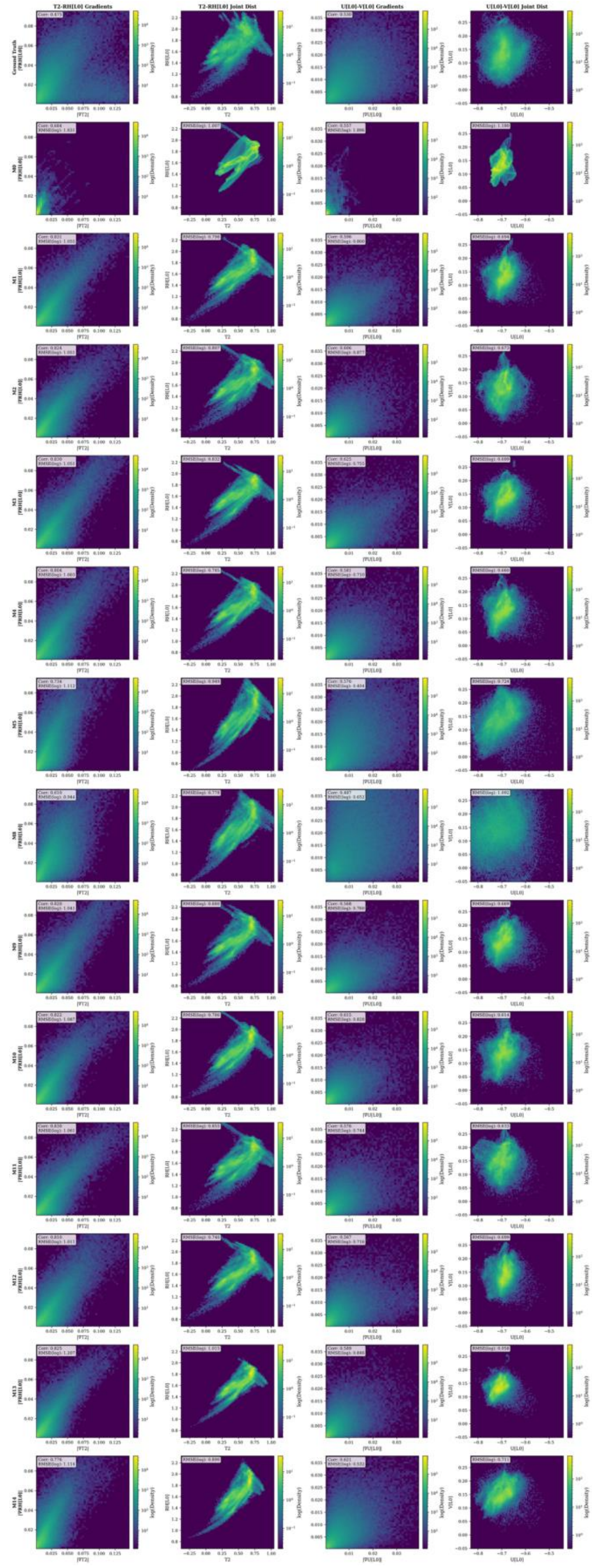

**Supplementary Figure S22: Multivariate consistency and spatial coupling validation.** Columns 1-2: T2-RH cross-variable gradients and joint distributions. Columns 3-4: U-V cross-

variable gradients and joint distributions. Comparison across 15 models with different weights for the loss function components with WRF target.

**5.9 Training Protocol**

Model training employs the AdamW optimizer [36] with initial learning rate $10^{-4}$ and weight decay regularization $10^{-4}$. The complete training hyperparameter configuration appears in Supplementary Table S7.

**Supplementary Table S7: Training hyperparameters and optimization configuration.**

| Parameter | Value | Description |
|---|---|---|
| Optimizer | AdamW | Adam with decoupled weight decay |
| Initial learning rate | $1\times10^{-4}$ | Maximum rate |
| Minimum learning rate | $1\times10^{-6}$ | Asymptotic rate from cosine annealing |
| Weight decay | $1\times10^{-4}$ | L2 regularization strength |
| Batch size (per GPU) | 2 | Samples per gradient computation |
| Gradient accumulation | 2 steps | Effective batch size multiplier |
| Effective batch size | 4 | Total samples per parameter update |
| Gradient clip norm | 1.0 | Maximum gradient magnitude |
| Number of epochs | 100 | Maximum training duration |
| Early stopping patience | 15 epochs | Tolerance before termination |
| Early stopping threshold | $1\times10^{-4}$ | Minimum improvement required |
| Checkpoint frequency | 5 epochs | Save interval |
| Data workers | 2 | Parallel data loading processes |

The AdamW formulation decouples weight decay from gradient-based parameter updates, improving generalization performance compared to standard Adam optimization. The relatively small per-GPU batch size of 2 samples reflects memory constraints imposed by processing 363×390 spatial grids across 53 input and 42 output channels with FP32. To improve gradient estimate stability without exceeding available GPU memory, training employs gradient accumulation over 2 steps as shown in Supplementary Table S7, yielding an effective batch size of 4 samples before parameter updates. Gradients are clipped to a maximum norm of 1.0 to prevent training instabilities that can arise when optimizing deep networks with attention mechanisms.

Learning rate scheduling follows a cosine annealing strategy defined as $\eta_t = \eta_{\min} + \frac{1}{2}(\eta_{\max} - \eta_{\min})(1 + \cos(\pi t/T))$ where $T = 100$ epochs represents the maximum training duration, $\eta_{\max} = 10^{-4}$ denotes the initial learning rate, and $\eta_{\min} = 10^{-6}$ specifies the minimum rate approached asymptotically. This schedule provides rapid initial convergence with high learning rates during early epochs, while enabling fine-scale parameter refinement in later epochs through gradual annealing toward the minimum learning rate.

Data loading operations employ 2 worker processes with memory pinning enabled to overlap CPU-GPU data transfer with GPU computation. Workers persist across training epochs, eliminating the overhead of subprocess initialization. Despite these optimizations, data loading occasionally creates bottlenecks in the training pipeline, suggesting that future work could benefit from faster storage systems or increased preprocessing parallelism to fully saturate GPU utilization.

Supplementary Figure S23 reports the training and validation loss of the nested one-, two-, four- and eight-month configurations, together with their in-sample and out-of-sample error.

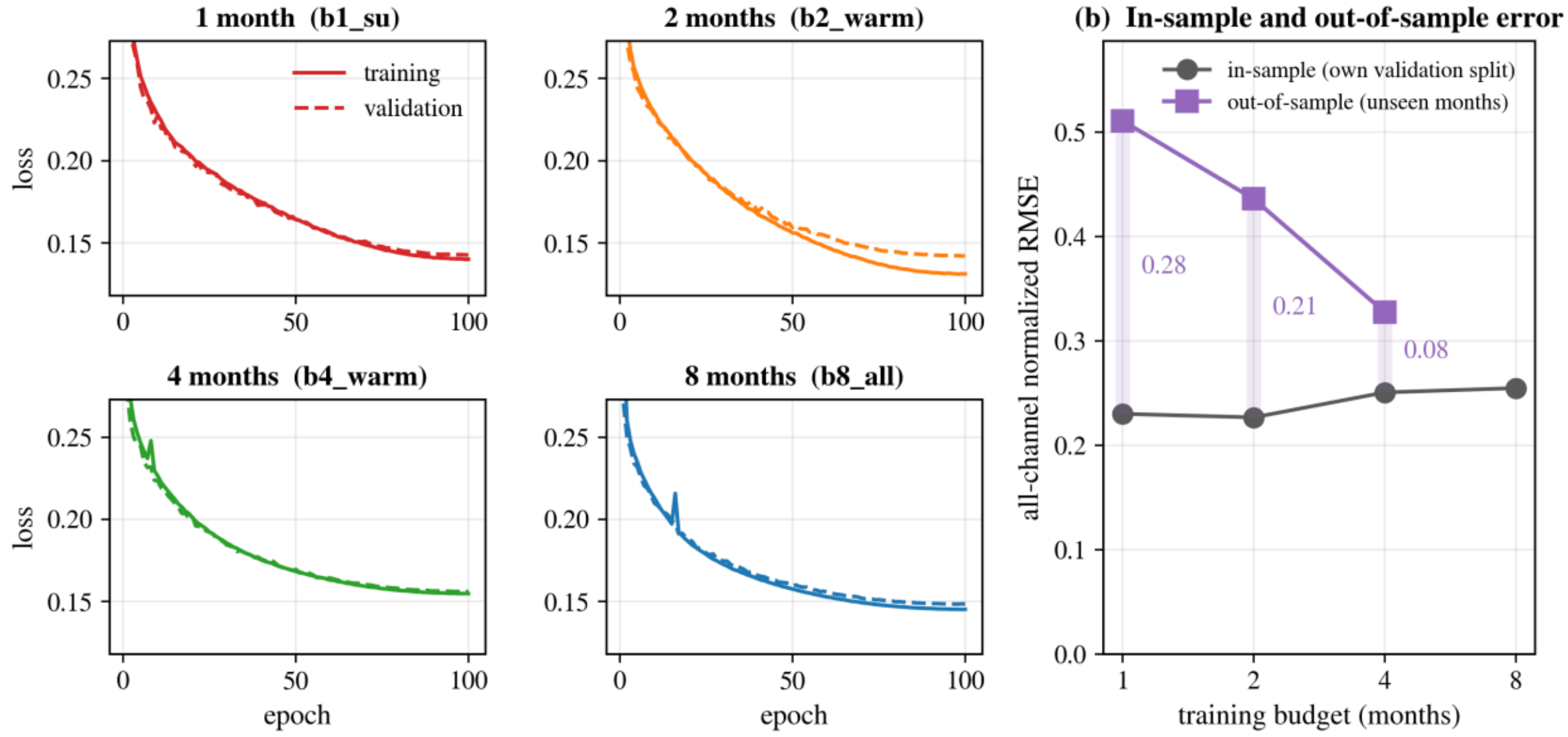

**Supplementary Figure S23: Training and validation loss across data budgets.** (a) Training (solid) and validation (dashed) loss over 100 epochs for the nested configurations of one, two, four and eight months (b1_su, b2_warm, b4_warm, b8_all); the two curves track each other throughout at every budget, and at the final epoch the validation loss exceeds the training loss by 1.8, 8.3, 0.8 and 2.2 % respectively. (b) The same models scored on their own validation split (in-sample) and on the four months withheld from every one-, two- and four-month configuration (out-of-sample), as all-channel normalized root-mean-square error, with the gap annotated; the eight-month configuration was trained on all eight months and so has no month in that common held-out set. In-sample error is flat across budgets while out-of-sample error falls from 0.51 to 0.33, so the larger error of the small-budget models on unseen months reflects climatological distance rather than a failure to fit the training data.

### 5.10 Baseline Model Specifications

Quadratic interpolation provides the simplest baseline by fitting a quadratic function between the input and its equivalent output variable. While computationally trivial and requiring no training, this approach provides only $C^0$ continuity (continuous but not differentiable) and severely smooths all spatial features, making it a natural lower bound on expected downscaling performance. We also trained a random forest model with 100 trees as another baseline model for basic machine learning applications

An architecturally identical U-Net using exclusively the L1 reconstruction loss was also used as a baseline. This model maintains identical network capacity (architecture specified in Supplementary Table S4), training procedures, learning rate schedules (Supplementary Table S7), and optimization hyperparameters, differing only in the loss-function specification, with weights [w1, w2, w3, w4, w5] = [1, 0, 0, 0, 0], and in omitting the static geographic inputs, to show the improvement their integration brings.

We also attempted a generative diffusion baseline to gauge whether a score-based model could match CASPER at the same data budget. Following the one-stage conditional formulation of denoising diffusion, we trained a denoiser of comparable capacity to predict the noise added to the 42-channel high-resolution field, conditioned on the same 53-channel coarse input, on the identical eight-month training set and normalization used for CASPER, and sampled with the standard 1000-step ancestral reverse process. The per-step denoising loss decreased steadily over training, but running the full reverse-diffusion chain did not yield physically valid fields: the generated 2 m temperature drifted far outside any realizable range, giving errors of order 100 K against the WRF target, so the model could not be scored on the test splits. This is consistent with the well-documented dependence of generative diffusion models on large and diverse training corpora for stable sampling, and with our central finding that score-based training is ill-suited to the small-data regime this study targets. We therefore report the diffusion attempt qualitatively and exclude it from the matched-budget comparison in Table 2. An ablation study presented in the extended results shows the performance of models with different weights for loss components. Comparison between this L1-only model and the full CASPER system reveals the specific performance gains attributable to gradient preservation, multi-scale structural, and statistical losses.

**Supplementary Table S8: Selection-strategy experiment.** Held-out 2 m temperature RMSE (K; mean ± s.d. over three seeds) when fine-tuning the one- and four-month base models on N samples chosen by each selection strategy (Supplementary Note 5.4).

| Base | Strategy | N | Bulk RMSE (K) | Tail RMSE (K) |
|---|---|---|---|---|
| 1 month | coverage | 8 | 4.88 ± 0.09 | 4.58 ± 0.65 |
| 1 month | coverage | 32 | 3.99 ± 0.40 | 3.18 ± 0.06 |
| 1 month | mindist | 8 | 8.47 ± 0.42 | 3.33 ± 0.10 |
| 1 month | mindist | 32 | 5.98 ± 0.41 | 3.35 ± 0.32 |
| 1 month | nearest | 8 | 5.24 ± 0.16 | 7.58 ± 0.64 |
| 1 month | nearest | 32 | 5.05 ± 0.05 | 5.86 ± 0.15 |
| 1 month | random | 8 | 5.82 ± 0.73 | 5.57 ± 1.07 |
| 1 month | random | 32 | 4.26 ± 0.22 | 3.76 ± 0.38 |
| 4 months | coverage | 8 | 3.68 ± 0.07 | 4.54 ± 0.13 |
| 4 months | coverage | 32 | 3.16 ± 0.08 | 2.51 ± 0.10 |
| 4 months | mindist | 8 | 4.45 ± 0.08 | 3.30 ± 0.19 |
| 4 months | mindist | 32 | 4.43 ± 0.17 | 2.91 ± 0.13 |
| 4 months | nearest | 8 | 4.34 ± 0.11 | 6.04 ± 0.20 |
| 4 months | nearest | 32 | 3.53 ± 0.15 | 3.92 ± 0.42 |
| 4 months | random | 8 | 4.03 ± 0.27 | 3.55 ± 0.78 |
| 4 months | random | 32 | 3.29 ± 0.09 | 3.10 ± 0.16 |